\documentclass[conference]{IEEEtran}
\usepackage{xparse}
\usepackage{xspace}
\usepackage{amsmath}
\usepackage{txfonts}
\usepackage{enumitem}
\usepackage{cleveref}
\usepackage{placeins}
\usepackage{cite}
\crefname{enumi}{}{}
\crefformat{section}{\S#2#1#3}
\crefname{appendix}{App.}{Apps.}
\crefname{figure}{Fig.}{Figs.}
\crefname{table}{Table}{Tables}
\crefname{equation}{Eqn.}{Eqns.}
\crefname{algorithm}{Alg.}{Algs.}
\crefname{subfigure}{Fig.}{Figs.}
\crefname{appsec}{App.}{Apps.}

\renewcommand\thesubsubsection{\thesubsection.\arabic{subsubsection}}
\usepackage{titlesec}
\titlespacing*{\subsubsection}{0pt}{3pt}{0pt}
\titleformat{\subsubsection}[runin]
            {\normalfont\bfseries}
            {\thesubsubsection}
            {5pt}
            {}
            [.{\hspace*{1em}}]
\usepackage{colortbl}
\usepackage{collcell}
\usepackage{fp}
\usepackage{graphicx}
\usepackage{pgfplots}
\pgfplotsset{compat=1.18}
\usepackage{tikz}
\usepgfplotslibrary{groupplots}
\usepackage{multirow}
\usepackage{caption}
\usepackage{subcaption}
\usepackage{booktabs}
\usepackage{xurl}
\usepackage{soul}
\usepackage{wrapfig}
\usepackage{stfloats}
\usepackage{amssymb}
\usepackage{nameref}
\usepackage[ruled,vlined]{algorithm2e}
\SetKwProg{Upon}{Upon}{ do}{end}
\SetInd{0.29em}{0.29em}
\crefname{algocf}{Algorithm}{Algorithms}
\Crefname{algocf}{Algorithm}{Algorithms}
\usetikzlibrary{patterns}
\usetikzlibrary{patterns.meta}
\usetikzlibrary{fit}
\usetikzlibrary{calc}
\usetikzlibrary{decorations.pathreplacing}

\DeclareMathOperator*{\argmin}{arg\,min}
\usepackage[T1]{fontenc}
\usepackage{PTSansNarrow}
\renewcommand{\mathsf}[1]{\textsf{#1}}
\begin{document}

\date{}

\title{Incentivizing Collaboration for Detection of Credential Database Breaches}

\author{
{\rm Mridu Nanda}\\
Duke University
\and
{\rm Michael K. Reiter}\\
Duke University
}

\maketitle

\NewDocumentCommand{\bidAlgoBoostrapLabel}{}{\textsf{P1a}}
\NewDocumentCommand{\bidAlgoLabel}{}{\textsf{P1b}}
\NewDocumentCommand{\rcvAlgoWarmupLabel}{}{\textsf{P2a$'$}}
\NewDocumentCommand{\rcvAlgoInitLabel}{}{\textsf{P2a}}
\NewDocumentCommand{\rcvAlgoLabel}{}{\textsf{P2b}}

\makeatletter
\newcommand{\algsteplabel}[2]{%
  \def\@currentlabel{#2}%
  \label{#1}%
}
\makeatother

\NewDocumentCommand{\ternary}{mmm}{%
  \left({#1 \mathbin{?} #2 \mathbin{:} #3}\right)%
}

\newcommand{\secs}{\ensuremath{\mathrm{s}}\xspace}
\newcommand{\percentile}{\textrm{pctile}\xspace}


\NewDocumentCommand{\setNotation}{g}{\ensuremath{\MakeUppercase{#1}}\xspace}
\NewDocumentCommand{\rvNotation}{g}{\ensuremath{\varmathbb{#1}}\xspace}
\NewDocumentCommand{\algNotation}{g}{\ensuremath{\mathcal{#1}}\xspace}
\NewDocumentCommand{\setSize}{g}{\ensuremath{\left|{#1}\right|}\xspace}
\NewDocumentCommand{\nats}{}{\ensuremath{\mathbb{N}}\xspace}
\NewDocumentCommand{\reals}{o o}%
  {\IfNoValueTF{#1}%
    {\ensuremath{\mathbb{R}}\xspace}%
    {\ensuremath{[{#1},{#2}]}\xspace}%
  }
\NewDocumentCommand{\prob}{g}{\ensuremath{\mathbf{Pr}\mathopen{}\left({#1}\right)\mathclose{}}\xspace}
\NewDocumentCommand{\expv}{g}{\ensuremath{\mathbf{E}\mathopen{}\left({#1}\right)\mathclose{}}\xspace}
\newcommand{\cset}[3]{\ensuremath{#1\{}{#2}\ensuremath{\;#1|} \ifmmode{\;}\fi {#3}\ensuremath{#1\}}\xspace}
\NewDocumentCommand{\median}{}{\ensuremath{\mathsf{med}}\xspace}
\NewDocumentCommand{\targetIndicator}{}{\ensuremath{\ast}\xspace}
\NewDocumentCommand{\attacker}{}{\ensuremath{a}\xspace}
\NewDocumentCommand{\propLabel}{}{\ensuremath{\mathsf{p}}\xspace}
\NewDocumentCommand{\exhaustiveLabel}{}{\ensuremath{\mathsf{e}}\xspace}
\NewDocumentCommand{\targetSite}{o}{\ensuremath{s^{\targetIndicator}\IfNoValueF{#1}{_{#1}}}\xspace}
\NewDocumentCommand{\site}{o}{\ensuremath{s\IfNoValueF{#1}{_{#1}}}\xspace}
\NewDocumentCommand{\siteAlt}{o}{\ensuremath{\hat{s}\IfNoValueF{#1}{_{#1}}}\xspace}
\NewDocumentCommand{\siteAltAlt}{o}{\ensuremath{\tilde{s}\IfNoValueF{#1}{_{#1}}}\xspace}
\NewDocumentCommand{\siteAltAltAlt}{o}{\ensuremath{\bar{s}\IfNoValueF{#1}{_{#1}}}\xspace}

\NewDocumentCommand{\user}{o}{\ensuremath{u\IfNoValueF{#1}{_{#1}}}\xspace}
\NewDocumentCommand{\siteSet}{o}%
   {\IfNoValueTF{#1}%
     {\ensuremath{\setNotation{S}}\xspace}%
     {\ensuremath{{#1}.\mathsf{sites}}\xspace}%
   }
\NewDocumentCommand{\userSet}{o}%
   {\IfNoValueTF{#1}%
     {\ensuremath{\setNotation{U}}\xspace}%
     {\ensuremath{{#1}.\mathsf{users}}\xspace}%
   }
\NewDocumentCommand{\vulnSet}{o}%
   {\IfNoValueTF{#1}%
     {\ensuremath{\setNotation{V}}\xspace}%
     {\ensuremath{{#1}.\mathsf{vulnUsers}}\xspace}%
   }
\NewDocumentCommand{\peerSet}{o}%
   {\IfNoValueTF{#1}%
     {\ensuremath{\siteSet \setminus \{\targetSite\}}\xspace}%
     {\ensuremath{\siteSet \setminus \{{#1}\}}\xspace}%
   }
\NewDocumentCommand{\auction}{o o}{\ensuremath{a\IfNoValueF{#1}{_{#1}}\IfNoValueF{#2}{[{#2}]}}\xspace}
\NewDocumentCommand{\pairedAuctionsSet}{o}{\ensuremath{\setNotation{A}\IfNoValueF{#1}{_{#1}}}\xspace}
\NewDocumentCommand{\indicatorFn}{g}{\ensuremath{\mathbf{1}\mathopen{}\left({#1}\right)\mathclose{}}\xspace}
\NewDocumentCommand{\configVar}{}{%
   \ensuremath{\theta}\xspace
 }
\NewDocumentCommand{\constrainedConfig}{}{%
 \ensuremath{\psi}\xspace
}
\NewDocumentCommand{\ultraConstrainedConfig}{}{%
 \ensuremath{\psi^*}\xspace
}
\NewDocumentCommand{\capacity}{g}{\ensuremath{{#1}.\mathsf{cap}}\xspace}
\NewDocumentCommand{\capacityCoeff}{g}{\ensuremath{#1.\mathsf{capC}}\xspace}
\NewDocumentCommand{\allocTo}{g g}{\ensuremath{{#1}.\mathsf{allocTo}\mathopen{}\left({#2}\right)\mathclose{}}\xspace}
\NewDocumentCommand{\weight}{g g}{\ensuremath{{#1}.\mathsf{weight}\mathopen{}\left({#2}\right)\mathclose{}}\xspace}
\NewDocumentCommand{\baselineWeight}{g g}{\ensuremath{{#1}.\mathsf{baseWt}\mathopen{}\left({#2}\right)\mathclose{}}\xspace}
\NewDocumentCommand{\blendedWeight}{g g}{\ensuremath{{#1}.\mathsf{blendWt}\mathopen{}\left({#2}\right)\mathclose{}}\xspace}
\NewDocumentCommand{\nmbrSites}{}{\ensuremath{n}\xspace}
\NewDocumentCommand{\siteIdx}{}{\ensuremath{i}\xspace}
\NewDocumentCommand{\siteIdxAlt}{}{\ensuremath{i'}\xspace}
\NewDocumentCommand{\allocationVal}{}{\ensuremath{k}\xspace}
\NewDocumentCommand{\allocationValAdj}{}{\ensuremath{k'}\xspace}
\NewDocumentCommand{\allocationValPrev}{}{\ensuremath{k''}\xspace}
\NewDocumentCommand{\aggressionLevel}{g}{\ensuremath{{#1}.\mathsf{aggression}}\xspace}
\NewDocumentCommand{\slack}{}{\ensuremath{\mathsf{slack}}\xspace}
\NewDocumentCommand{\cutLine}{g}{\ensuremath{{#1}.\mathsf{cutline}}\xspace}
\NewDocumentCommand{\foresight}{g}{\ensuremath{{#1}.\mathsf{foresight}}\xspace}
\NewDocumentCommand{\lookahead}{g}{\ensuremath{{#1}.\mathsf{lookahead}}\xspace}
\NewDocumentCommand{\PSILabel}{}{\ensuremath{\mathsf{psi}}\xspace}
\NewDocumentCommand{\PSICALabel}{}{\ensuremath{\mathsf{psica}}\xspace}

\NewDocumentCommand{\risk}{g o o}{%
  \IfNoValueTF{#2}%
    {\ensuremath{{#1}.\mathsf{risk}}\xspace}%
    {\ensuremath{{#1}.\mathsf{risk}\mathopen{}({#2}, {#3})\mathclose{}}\xspace}%
}

\NewDocumentCommand{\riskPerBid}{g g}{%
    {\ensuremath{{#1}.\mathsf{risk}\mathopen{(}{#2}\mathclose{)}}\xspace}%
}

\NewDocumentCommand{\maxRisk}{}{\ensuremath{\mathsf{maxRisk}}\xspace}

\NewDocumentCommand{\normRisk}{g g g}{%
    {\ensuremath{{#1}.\mathsf{normRisk}\mathopen{(}{#2}, {#3}\mathclose{)}}\xspace}%
}

\NewDocumentCommand{\normRiskPerBid}{g g}{%
    {\ensuremath{{#1}.\mathsf{normRisk}\mathopen{(}{#2}\mathclose{)}}\xspace}%
}

\NewDocumentCommand{\costPerBid}{g g}{\ensuremath{{#1}.\mathsf{cost}({#2})}\xspace}
\NewDocumentCommand{\normCostPerBid}{g g}{\ensuremath{{#1}.\mathsf{normCost}({#2})}\xspace}
\NewDocumentCommand{\cost}{g}{\ensuremath{{#1}.\mathsf{costTotal}}\xspace}
\NewDocumentCommand{\normCost}{g}{\ensuremath{{#1}.\mathsf{normCostTotal}}\xspace}

\NewDocumentCommand{\dodgeEvent}{g g g g}{\ensuremath{\mathsf{dodge}\mathopen{}({#1}, {#2}, {#3}, {#4})\mathclose{}}\xspace}
\NewDocumentCommand{\attackerGainRV}{g
g}{\ensuremath{\rvNotation{L}_{{#1},{#2}}}\xspace}
\NewDocumentCommand{\avgRisk}{g g}{\ensuremath{{#1}.\mathsf{avgRisk}\mathopen{}({#2})\mathclose{}}\xspace}
\NewDocumentCommand{\avgAllocFrom}{g
g}{\ensuremath{{#1}.\mathsf{allocFrom}\mathopen{}({#2})\mathclose{}}\xspace}
\NewDocumentCommand{\smoothingFactor}{g}{\ensuremath{{#1.\mathsf{smf}}}\xspace}
\NewDocumentCommand{\popularity}{}{\ensuremath{\targetSite.\mathsf{pop}}\xspace}
\NewDocumentCommand{\id}{g}{\ensuremath{{#1}.\mathsf{id}}\xspace}
\NewDocumentCommand{\intersectionSize}{}{\ensuremath{n'}\xspace}
\NewDocumentCommand{\minSlotIntersectionVal}{}{\ensuremath{m}\xspace}
\NewDocumentCommand{\nmbrUsers}{}{\ensuremath{\ell}\xspace}
\NewDocumentCommand{\nmbrBids}{}{\ensuremath{r_{\mathsf{max}}}\xspace}
\NewDocumentCommand{\bidIdx}{}{\ensuremath{r}\xspace}
\NewDocumentCommand{\bidIdxAlt}{}{\ensuremath{r'}\xspace}
\NewDocumentCommand{\firstbidIdx}{}{\ensuremath{\bidIdx_{\ref{alg:prop-rsp:next-bid}}}\xspace}
\NewDocumentCommand{\secondbidIdx}{}{\ensuremath{\bidIdx_{\ref{alg:prop-rsp:next-bid},
2}}\xspace}
\NewDocumentCommand{\biddingSeq}{}{\ensuremath{\tau}\xspace}
\NewDocumentCommand{\stuffingAttempts}{}{\ensuremath{f}\xspace}
\NewDocumentCommand{\prop}{}{\ensuremath{\mathsf{proportional}}\xspace}
\NewDocumentCommand{\exhaustive}{}{\ensuremath{\mathsf{exhaustive}}\xspace}
\NewDocumentCommand{\Exhaustive}{}{\ensuremath{\mathsf{Exhaustive}}\xspace}
\NewDocumentCommand{\Greedy}{}{Greedy\xspace}
\NewDocumentCommand{\greedy}{}{greedy\xspace}
\NewDocumentCommand{\siteSize}{m}{\ensuremath{\mathsf{#1}}\xspace}

\NewDocumentCommand{\fracImprovement}{g}{
  \ensuremath{
    \mathsf{fracAdv}
    \IfValueT{#1}{_{#1}}
    \mathopen{}\mathclose{}
  }
  \xspace
}
\NewDocumentCommand{\absImprovement}{g}{
  \ensuremath{
    \mathsf{absAdv}
    \IfValueT{#1}{_{#1}}
    \mathopen{}\mathclose{}
  }
  \xspace
}
\NewDocumentCommand{\improvementFreq}{g}{
  \ensuremath{
    \mathsf{advFreq}
    \IfValueT{#1}{_{#1}}
    \mathopen{}\mathclose{}
  }
  \xspace
}

\NewDocumentCommand{\runtime}{g}{\textsf{time}\ensuremath{\mathopen{(}\text{#1})\mathclose{}}\xspace}
\NewDocumentCommand{\monReqGenOp}{}{\textsf{AmnesiaReqGen}\xspace}

\NewDocumentCommand{\boolTrue}{}{\textit{true}\xspace}
\NewDocumentCommand{\boolFalse}{}{\textit{false}\xspace}
\NewDocumentCommand{\pValue}{}{\ensuremath{p}\xspace}
\NewDocumentCommand{\effectSize}{}{\ensuremath{\eta^2}\xspace}
\NewDocumentCommand{\hStat}{}{\ensuremath{H}\xspace}
\NewDocumentCommand{\stirlingTerm}{}{\ensuremath{g}\xspace}
\NewDocumentCommand{\StirlingApprox}{}{\ensuremath{S}\xspace}
\NewDocumentCommand{\ErrorTerm}{o}{\ensuremath{E_{\mathrm{total}}\IfValueT{#1}{^{#1}}}\xspace}
\NewDocumentCommand{\mathEpsilon}{}{\ensuremath{\epsilon}\xspace}
\NewDocumentCommand{\mathDelta}{}{\ensuremath{\delta}\xspace}

\ExplSyntaxOn
\cs_new:Npn \intensity #1
   {\fp_eval:n {\MinIntensity + (1.0-\MinIntensity) * (((\MaxNumber-{#1})/(\MaxNumber-\MinNumber)))}
   }
\ExplSyntaxOff
\FPeval{\MinIntensity}{0.5}   
\FPeval{\MinNumber}{0.0}
\FPeval{\MaxNumber}{0.0}
\newcommand{\ApplyGradientX}[1]{\cellcolor[gray]{\intensity{#1}}{\raggedleft ${#1}$}}
\newcolumntype{X}{>{\collectcell\ApplyGradientX}c<{\endcollectcell}}

\ExplSyntaxOn
\cs_new:Npn \intensityA #1
   {\fp_eval:n {\MinIntensity + (1.0-\MinIntensity) * (((\MaxNumberA-{#1})/(\MaxNumberA-\MinNumberA)))}
   }
\cs_new:Npn \intensityB #1
   {\fp_eval:n {\MinIntensity + (1.0-\MinIntensity) * (((\MaxNumberB-{#1})/(\MaxNumberB-\MinNumberB)))}
   }
\ExplSyntaxOff

\FPeval{\MinIntensity}{0.5}   
\FPeval{\MinNumberA}{0.339}
\FPeval{\MaxNumberA}{0.602}
\FPeval{\MinNumberB}{0.068}
\FPeval{\MaxNumberB}{0.134}

\newcommand{\ApplyGradientA}[1]{\cellcolor[gray]{\intensityA{#1}}{\raggedleft ${#1}$}}
\newcommand{\ApplyGradientB}[1]{\cellcolor[gray]{\intensityB{#1}}{\raggedleft ${#1}$}}
\newcolumntype{A}{>{\collectcell\ApplyGradientA}c<{\endcollectcell}}
\newcolumntype{B}{>{\collectcell\ApplyGradientB}c<{\endcollectcell}}

\ExplSyntaxOn
\cs_new:Npn \intensityC #1
   {\fp_eval:n {\MinIntensity + (1.0-\MinIntensity) * (((\MaxNumberC-{#1})/(\MaxNumberC-\MinNumberC)))}
   }
\cs_new:Npn \intensityD #1
   {\fp_eval:n {\MinIntensity + (1.0-\MinIntensity) * (((\MaxNumberD-{#1})/(\MaxNumberD-\MinNumberD)))}
   }
\ExplSyntaxOff

\FPeval{\MinNumberC}{0.620}
\FPeval{\MaxNumberC}{0.681}
\FPeval{\MinNumberD}{0.213}
\FPeval{\MaxNumberD}{0.263}

\newcommand{\ApplyGradientC}[1]{\cellcolor[gray]{\intensityC{#1}}{\raggedleft ${#1}$}}
\newcommand{\ApplyGradientD}[1]{\cellcolor[gray]{\intensityD{#1}}{\raggedleft ${#1}$}}
\newcolumntype{C}{>{\collectcell\ApplyGradientC}c<{\endcollectcell}}
\newcolumntype{D}{>{\collectcell\ApplyGradientD}c<{\endcollectcell}}

\begin{abstract}
Decoy passwords, or ``honeywords,'' alert a site to its breach if
entered in a login attempt on that site.  However, an attacker can
identify a user-chosen password from among the decoys, without
alerting the site to its breach, via credential stuffing, i.e.,
entering the stolen passwords at another site where a user reused her
password.  Prior work thus proposed that sites monitor for the entry
of their honeywords at other sites, but the incentives for sites to
participate in this monitoring remain unclear. In this paper, we
propose and evaluate an algorithm by which sites can exchange
monitoring favors. Through a model-checking analysis, we show that a
site can improve its ability to detect its own breach when it
increases the monitoring effort it expends for others.  We quantify
how key parameters impact detection effectiveness and their
implications for deploying a monitoring ecosystem. Finally, we
evaluate our algorithm on a breached credential dataset, demonstrating
effectiveness at scale.
\end{abstract}

\section{Introduction}
\label{sec:introduction}
According to Verizon~\cite[Fig.~15]{Verizon:2024:DBIR} and IBM
\cite[Fig.~7]{IBM:2024:Cost}, the most prevalent initial attack vector
causing data breaches continues to be stolen credentials.  Moreover,
the sources of these stolen credentials are often themselves data
breaches---over 20\% of all data breaches compromise
credentials~\cite[Fig.~24]{Verizon:2024:DBIR}. Data breaches are
notoriously slow to be identified (207 days by one recent
estimate~\cite{OneCloud:2024:Detection}) and those stemming from
stolen credentials are even harder to discover, requiring an average
of 229 days after initial compromise~\cite[Fig.~8]{IBM:2024:Cost}.
These trends will likely persist, given the ubiquity of passwords as a
primary authentication mechanism~\cite{Hunt:2018:Killer}.

Honeywords~\cite{Juels:2013:Honeywords} seek to shrink the window
between the attacker's use of breached passwords and the defender's
realization that it has been breached.  Honeywords are decoy passwords
created by the defender and stored alongside user-chosen passwords in
its credential database; login attempts using honeywords alert the
site to its breach, since legitimate users do not know them. An
attacker who breaches the password database and attempts to harvest
its accounts can sidestep detection only if it can determine which of
the passwords associated with each account is the user-chosen one.
Methods to select honeywords to make this difficult for the attacker
have been the subject of much research
(e.g.,~\cite{Juels:2013:Honeywords, Erguler:2016:Flatness,
Akshima:2019:Honeywords, Dionysiou:2021:Honeygen,
Wang:2022:Honeywords, Chakraborty:2022:Honeyword,
Huang:2024:Exposed}).

Regardless of the honeyword-generation method, however, a reliable way
for an attacker to separate the user-chosen password from the
honeywords is to attempt to use these passwords at another site where
the same user has an account.  
If the user reused her password (or a similar one) there--—as users
often do~\cite{Radwan:2025:Password, Lastpass:2022:Psychology,
Kim:2024:PassREfinder}, even despite password-manager
support~\cite{Mayer:2022:Managers}, differing composition
policies~\cite{Seitz:2017:Differences,Nisenoff:2023:Two}, and forced
resets after breaches~\cite{Habib:2018:User, Pal:2019:Beyond}---then
the password that works at the remote site will almost certainly be
the user-chosen password at the breached site. Subsequent
honeyword-system designs (e.g.,~\cite{Wang:2021:Amnesia,
Wang:2024:Bernoulli}) have thus developed methods by which a site can
remotely monitor the login attempts at other sites for entry of its
honeywords.  This monitoring, however, consumes nontrivial resources
at sites where monitoring occurs and can exacerbate the load induced
on those sites by credential-stuffing campaigns, which can reach
denial-of-service volumes in some cases
(e.g.,~\cite{Lemos:2021:Stuffing}).

Our credential ecosystem is therefore held captive by misaligned
incentives: remote monitoring of login attempts for honeyword use is
\textit{necessary} to overcome a key vulnerability of honeywords---and
thus to unlock their adoption---and yet remote monitoring requires
that each site invest potentially significant resources \textit{to
protect others}.  In this paper, we offer a way out.  The key insight
of this work is that the dependence between sites (reused passwords)
that poses a risk to one site's breach-detection capability is
symmetric: common users at the two sites provide accounts at each site
whose honeywords can be sidestepped by stuffing credentials at the
other, without the risk of alerting either site to its breach.  In
this paper we leverage this insight to develop a simple and scalable
algorithm to support an ecosystem of sites, each self-interested, to
barter monitoring favors so as to enable each site to improve its
\textit{own} breach-detection capability by increasing the amount of
monitoring it performs \textit{for others}.  By aligning incentives,
we transform a perceived altruistic burden into a self-interested
gain, opening a practical path to honeyword adoption.

The technical challenges to developing such an algorithm are several.
For example, the algorithm must accommodate sites with varying
resource constraints so as to not dissuade smaller sites from
participating in the ecosystem. It must also support decentralized
participation based solely on local information, since each site must
retain control over its resource-allocation decisions and protect
sensitive internal information. And most importantly, the algorithm
should effectively elevate the ability of a site to detect its own
breach when its credentials are stuffed elsewhere---even though
sites cannot know when attacks will occur, where the attacker will
stuff, or how aggressive an attacker may be.  We construct a bidding
algorithm that, we show, achieves these goals and, moreover, yields
risk for each site comparable to the best (myopic) bidding strategy it
could adopt.  We further show that decreasing risk according to our
metric translates to improved security for sites, against an attacker
who attempts to harvest the accounts of a previously breached site
through strategic stuffing attempts at peers.

Our strategy to evaluate the efficacy of our algorithm begins with
model-checking across a wide array of parameters including user
account distributions, site strategies, and attacker aggression. Model
checking exhaustively explores site and attacker interactions to
identify worst-case outcomes, and so suffices to characterize
incentives under strategic behavior. This analysis revealed the key
factors that drive our algorithm's near-optimal performance and
resilience to strategic attackers, offering guidance for real-world
deployment (e.g., preventing sites from predicting future bidders). We
complement our model-checking analysis by simulating real-world
conditions using user account distributions and password reuse rates
from a previously breached dataset covering $\approx$ 8000 sites and
$\approx$74 million accounts.

In summary the contributions of this work are:
\begin{itemize}[nosep,leftmargin=1em,labelwidth=*,align=left]
    \item \textbf{Algorithm design:} We design the first algorithm for
      collaboratively detecting credential database breaches by
      bartering monitoring resources across sites
      (\cref{sec:auction:prop-response}), providing the economic basis
      for honeyword deployment at scale.
    \item \textbf{Incentive and security properties:} Using model
      checking, we show that our algorithm offers near-optimal risk
      reduction under diverse conditions (\cref{sec:auction:eval}). We
      also show this risk reduction directly translates to stronger
      protection against an attacker who tries to evade detection by
      stuffing breached credentials at peer sites
      (\cref{sec:attacker:eval}), and that the ecosystem is
      self-sustaining: sites are incentivized to invest in protecting
      \textit{others} to boost their \textit{own} security, creating
      cyclic improvements (\cref{sec:attacker:larger}). While the
      highest at-risk sites drive early gains, our algorithm
      incentivizes broad participation, so sites with differing
      resource constraints achieve comparable protection.
    \item \textbf{Real-world validation:} We evaluate our algorithm
      using a publicly leaked breach dataset, validating security
      through simulations parameterized by real user account
      distributions and password-reuse rates (\cref{sec:data-driven}).
      In this way, we demonstrate the algorithm's effectiveness under
      realistic conditions.
    \item \textbf{System design insights:} We use the model-checking
      findings to inform the architecture of a credential
      breach-detection ecosystem (\cref{sec:discussion}).
\end{itemize}

\section{Related Work}

\subsection{Breach Detection}
Breach discovery today works primarily by scanning for breached
datasets across various sources, such as the dark web, where the data
might be advertised for sale or simply exposed.  Once a breach is
discovered, breach alerting services help affected users and the
breached organizations become aware of their exposure. While such
compromised credential checking services
(e.g.,~\cite{Hunt:HaveIBeenPwned, Pullman:2019:Checkup,
Lauter:2021:Monitor, Pal:2022:MightIGetPwned}) have seen wide
deployment, they do not \textit{discover} breaches. Credential
database breaches that are not advertised or exposed, but instead are
used directly by the attackers who breach them to harvest accounts,
remain invisible to these defenses.

Honey accounts~\cite{DeBlasio:2017:Tripwire, Terry:2025:Honey} and
honeywords are deceptive techniques to leverage an attacker's
account-harvesting efforts to discover credential database breaches
(indicated by honey-account or honeyword use).  A honey account must
be difficult to distinguish from real accounts to yield a high true
detection rate, and ensuring a quantifiable rate of false detections
\textit{for database breaches} (versus another form of attack on its
password, e.g., online guessing) remains a challenge, especially when
the account's password is shared with another site (as in the Tripwire
study~\cite{DeBlasio:2017:Tripwire}).  True- and false-detection rates
for honeywords have received considerably more study
(e.g.,~\cite{Juels:2013:Honeywords, Erguler:2016:Flatness,
Akshima:2019:Honeywords, Dionysiou:2021:Honeygen,
Wang:2022:Honeywords, Chakraborty:2022:Honeyword, Wang:2024:Bernoulli,
Huang:2024:Exposed}).  Some works have proposed to monitor attacker
efforts to reduce the true-detection rate of honeywords by first
trying them in login attempts at other sites~\cite{Wang:2021:Amnesia,
Wang:2024:Bernoulli}, though none have tackled how to motivate sites
to support this monitoring, which is our focus here.

\subsection{Resource Allocation}
\label{sec:related-work:resource-allocation}
Our goal is to devise a mechanism for allocating computational
resources so that when a site invests resources to secure its peers
(i.e., by enabling remote monitoring) it gains protection in return.
Such allocations cannot be dictated by a centralized ``social
planner,'' as a planner would require access to a site's private, changing
preferences and constraints, and would impose decisions on how
a site should expend its own computational resources when enabling
remote monitoring for peers—decisions that sites are neither willing
nor able to delegate. Instead, we aim to devise a peer-to-peer
mechanism that allows sites to directly trade computational resources. 

This goal echoes prior work on P2P content distribution, notably the
BitTorrent protocol \cite{Cohen:2003:BitTorrent}, where agents
strategically allocate upload bandwidth to maximize their own download
speed, creating an incentive-driven exchange. Most relevant to our
work is Levin et al.~\cite{Levin:2008:BitTorrent}, which models
BitTorrent as an auction and presents \textit{proportional response
strategy} as an alternative auction-clearing mechanism, in which a
client uploads content to a peer at a rate proportional to the rate at
which the peer provided content to it. Instead of exchanging
bandwidth, our sites barter monitoring favors, providing
proportionally more favors (within resource constraints) to a peer
from whom they receive more. Unlike Levin et al., which study a
performance-oriented setting with interchangeable resources and
tolerable slowdowns, our setting is security-sensitive, where brief
monitoring lapses can expose vulnerabilities and monitoring favors are
not equally valuable. 

Proportional response is known to converge to \emph{market
equilibrium} in some settings, including Fisher markets (fixed-budget
buyers, dedicated sellers) with
synchronous~\cite{Zhang:2011:Proportional} and
asynchronous~\cite{Kolumbus:2023:Asynchronous} updates, and exchange
economies (each participant is both a buyer and seller) under
synchronous updates with homogeneous valuations in a bartering
model~\cite{Wu:2007:PropResponse} and with heterogeneous valuations in
a money-based model~\cite{Branzei:2021:Proportional}. The closest
setting to ours---an exchange economy with heterogeneous valuations in
a bartering model---does not converge under synchronous
updates~\cite{Branzei:2021:Proportional}, let alone in the
asynchronous bidding we consider.  Regardless, convergence to market
equilibrium is not the right goal here. Since an attacker can exploit
allocations at any point, security must be evaluated throughout, as we
do here with our model-checking analysis (\cref{sec:attacker}).

While market equilibrium indicates that the economy has cleared,
\emph{Nash equilibrium} indicates that no site can do better by
changing strategy---an orthogonal property \cite[Sec.
1.3]{Osborne:1994:Game}. 
Levin, et al.~\cite{Levin:2008:BitTorrent}
show that proportional response fails this notion, however: a site
that concentrates effort on higher-return peers does better than one
that spreads allocations proportionally. As such, following Levin, et
al.~\cite{Levin:2008:BitTorrent}, we empirically bound a site's gain
when it unilaterally deviates to a myopic best-response
strategy~(\cref{sec:auction}).


\subsection{Cooperative Security}
Interdependent security models, introduced by Kunreuther and
Heal~\cite{Kunreuther:2003:Interdependent}, examine how a
self-interested agent's security decisions affect peers. One class
focuses on defender-defender interactions, where agents balance
reducing their own risk with minimizing security investment costs
through strategic interaction with peers~\cite{Lelarge:2008:Local,
Miura:2008:Security, Jiang:2010:Bad, Abdallah:2022:Tasharok}. Another
class includes attacker-defender dynamics, analyzing how adversarial
behavior shapes security decisions~\cite{Nguyen:2009:Stochastic,
Lou:2017:Multidefender, Hota:2018:Game, Abdallah:2020:Behavioral,
Oruganti:2023:Impact}. Like these models, our work assumes sites are
rational but selfish defenders: they make strategic security
investments to maximize their own security while minimizing costs, but
do not engage in overtly malicious behavior or collusion. However,
unlike these models---which assume agents can independently achieve
their security goals, perhaps with greater investment---detecting the
remote entry of honeywords requires collaboration as a baseline.

A related but distinct area is collaborative intrusion detection,
where agents improve the accuracy of their own intrusion detection
assessment by sharing alerts and information with peers. Prior work
has explored system-level mechanisms for exchanging alerts
\cite{Wu:2003:CIDS, Yegneswaran:2004:Domino, Janakiraman:2003:Indra,
Fung:2016:Facid, Duma:2006:Trust}. In contrast, our work builds on
existing remote monitoring mechanisms \cite{Wang:2021:Amnesia,
Wang:2024:Bernoulli} and focuses on the incentives of exchanging
monitoring favors. Zhu et al.~\cite{Zhu:2012:GUIDEX} and
Fung~\cite{Fung:2013:IDN} also consider incentive-driven resource
sharing, but do not quantify the resulting security improvements,
which is central to our approach.

\section{Inter-Organization Model}
\label{sec:auction}
We consider a collection \siteSet of $\setSize{\siteSet} = \nmbrSites$
\textit{sites}, and a collection \userSet of $\setSize{\userSet} =
\nmbrUsers$ \textit{users}.\footnote{See \cref{tab:notation} after the
appendices for a summary of definitions.} Each user has an
\textit{account} at one or more of the sites.  We use $\userSet[\site]
\subseteq \userSet$ to denote the set of users with accounts at site
$\site \in \siteSet$, and $\siteSet[\user] \subseteq \siteSet$ to
denote the set of sites at which user $\user \in \userSet$ has
accounts.  Naturally, $\user \in \userSet[\site]$ if and only if
$\site \in \siteSet[\user]$.  Each account is protected using a
password selected by its user.

The password database at site \site includes, for each account, a
user-chosen password as well as a number of \textit{honeywords}, of
which the user is unaware. The honeywords for an account together with
the user-chosen password are called the \textit{sweetwords} for the
account.  Honeywords exist to alert \site to the breach of its
password database by an attacker, in that a login attempt at \site
using a honeyword for the attempted account is evidence of the breach
of \site's database.  For this reason, an attacker who breaches
\site's password database instead stuffs the sweetwords for an account
at the same user's accounts at \textit{other} sites, in the hopes of
determining the user-chosen password due to its reuse at those other
sites~\cite{Das:2014:Tangled, Pearman:2017:Habitat, Wang:2018:Domino}.

To counter this threat, a site \site can ask another site \siteAlt to
monitor for the entry of \site's honeywords in login attempts at
\siteAlt.  The mechanics of how this monitoring is done without
exposing \site's sweetwords to \siteAlt, and without exposing to \site
any other passwords used in login attempts at \siteAlt, are described
in prior works~\cite{Wang:2021:Amnesia, Wang:2024:Bernoulli} and are
not our concern here.  Rather, we abstract this process as follows:
\site poses a \textit{monitoring request} to \siteAlt, which names an
account for which logins should be monitored.  If \siteAlt accepts
this monitoring request, then any incorrect login attempt to the
account named in the request will generate a \textit{monitoring
response} to the site \site that created it. If the password used in
that login attempt at \siteAlt is a honeyword for the same user's
account at \site, then \site learns the honeyword used and can treat
it as if it were attempted locally, for the sake of breach detection.

Each site \site has a \textit{monitoring capacity} $\capacity{\site}
\in \nats$, which is the number of \textit{monitoring slots} that each
host one monitoring request. Each site is rational in the sense that
desires to trade its slots for those at other sites that most
effectively help it detect its own database breach. To do so, each
site occasionally issues a \textit{bid}.

Formally, when issuing a bid, site \site chooses nonnegative integers
$\allocTo{\site}{\siteAlt} \in \nats$ for each peer $\siteAlt \in
\peerSet[\site]$ such that $\sum_{\siteAlt \in \peerSet[\site]}
\allocTo{\site}{\siteAlt} \le \capacity{\site}$. The bid commits \site
to host monitoring requests in up to $\allocTo{\site}{\siteAlt}$ slots
for \siteAlt and simultaneously solicits up to that many monitoring
requests from \siteAlt (see \cref{fig:bidding-model}). Sites issue bids sequentially, and this
bidding sequence is indexed by \textit{bid index} \bidIdx. A site
\site can respond to bids issued by peers $\siteAlt \in
\peerSet[\site]$ by issuing a new bid when \site next appears in the
bidding sequence; the allocations $\allocTo{\siteAlt}{\site}$ visible
to \site at \bidIdx are those included by bids with index $< \bidIdx$.

\begin{figure}[t]
\centering
\begin{tikzpicture}[scale=0.65, transform shape]
    
    \draw[->, thick] (0, 0.95) -- (4.2, 0.95);
    \node[anchor=south west, font=\small] at (0, 0.95) {time};
    
    \node[anchor=east, font=\small] at (0, 0.6) {$\site[1]$};
    \node[anchor=east, font=\small] at (0, 0.3) {$\site[2]$};
    \node[anchor=east, font=\small] at (0, 0.0) {$\site[3]$};
    
    \draw[gray] (0, 0.6) -- (4.0, 0.6);
    \draw[gray] (0, 0.3) -- (4.0, 0.3);
    \draw[gray] (0, 0.0) -- (4.0, 0.0);
    
    \fill (0.5, 0.6) circle (2pt);
    \fill (1.1, 0.3) circle (2pt);
    \fill (1.7, 0.0) circle (2pt);
    \fill (2.3, 0.3) circle (2pt);
    \fill (2.9, 0.0) circle (2pt);
    \fill (3.5, 0.6) circle (2pt);
    
    \node[anchor=base west, font=\small] at (-0.5, -0.7) {$\slack = 1$};
    \node[anchor=base west, font=\small] at (1.0, -0.7) {$\biddingSeq = \langle \site[1], \site[2], \site[3], \site[2], \site[3], \site[1] \rangle$};
    
    \draw[dashed, thick, gray] (4.8, 1.4) -- (4.8, -0.9);
    
    \draw[->, thick] (5.4, 0.95) -- (9.6, 0.95);
    \node[anchor=south west, font=\small] at (5.4, 0.95) {time};
    
    \node[anchor=east, font=\small] at (5.4, 0.6) {$\site[1]$};
    \node[anchor=east, font=\small] at (5.4, 0.3) {$\site[2]$};
    \node[anchor=east, font=\small] at (5.4, 0.0) {$\site[3]$};
    
    \draw[gray] (5.4, 0.6) -- (9.4, 0.6);
    \draw[gray] (5.4, 0.3) -- (9.4, 0.3);
    \draw[gray] (5.4, 0.0) -- (9.4, 0.0);
    
    \fill (5.9, 0.6) circle (2pt);   
    \fill (6.3, 0.3) circle (2pt);   
    \fill (6.9, 0.6) circle (2pt);   
    \fill (7.3, 0.3) circle (2pt);   
    \fill (7.9, 0.0) circle (2pt);   
    \fill (8.5, 0.6) circle (2pt);   
    \fill (9.1, 0.3) circle (2pt);   
    
    \node[anchor=base west, font=\small] at (5.0, -0.7) {$\slack = 2$};
    \node[anchor=base west, font=\small] at (6.5, -0.7) {$\biddingSeq = \langle \site[1], \site[2], \site[1], \site[2], \site[3], \site[1], \site[2] \rangle$};
    
\end{tikzpicture}
\vspace{0.5em}
\hrule
\vspace{0.5em}

\begin{minipage}{0.95\columnwidth}
\small
\textbf{At bid index $\bidIdx$ ($\site = \biddingSeq[\bidIdx]$):} \textbf{foreach} $\siteAlt \in \peerSet[\site]$ \textbf{do}\\[0.2em]
\hspace*{1.5em}$\site$ decides $\allocTo{\site}{\siteAlt}$ using bidding strategy\\
\hspace*{1.5em}$\site$ commits $\allocTo{\site}{\siteAlt}$ to $\siteAlt$\\
\hspace*{1.5em}$\siteAlt$ returns $\allocTo{\site}{\siteAlt}$ monitoring requests to $\site$
\end{minipage}

\caption{An auction with $\nmbrSites = 3$ sites. Dots indicate when
each site bids, producing bidding sequence $\biddingSeq$. The $\slack$
parameter characterizes bidding sequences by bounding the max
difference in the number of bids across sites (see
\cref{sec:auction:sequence}). At each bid index $\bidIdx$, the site
$\site = \biddingSeq[\bidIdx]$ executes the protocol.}
\label{fig:bidding-model}
\end{figure}

The focus of this paper is to develop a \textit{bidding strategy} that
rational sites can use to trade monitoring slots with peers. The
proposed bidding strategy should have several desirable properties to
incentivize adoption.  It should reward reciprocity, so sites that
contribute more monitoring capacity receive more slots in return. It
should ensure fairness by allocating similar numbers of slots to
similarly at-risk peers, and by including smaller or less popular
sites. Crucially, the strategy must be locally computable, allowing
each site to operate independently based on its own information, and
scalable to enable deployment across large networks of sites.

\subsection{Site Threat Model}
\label{sec:auction:model}
We assume that each site \site is rational in wishing to use its bids
to provide itself the best chance of detecting a breach of its own
password database. Aside from manipulating the allocations
$\{\allocTo{\site}{\siteAlt}\}_{\siteAlt \in \peerSet[\site]}$ to
accomplish this, however, we assume each site \site respects
monitoring requests that peers deploy to it in accordance with those
allocations. (Prior work~\cite{Wang:2021:Amnesia} also discussed how
\siteAlt could audit \site to ensure it does so.)  That is, sites will
not deliberately ignore monitoring requests, which would be
reputation-damaging given the known peer identities typical of
threat-intelligence sharing communities
\cite{Meta:2025:ThreatExchange, CTA:2025:Membership,
RENISAC:Membership}. However, sites remain strategically
self-interested: they will bid to maximize the security benefit they
receive while minimizing the monitoring burden they bear---a threat
model standard in market mechanism design (e.g., stock exchanges, ad
auctions~\cite{IAB:2025:ADMAP}) and interdependent security models (e.g.,
\cite{Miura:2008:Security}) where participants follow rules while
optimizing individual returns. This creates potential for misaligned
incentives, e.g., all sites preferring to be monitored rather than to
monitor others, which our mechanism design addresses. 

Each site \site knows the set \siteSet of sites, as well as its own
users \userSet[\site]. However, \site is not privy to the allocations
that other peers receive, nor does it know other peers' capacities. We
consider two possibilities regarding how much information \site has
about the accounts at a peer \siteAlt.  Either \site knows the
membership of $\userSet[\site] \cap \userSet[\siteAlt]$, which we
presume it would learn by running a private set intersection
(\PSILabel) protocol~\cite{Pinkas:2018:PSI} with \siteAlt, or \site
knows only $\setSize{\userSet[\site] \cap \userSet[\siteAlt]}$, which
it could learn by running a PSI cardinality (\PSICALabel) protocol
(e.g.,~\cite{Kissner:2005:PSICA, DeCristofaro:2012:PSICA,
Debnath:2015:PSICA, Egert:2015:PSICA, Davidson:2017:PSICA}) with
\siteAlt. For simplicity, we assume that all sites have the same
privacy level, and so use the same protocol type with all of their
peers. Cross-organization private set computation over user
identifiers has been deployed at industry scale between separate
businesses~\cite{Ion:2020:Deploying, IAB:2025:ADMAP}, though
deployment might require user consent as per legislation like GDPR and
CCPA. In the rest of the main text, we focus on the \PSILabel setting,
and defer the discussion of \PSICALabel to
\cref{sec:prop-response-receiving-psica}.

\subsection{Risk}
\label{sec:auction:risk}

The allocations received from another site \siteAlt are valuable to
site \site, since they provide an opportunity to deploy that many
monitoring requests to \siteAlt.  But not all peer allocations equally
enhance \site's ability to detect its own breach. For example, if
\site and \siteAlt share no users, then allocations from \siteAlt are
not useful to \site; the attacker will never stuff sweetwords stolen
from \site at \siteAlt, as they provide no help in harvesting accounts
at \site. Thus, the extent to which peer allocations reduce \site's
risk depends entirely on the mechanics of credential stuffing, which
hinge on user overlap and password reuse.

Site \site measures the usefulness of an allocation
\allocTo{\siteAlt}{\site} by the amount of defense it provides for the
users that it shares with \siteAlt, since those are the only users
whose accounts will be stuffed at \siteAlt to harvest accounts at
\site. More precisely, consider an attacker stuffing \stuffingAttempts
accounts at \siteAlt for users in $\userSet[\site] \cap
\userSet[\siteAlt]$ (whose membership the attacker knows, as assumed
in \cref{sec:attacker:model}) after \site has deployed $\allocationVal
= \allocTo{\siteAlt}{\site}$ monitor requests to \siteAlt (for users
unknown to the attacker). Let
\dodgeEvent{\site}{\siteAlt}{\allocationVal}{\stuffingAttempts} be the
event that none of the monitor requests deployed by \site to \siteAlt
was for one of the \stuffingAttempts users whose accounts the attacker
stuffs at \siteAlt. We define a random variable
\attackerGainRV{\stuffingAttempts}{\allocationVal}, termed the
\textit{defender loss}
when stuffing these \stuffingAttempts accounts, by
\[
\attackerGainRV{\stuffingAttempts}{\allocationVal} =
\left\{\begin{array}{ll}
  \stuffingAttempts & \mbox{if $\dodgeEvent{\site}{\siteAlt}{\allocationVal}{\stuffingAttempts}$} \\
  0 & \mbox{otherwise}
  \end{array}\right.
\]
That is, if the \stuffingAttempts stuffing attempts at \siteAlt do not
overlap with the (up to) \allocationVal accounts monitored by \site,
then the attacker gains the accounts of the \stuffingAttempts users it
stuffed at \siteAlt. Otherwise, the attacker's stuffing attempts
risked alerting \site to its breach, and so we estimate the attacker's
gain as nothing. We call this the defender loss because it models the
attack from $s$'s perspective. When $s$ is alerted to its breach, the
attacker gains nothing—equivalent to not stuffing—since $s$ can prompt
remedial actions (e.g., password resets) that would invalidate reused
credentials.

The \textit{risk} that \site incurs from an allocation
$\allocationVal = \allocTo{\siteAlt}{\site}$ is:
\begin{align}
\risk{\site}[\siteAlt][\allocationVal]
& = \max_{0 \leq \stuffingAttempts \leq \intersectionSize} \expv{\attackerGainRV{\stuffingAttempts}{\allocationVal}}
\label{eqn:risk}
\end{align}
where $\intersectionSize = \setSize{\userSet[\site] \cap
  \userSet[\siteAlt]}$.  Since \intersectionSize of \site's accounts
are vulnerable to stuffing attempts at \siteAlt, it will be useful
to also define risk as a fraction of \intersectionSize, i.e.,
\begin{align}
  \normRisk{\site}{\siteAlt}{\allocationVal} & = \frac{\risk{\site}[\siteAlt][\allocationVal]}{\setSize{\userSet[\siteAlt] \cap \userSet[\site]}}
  \label{eqn:normRisk}
\end{align}

Since $\expv{\attackerGainRV{\stuffingAttempts}{\allocationVal}} =
\stuffingAttempts \times
\prob{\dodgeEvent{\site}{\siteAlt}{\allocationVal}{\stuffingAttempts}}$,
to compute \cref{eqn:risk}, \site needs to quantify
\prob{\dodgeEvent{\site}{\siteAlt}{\allocationVal}{\stuffingAttempts}}.
While \site cannot predict the attacker's choice of \stuffingAttempts,
as it depends on which accounts have already been harvested, the
accuracy with which it can compute
\prob{\dodgeEvent{\site}{\siteAlt}{\allocationVal}{\stuffingAttempts}}
for any fixed \siteAlt, \allocationVal, \stuffingAttempts depends on
what it knows about $\userSet[\site] \cap \userSet[\siteAlt]$.  In the
\PSILabel case, where \site knows the membership of $\userSet[\site]
\cap \userSet[\siteAlt]$, \site can compute
\prob{\dodgeEvent{\site}{\siteAlt}{\allocationVal}{\stuffingAttempts}}
exactly, as
\begin{align}
  \prob{\dodgeEvent{\site}{\siteAlt}{\allocationVal}{\stuffingAttempts}}
  & = \frac{{\intersectionSize - \stuffingAttempts \choose \min\{\intersectionSize, \allocationVal\}}}{{\intersectionSize \choose \min\{\intersectionSize, \allocationVal\}}}
\label{eqn:psiCatchProb}
\end{align}
\noindent
provided that \site deploys monitor requests to \siteAlt for
$\min\{\intersectionSize, \allocationVal\}$ accounts in
$\userSet[\site] \cap \userSet[\siteAlt]$ chosen uniformly at random.
The numerator is the number of ways that \site could deploy requests
for $\min\{\intersectionSize, \allocationVal\}$ accounts from the
$\intersectionSize - \stuffingAttempts$ that the attacker does not
stuff, whereas the denominator is the number of ways it could deploy
requests for $\min\{\intersectionSize, \allocationVal\}$ accounts from
all \intersectionSize.  

Whenever \site receives a new allocation $\allocationVal =
\allocTo{\siteAlt}{\site}$ from \siteAlt, we assume that \site deploys
not only \allocationVal monitors to \siteAlt, but also redeploys
monitors to each other site, in accordance with the allocation last
received from each. We quantify the risk that \site incurs at bid
index \bidIdx in an auction of \nmbrBids bids (i.e., $1 \le \bidIdx
\le \nmbrBids$) as
\begin{align}
  \riskPerBid{\site}{\bidIdx} & = \sum_{\siteAlt \in \peerSet[\site]} \risk{\site}[\siteAlt][\allocationVal]
  \label{eqn:perBidRisk} \\
  \normRiskPerBid{\site}{\bidIdx} & = \sum_{\siteAlt \in \peerSet[\site]} \normRisk{\site}{\siteAlt}{\allocationVal}
  \label{eqn:perBidNormRisk}
\end{align}
where $\allocationVal = \allocTo{\siteAlt}{\site}$ is the allocation
to \site specified by the most recent bid issued by \siteAlt with bid
index $\leq \bidIdx$. 

\subsection{Bidding sequence}
\label{sec:auction:sequence}

We assume that before placing a bid, a site knows all allocations
previously received from peers. We are agnostic to the bidding order
(\cref{sec:discussion} gives examples) and instead characterize
sequences by \slack, which bounds the difference between the most and
fewest bids by any site. When $\slack = \infty$, there is no
constraint; when $\slack = 1$, a site may only bid again after all
others have bid at least as many times.

\subsection{Exhaustive bidding strategy}
\label{sec:auction:exhaustive}
Our goal is to develop a practical \textit{bidding strategy} for each
site \site, which will be our focus in
\cref{sec:auction:prop-response}. A ``practical'' strategy is one that
is computationally efficient and leverages only on the history locally
visible to the site issuing the bid. We aim for the practical bidding
strategy to remain competitive with an \exhaustive, though
impractical, strategy that always selects the ``best'' next move. We
will evaluate the practical bidding strategy by comparing the risk to
a particular site \targetSite when all sites use it, versus when only
\targetSite switches to an \exhaustive strategy (with others remaining
practical; c.f.,~\cite{Levin:2008:BitTorrent}). If the difference is
modest, we deem the practical strategy adequate. The \exhaustive
strategy is defined by the following parameters:

\begin{itemize}[nosep,leftmargin=1em,labelwidth=*,align=left]
\item \cutLine{\targetSite}: When $\cutLine{\targetSite} = \boolTrue$,
  \targetSite is excluded from the dictated bidding sequence and
  instead can choose to bid when it wants (subject to the \slack
  constraint), essentially ``cutting in line''.  If
  $\cutLine{\targetSite} = \boolTrue$ and \targetSite bids, then some
  other \site must bid before \targetSite is allowed to bid again,
  lest \targetSite defer others' bids indefinitely.  If
  $\cutLine{\targetSite} = \boolFalse$, then it must wait its turn in
  the bidding sequence to bid.
  \item \foresight{\targetSite}: A natural number that indicates the
  number of forthcoming bidders that \targetSite can predict
  correctly.  For example, if sites bid in a round-robin fashion, then
  \targetSite could predict the full sequence of bidders in advance.
  It could then use this information in determining what allocations
  to specify in its next bid if it is the next bidder or, if
  $\cutLine{\targetSite} = \boolTrue$ and \slack allows, it chooses to
  bid next.
\item \lookahead{\targetSite}: A natural number that indicates the
  depth beyond the next \foresight{\targetSite} bidders, to which
  \targetSite analyzes all possible bidding sequences. That is,
  \targetSite can explore all possible bidding sequences of length
  \lookahead{\targetSite}, beginning after the \foresight{\targetSite}
  bidders that it knows, to inform the allocations it
  specifies in its next bid or, if $\cutLine{\targetSite} = \boolTrue$
  and \slack allows, its choice of whether to insert a bid.  In this
  exploration, \targetSite considers every other possible bidder
  (subject to \slack) as equiprobable.  We always set
  $\lookahead{\targetSite} \ge 1$.
\end{itemize}
\Cref{fig:exhaustive_expl} illustrates these parameters. At
$\bidIdx=2$, $\targetSite$ predicts the next $\foresight{\targetSite}$
bidders and evaluates $\lookahead{\targetSite}$-length continuations.
If \targetSite cuts, it cannot bid at $\bidIdx=3$; if \targetSite
waits, \slack becomes tight, forcing it to bid at $\bidIdx=3$ and shifting
the subsequent sequence.

\begin{figure}[htbp]
    \centering
    \begin{tikzpicture}[
    label/.style={font=\small, align=left, anchor=east},
    note/.style={font=\scriptsize, color=black!60, align=left}
]

\node[label] at (-2.3, 0.5) {$\bidIdx  = 2$:};

\node[font=\normalsize, color=blue!55!black] at (-1.7, 0.5) {$\site_{1}$};
\node[font=\normalsize, color=blue!55!black] at (-1.0, 0.5) {$\site_{2}$};

\node[font=\normalsize, color=blue] at (-0.3, 0.5) {$\site_{3}$};
\node[font=\normalsize, color=blue] (root) at (0.5, 0.5) {$\site_{2}$};

\draw[decorate, decoration={brace, amplitude=2pt, mirror}, thin, blue!55!black] 
    (-1.8, 0.12) -- (-0.9, 0.12) node[midway, below=2pt, font=\tiny, blue!55!black] {past};
\draw[decorate, decoration={brace, amplitude=2pt, mirror}, thick, blue] 
    (-0.4, 0.12) -- (0.5, 0.12) node[midway, below=2pt, font=\tiny] {\foresight{\targetSite}};

\fill[purple!20, opacity=0.4] (-0.5, 0.1) rectangle (2.8, 1.2);

\node[font=\small, align=center] at (-1.9, 1.1) {$\targetSite$: cut or wait?};
\draw[-{stealth}, line width=0.8pt, black] (-1.9, 0.9) -- (-1.9, 0.75) -- (-0.65, 0.75) -- (-0.65, 0.6);

\node[font=\normalsize, color=red] (l1s1) at (1.5, 0.9) {$\site_{1}$};
\node[font=\normalsize, color=red] (l1s3) at (1.5, 0.65) {$\site_{3}$};
\node[font=\normalsize, color=red] (l1ss) at (1.5, 0.4) {$\targetSite$};

\draw[->, red, very thin] (root) -- (l1s1);
\draw[->, red, very thin] (root) -- (l1s3);
\draw[->, red, very thin] (root) -- (l1ss);

\node[font=\normalsize, color=red] (l2s3) at (2.4, 1.05) {$\site_{3}$};
\node[font=\normalsize, color=red] (l2ss) at (2.4, 0.8) {$\targetSite$};

\draw[->, red, very thin] (l1s1) -- (l2s3);
\draw[->, red, very thin] (l1s1) -- (l2ss);

\draw[->, red, very thin] (l1s3) -- (2.2, 0.65);
\fill[red!30] (l1s3.east) -- (2.2, 0.75) -- (2.2, 0.55) -- cycle;
\draw[->, red, very thin] (l1ss) -- (2.2, 0.4);
\fill[red!30] (l1ss.east) -- (2.2, 0.5) -- (2.2, 0.3) -- cycle;

\draw[decorate, decoration={brace, amplitude=2pt, mirror}, thick, red] 
    (1.0, 0.12) -- (2.7, 0.12) node[midway, below=2pt, font=\tiny] {\lookahead{\targetSite}};

\begin{scope}[xshift=-1.5cm, yshift=-1.3cm]

\node[label] at (-0.8, 0.5) {If $\targetSite$ cut at $\bidIdx = 2$, \\ then at $\bidIdx  = 3$:};

\node[font=\normalsize, color=blue!55!black] at (-0.2, 0.5) {$\site_{1}$};
\node[font=\normalsize, color=blue!55!black] at (0.5, 0.5) {$\site_{2}$};
\node[font=\normalsize, color=blue!55!black] at (1.2, 0.5) {$\targetSite$};

\node[font=\normalsize] at (2.0, 0.5) {$\site_{3}$};
\node[font=\normalsize] at (2.8, 0.5) {$\site_{2}$};
\node[font=\normalsize] at (3.6, 0.5) {\ldots};

\draw[decorate, decoration={brace, amplitude=2pt, mirror}, thin, blue!55!black] 
    (-0.3, 0.12) -- (1.3, 0.12) node[midway, below=2pt, font=\tiny, blue!55!black] {past};

\end{scope}

\begin{scope}[xshift=0.2cm, yshift=-2.4cm]

\node[label] at (-2.5, 0.5) {If $\targetSite$ waited at $\bidIdx = 2$, \\ then at $\bidIdx  = 3$:};

\node[font=\small, align=center] at (0.8, 1.1) {$\targetSite$ must cut};
\draw[-{stealth}, line width=0.8pt, black] (0.8, 0.9) -- (0.8, 0.75) -- (-0.1, 0.75) -- (-0.1, 0.6);

\node[font=\normalsize, color=blue!55!black] at (-1.9, 0.5) {$\site_{1}$};
\node[font=\normalsize, color=blue!55!black] at (-1.2, 0.5) {$\site_{2}$};
\node[font=\normalsize, color=blue!55!black] at (-0.5, 0.5) {$\site_{3}$};

\node[font=\normalsize, color=blue] at (0.3, 0.5) {$\site_{2}$};
\node[font=\normalsize, color=blue] (root_bottom) at (1.1, 0.5) {$\site_{1}$};  
\fill[purple!20, opacity=0.4] (0.15, 0.1) rectangle (3.2, 0.85);

\draw[decorate, decoration={brace, amplitude=2pt, mirror}, thin, blue!55!black] 
    (-2.0, 0.12) -- (-0.2, 0.12) node[midway, below=2pt, font=\tiny, blue!55!black] {past};
\draw[decorate, decoration={brace, amplitude=2pt, mirror}, thick, blue] 
    (0.2, 0.12) -- (1.3, 0.12) node[midway, below=2pt, font=\tiny] {\foresight{\targetSite}};

\node[font=\normalsize, color=red] (l3s3) at (2.1, 0.7) {$\site_{3}$};
\node[font=\normalsize, color=red] (l3ss) at (2.1, 0.45) {$\targetSite$};

\draw[->, red, very thin] (root_bottom) -- (l3s3);
\draw[->, red, very thin] (root_bottom) -- (l3ss);

\node[font=\normalsize, color=red] (l4ss_from_s3) at (2.9, 0.7) {$\targetSite$};
\draw[->, red, very thin] (l3s3) -- (l4ss_from_s3);

\node[font=\normalsize, color=red] (l4s3) at (2.9, 0.38) {$\site_{3}$};
\draw[->, red, very thin]
  ($(l3ss.east |- l4s3.west)$) -- (l4s3.west);

\draw[decorate, decoration={brace, amplitude=2pt, mirror}, thick, red] 
    (1.7, 0.12) -- (3.2, 0.12) node[midway, below=2pt, font=\tiny] {\lookahead{\targetSite}};

\end{scope}

\end{tikzpicture}    
    \caption{An auction with $\nmbrSites = 4$ and $\slack = 1$ from
      the perspective of a site $\targetSite$ using the \exhaustive
      strategy. Here $\cutLine{\targetSite} = \boolTrue$,
      $\foresight{\targetSite} = 2$, and $\lookahead{\targetSite} =
      2$.  The shaded region highlights the next
      $\foresight{\targetSite} + \lookahead{\targetSite}$ bidders
      considered by $\targetSite$. \textbf{Top row:} At bid index
      $\bidIdx = 2$, $\targetSite$ decides whether to cut in line or
      wait, using $\foresight{\targetSite}$ to correctly predict the
      next two bidders and $\lookahead{\targetSite}$ to explore
      continuations. \textbf{Middle row:} If $\targetSite$ cuts and
      bids at $\bidIdx = 2$, the \slack constraint is satisfied, but
      consecutive cuts are prohibited, preventing $\targetSite$ from
      bidding again at $\bidIdx = 3$. \textbf{Bottom row:} If
      $\targetSite$ waits at $\bidIdx = 2$, the \slack constraint
      becomes tight, forcing $\targetSite$ to bid at $\bidIdx = 3$.
      The difference in subsequent bidders between the middle and
      bottom rows reflects how cutting or waiting at $\bidIdx = 2$
      shifts the underlying bidding sequence.}
    \label{fig:exhaustive_expl}
\end{figure}

For the \exhaustive \targetSite to make the ``best'' allocation after
looking forward $\foresight{\targetSite} + \lookahead{\targetSite}$
bids, we equip it with access to each peer's capacities and
allocations specified by peers' bids---information unavailable to
other sites. We limit the \exhaustive \targetSite to allocating in
fixed proportions of \capacity{\targetSite} (e.g., integer multiples
of $\lfloor\frac{\capacity{\targetSite}}{20}\rfloor$); otherwise, the
number of possible allocations would grow combinatorially with
\capacity{\targetSite}. Given fixed $\cutLine{\targetSite}$,
$\foresight{\targetSite}$, and $\lookahead{\targetSite}$, \targetSite
evaluates the tree of possible outcomes based on the foreseeable
\foresight{\targetSite} bidders followed by equiprobable
\lookahead{\targetSite} bidders, considering all potential bidding
positions if $\cutLine{\targetSite} = \boolTrue$. If it is
\targetSite's turn or if $\cutLine{\targetSite} = \boolTrue$, \slack
permits \targetSite to bid, and bidding now is optimal, then
\targetSite issues a bid that specifies the allocation minimizing
\risk{\targetSite} per \cref{eqn:perBidRisk}.

\subsection{Proportional response bidding strategy}

\label{sec:auction:prop-response}
Now we propose a practical bidding strategy.  We say the strategy is
``practical'' in that, unlike the \exhaustive strategy described in
\cref{sec:auction:exhaustive}, this strategy can be computed
efficiently\footnote{Please see \cref{sec:performance} for a
performance analysis of the proposed strategy.} with only local
information, namely a site \site's own capacity, the sequence of
allocations \site has received so far, and either
$\setSize{\userSet[\site] \cap \userSet[\siteAlt]}$ or
$\userSet[\site] \cap \userSet[\siteAlt]$ for $\siteAlt \in \peerSet$
depending on the \PSICALabel or \PSILabel setting, respectively.

The practical bidding strategy we explore is \textit{proportional
  response}, denoted \prop.  The \prop strategy has two steps:
  \cref{alg:prop-rsp-bid} specifies how a site computes the
  allocations it specifies in its bid and
  \cref{alg:prop-rsp-receive-psi} specifies how \site updates its
  state upon receiving an allocation in the \PSILabel setting. The
  corresponding update procedure for the \PSICALabel setting appears
  in \cref{sec:prop-response-receiving-psica}.

\begin{algorithm}[t!]
\caption{Proportional Response: Bidding}
\label{alg:prop-rsp-bid}

\Upon{site $\site$ bidding}{
  \textbf{\bidAlgoBoostrapLabel:} \If{some $\siteAltAlt \in \siteSet$
    has not yet placed a bid}{
    \algsteplabel{alg:prop-rsp:init-bid}{\bidAlgoBoostrapLabel}
    \ForEach{$\siteAlt \in \peerSet[\site]$}{ $\displaystyle
      \baselineWeight{\site}{\siteAlt} \gets \frac{1}{\nmbrSites-2} \left(1 -
      \frac{\risk{\site}[\siteAlt][1]}{\sum_{\siteAltAlt \in
          \peerSet[\site]} \risk{\site}[\siteAltAlt][1]}\right)$
      \label{line:baseline-weight}
      
      \vspace{2mm}
      
      $\allocTo{\site}{\siteAlt} \gets \left\lfloor \capacity{\site}
      \times \baselineWeight{\site}{\siteAlt} \right\rfloor$
      \label{line:baseline-alloc}
    }
    $\siteAltAltAlt \gets \argmin_{\siteAlt \in \peerSet[\site]} \baselineWeight{\site}{\siteAlt}$
    \label{line:lowest-baseline}
  }
  
      \textbf{\bidAlgoLabel:} \Else{%
        \algsteplabel{alg:prop-rsp:next-bid}{\bidAlgoLabel}%

        $\maxRisk \gets \max_{\siteAlt \in \peerSet[\site]}
        \avgRisk{\site}{\siteAlt}$
        \label{line:max-risk}

        \vspace{2mm}

        \ForEach{$\siteAlt \in \peerSet[\site]$}{ $
          \weight{\site}{\siteAlt} \gets \tfrac{1}{\nmbrSites-2} \left(1 -
          \tfrac{\avgRisk{\site}{\siteAlt}}{\sum_{\siteAltAlt \in
              \peerSet[\site]} \avgRisk{\site}{\siteAlt}}\right)$
          \label{line:prop-weight}

          $\displaystyle
          \blendedWeight{\site}{\siteAlt} \gets \left(\begin{array}{@{}l@{}}
            \left(\frac{1}{1+\maxRisk}\right) \times
            \baselineWeight{\site}{\siteAlt} \\
            +~\left(\frac{\maxRisk}{1+\maxRisk}\right) \times
            \weight{\site}{\siteAlt} \end{array}\right)$
          \label{line:blended-weight}

          \vspace{2mm}

        $\allocTo{\site}{\siteAlt} \gets \left\lfloor \capacity{\site}
          \times \blendedWeight{\site}{\siteAlt} \right\rfloor$
          \label{line:blended-alloc}
        }
      }
  $\allocTo{\site}{\siteAltAltAlt} \mathrel{+}=  \capacity{\site} - \sum_{\siteAlt \in \peerSet[\site]} \allocTo{\site}{\siteAlt}$
  \label{line:assign-leftover}
}
\end{algorithm}

\begin{algorithm}[t!]
\caption{Proportional Response: Receiving (\PSILabel)}
\label{alg:prop-rsp-receive-psi}
$\avgAllocFrom{\site}{\siteAlt} \gets \bot$ \\

\vspace{1ex}

\Upon{site \site receiving allocation \allocationVal from \siteAlt}{

  $\allocationValAdj \gets \min\{\allocationVal,
  \setSize{\userSet[\site] \cap \userSet[\siteAlt]}\}$
  
  \textbf{\rcvAlgoInitLabel:} \If{some $\siteAltAlt \in \siteSet$ has
    not yet placed a bid}{
    \algsteplabel{alg:prop-rsp:init-rcv-alloc}{\rcvAlgoInitLabel}
    $\avgRisk{\site}{\siteAlt} \gets
    \risk{\site}[\siteAlt][\allocationValAdj]$
    \label{line:init-rcv-alloc}
  } \textbf{\rcvAlgoLabel:} \Else{
    \algsteplabel{alg:prop-rsp:rcv-alloc}{\rcvAlgoLabel}
    $\allocationValPrev \gets \ternary{\avgAllocFrom{\site}{\siteAlt} =
      \bot}{\allocationValAdj}{\avgAllocFrom{\site}{\siteAlt}}$
    
    $\avgAllocFrom{\site}{\siteAlt} \gets \smoothingFactor{\site} \times
    \allocationValAdj + (1-\smoothingFactor{\site}) \times
    \allocationValPrev$
    
    $\avgRisk{\site}{\siteAlt} \gets
    \risk{\site}[\siteAlt][\avgAllocFrom{\site}{\siteAlt}]$
    \label{line:exp-smoothing}
  }
  \label{line:avg-risk}
}
\end{algorithm}

Below we identify properties essential for \prop in our setting. Some,
such as capacity feasibility, are standard in proportional response,
while others, such as allocation stability, address failure modes
specific to our model. These properties motivate the design choices
embodied in \cref{alg:prop-rsp-bid,alg:prop-rsp-receive-psi}; formal
statements and proofs appear in \cref{sec:prop-response-proofs}.
However, they are insufficient to characterize the risk or security
equilibria that arise under strategic interaction. We therefore
evaluate emergent behavior via model checking in subsequent sections.

\subsubsection{Allocation feasibility: A site must respect its
capacity constraint when constructing bids} This standard requirement
serves as a baseline for the properties that follow.

\textit{Enforcement in \cref{alg:prop-rsp-bid}:} In
\ref{alg:prop-rsp:init-bid}, \site computes
$\baselineWeight{\site}{\siteAlt}$, and in
\ref{alg:prop-rsp:next-bid}, \site computes
$\weight{\site}{\siteAlt}$; both are normalized to sum to one over
\site's peers. In \ref{alg:prop-rsp:next-bid}, the blended weights
$\{\blendedWeight{\site}{\siteAlt}\}_{\siteAlt}$ are defined as convex
combinations of $\baselineWeight{\site}{\siteAlt}$ and
$\weight{\site}{\siteAlt}$, and therefore also sum to one over \site's
peers. After scaling by $\capacity{\site}$ and flooring, the total
number of allocated slots is at most $\capacity{\site}$. Any leftover
capacity due to flooring is assigned to the peer $\siteAltAltAlt$
posing the greatest risk to \site, so that the final allocated
capacity equals $\capacity{\site}$.

\subsubsection{Capacity exhaustion: A site must fully utilize its
available capacity} Leaving capacity idle weakens reciprocity signals
that drive reallocation between peers.

\textit{Enforcement in \cref{alg:prop-rsp-bid}:} Any capacity left
unallocated after flooring (at most $\nmbrSites - 1$ slots) is
assigned to the peer posing the greatest risk to \site, ensuring full
utilization of $\capacity{\site}$.

\subsubsection{Monotonicity in risk: Allocations must be nonincreasing
  in risk} A site cannot directly determine how many of its users'
  accounts would be compromised at a peer under an attack, because
  that cost depends on the attacker's unknown strategy and timing (see
  \cref{sec:attacker}). Instead, a site uses risk as an observable
  proxy and allocates capacity in a manner that is nonincreasing as a
  function of risk.

\textit{Enforcement in \cref{alg:prop-rsp-bid}:} The weights \site
assigns to each peer are nonincreasing in that peer's risk: in
\ref{alg:prop-rsp:init-bid}, $\baselineWeight{\site}{\siteAlt}$
decreases as $\risk{\site}[\siteAlt][1]$ increases, and in
\ref{alg:prop-rsp:next-bid}, $\weight{\site}{\siteAlt}$ decreases as
$\avgRisk{\site}{\siteAlt}$ increases. Since
$\blendedWeight{\site}{\siteAlt}$ is a convex combination of these
weights, this ordering is preserved in the resulting allocations.
Assigning the leftover capacity (due to flooring) can relax the
ordering by at most $\nmbrSites - 1$ slots.

\subsubsection{Allocation smoothness: Similar normalized risk should
induce similar allocations} Small changes in normalized risk should
not induce large changes in allocations. Unsmooth allocation rules
allow minor risk perturbations to trigger abrupt reallocations,
leading to unstable behavior that can be exploited by a site using the
\exhaustive strategy through free-riding. Smooth allocations instead
support reciprocal escalation, since incremental increases in
allocation induce proportional and immediately observable increases in
the allocations received by peers.

\textit{Enforcement in \cref{alg:prop-rsp-bid}:} In
\ref{alg:prop-rsp:init-bid} and \ref{alg:prop-rsp:next-bid}, \site
computes $\baselineWeight{\site}{\siteAlt}$ and
$\weight{\site}{\siteAlt}$, respectively, as linear functions of each
peer's risk, normalized by the total risk incurred across its peers.
Because all peer weights in a bid share this normalization, small
differences in normalized risk translate into correspondingly small
differences in weights, and thus in the resulting allocations after
scaling by $\capacity{\site}$ and flooring.

\subsubsection{Allocation stability: Allocations must be robust when
many peers have zero risk} When many peers pose zero average risk and
only a few pose low but nonzero average risk, normalization by total
average risk amplifies vanishing differences. Zero-risk peers receive
uniform weight, while low-risk peers receive negligible weight despite
non-negligible contributions. This instability is exploitable: a site
posing negligible risk may receive little or no reciprocal allocation,
making its users attractive targets for stuffing at peers.

\textit{Enforcement in \cref{alg:prop-rsp-bid}:} Blended weight
$\blendedWeight{\site}{\siteAlt}$ is defined as a convex combination
of $\baselineWeight{\site}{\siteAlt}$ and $\weight{\site}{\siteAlt}$,
with the interpolation controlled by $\maxRisk$. So, when most peers
have zero risk and the remaining risks are small, the interpolation
anchors allocations to $\baselineWeight{\site}{\siteAlt}$, reflecting
bootstrapped risk estimates $\risk{\site}[\siteAlt][1]$. Then,
assuming peers have similar $\risk{\site}[\siteAlt][1]$ (and thus
similar $\baselineWeight{\site}{\siteAlt}$), they receive similar
allocations despite amplified differences in
$\weight{\site}{\siteAlt}$.

\subsubsection{Risk persistence: Risk estimates must reflect sustained
peer allocations} If a site's risk estimate depends only on the
previous allocation from a peer, that peer can allocate many slots to
the site just before the site bids, then stop allocating to the site
(free-riding on the slots the site allocated to it) until it
anticipates the site to bid again.

\textit{Enforcement in \cref{alg:prop-rsp-receive-psi}:} The
exponential smoothing rule uses \smoothingFactor{\site} to ensure that
once $\siteAlt$ stops allocating to $\site$, its influence on
$\site$'s future bids decays exponentially, so only sustained peer
allocations affect reciprocity.

\subsection{Evaluation}
\label{sec:auction:eval}
We evaluate the \prop strategy to understand whether: (1) a site can
reduce its risk by increasing its capacity; (2) unpopular sites incur
comparable risk to their popular peers; and (3) the \prop strategy
incurs comparable risk to the \exhaustive strategy. To this end, we
conduct model-checking experiments over the parameter space defining
the \exhaustive strategy, \prop strategy, and bidding sequences. Per
auction, one site \targetSite uses either the \exhaustive or \prop
strategy (denoted \targetSite[\exhaustiveLabel] or
\targetSite[\propLabel], respectively) while all peers
\peerSet[\targetSite] use the \prop strategy. We allow \targetSite to
set its parameters (marked with a ``$\targetIndicator$'')
independently of its peers. To examine \risk{\targetSite} across
varying environments, we define:

\begin{itemize}[nosep,leftmargin=1em,labelwidth=*,align=left]
\item \popularity: This parameter is defined over \reals[0][1] for
  \targetSite and controls how \siteSet[\user] is determined per
  $\user \in \userSet$. By varying \popularity we generate a range of
  user placements that model a \targetSite with varying popularity.
  Assigning each site a unique site identifier $\id{\site} \in \{1,
  \ldots, \nmbrSites\}$ such that $\id{\targetSite} = 1$, we set
  \[
  \Pr[\site \in \siteSet[\user]] = (1 - \popularity) \times \frac{\id{\site}}{\nmbrSites} + \popularity \times \left(1 - \frac{\id{\site}-1}{\nmbrSites}\right)
  \]
  So, when $\popularity = 1$, \targetSite is the most popular site:
  $\Pr[\targetSite \in \siteSet[\user]] = 1$, versus $\Pr[\site \in
    \siteSet[\user]] = 1/\nmbrSites$ for the least popular site \site
  with $\id{\site} = \nmbrSites$.  When $\popularity = 0$, \targetSite
  is the least popular site: $\Pr[\targetSite \in \siteSet[\user]] =
  1/\nmbrSites$, versus $\Pr[\site \in \siteSet[\user]] = 1$ for the
  most popular site \site with $\id{\site} = \nmbrSites$.  When
  $\popularity = 1/2$, sites are equally popular ($ \forall \site:
  \Pr[\site \in \siteSet[\user]] = \frac{\nmbrSites +
    1}{2\nmbrSites}$).
\item \capacityCoeff{\site}: This parameter is the \textit{capacity
  coefficient}, defined over [0,1], which determines the capacity
  \capacity{\site} of each site $\site \in \peerSet$ as
  $\capacity{\site} \gets \capacityCoeff{\site} \times
  \setSize{\userSet[\site]}$. This reflects the assumption that more
  popular sites (i.e., sites with more users) may have more resources
  to monitor logins for others. \targetSite can vary
  \capacityCoeff{\targetSite} independently.
\end{itemize}

We conduct model-checking experiments with $\nmbrSites = 4$ and
$\nmbrUsers = 400$, modeling each auction as a Markov decision
process. We built a model checker following the approach of
PRISM~\cite{Kwiatkowska:2011:Prism} with application-specific
optimizations to scale to our parameter space (full details in
\cref{sec:model_checking}). To implement the \exhaustive strategy we
varied \foresight{\targetSite[\exhaustiveLabel]} and
\lookahead{\targetSite[\exhaustiveLabel]} so that
$\foresight{\targetSite[\exhaustiveLabel]} +
\lookahead{\targetSite[\exhaustiveLabel]} \leq 3$, for scalability
(see \cref{sec:model_checking:sites} for more details), and
$\lookahead{\targetSite[\exhaustiveLabel]} > 1$, so
\targetSite[\exhaustiveLabel] will always consider future bidders when
deciding its allocation. We varied
$\cutLine{\targetSite[\exhaustiveLabel]} \in \{\boolTrue,
\boolFalse\}$.  
To implement the \prop strategy for $\site \in \peerSet$ we varied
$\smoothingFactor{\site} \in \{1, 0.75, 0.5, 0.25 \}$. Additionally,
for \targetSite[\propLabel] we varied $\smoothingFactor{\site} \in
\{1, 0.75, 0.5, 0.25 \}$.

We evaluated the bidding strategies over auctions of length $\nmbrBids
= 10$, focusing on the post-initialization portion of the bidding
sequence. Let $\firstbidIdx$ denote the first bid index when any
$\site \in \peerSet[\targetSite[\exhaustiveLabel]]$ executes
\ref{alg:prop-rsp:next-bid} in \cref{alg:prop-rsp-bid}. All results
reported here (and in \cref{sec:attacker:eval}) consider bid indices
$\bidIdx \in [\firstbidIdx, \firstbidIdx + \nmbrBids)$. We generated
bidding sequences by varying $\slack \in \{1,2,3,\infty\}$ and
$\cutLine{\targetSite[\exhaustiveLabel]} \in
\{\boolTrue,\boolFalse\}$, generating 30 distinct sequences over
$[\firstbidIdx, \firstbidIdx + \nmbrBids)$. We varied \popularity,
$\capacityCoeff{\site}$, and $\capacityCoeff{\targetSite}$ over five
values in $[0,1]$, and for each setting generated 30 independent user
configurations (and resulting capacities). These configurations were
reused across both \exhaustive and \prop auctions. In total, our
evaluation includes approximately $500{,}000{,}000$ auctions, ensuring
broad coverage of plausible scenarios.

Since increasing \popularity increases \setSize{\userSet[\targetSite]}
and so the at-risk users \setSize{\userSet[\targetSite] \cap
\userSet[\site]} at each \site, we compare
\riskPerBid{\targetSite}{\bidIdx} across different \popularity values
by examining \normRiskPerBid{\targetSite}{\bidIdx}
(\cref{eqn:perBidNormRisk}). We use the Kruskal-Wallis \hStat test to
assess how parameter settings affect median
\normRiskPerBid{\targetSite}{\bidIdx} and a Dunn test with Bonferroni
correction to assess the direction and significance of effects
(\pValue-values in parentheses)~\cite{Datatab:2025:Kwh}. We sample 30
bidding sequences and 30 configurations to ensure enough observations
for valid Kruskal-Wallis inference \cite{Lomuscio:2021:Kwh}. Our
findings follow.

\begin{wrapfigure}{r}{0.5\columnwidth}
  \vspace{-3ex}
  \centering
  \resizebox{0.5\columnwidth}{!}{
    \input{user_param_psi.tex}
  }
  \caption{\normRiskPerBid{\targetSite}{\bidIdx} by \popularity and
    \capacityCoeff{\targetSite}.  Boxes span 25--75 \percentile;
    whiskers span 5--95 \percentile; diamonds are means; red lines
    are medians.}
  \label{fig:varying_target_cap_and_user_param}
  \vspace{-2ex}
\end{wrapfigure}

\subsubsection{Sites have incentive to increase their capacities}
\label{sec:auction:eval:incentive}
\cref{fig:varying_target_cap_and_user_param} illustrates that for a
fixed \popularity, the distribution of
\normRiskPerBid{\targetSite}{\bidIdx} decreased ($\pValue < 10^{-8}$)
as \targetSite increased \capacityCoeff{\targetSite}. This shows
\targetSite's incentive to increase its capacity to help
\textit{others} to reduce its \textit{own} risk, regardless of its
popularity or strategy.

\subsubsection{Unpopular sites still receive protection} 
\label{sec:auction:eval:popularity}
\cref{fig:varying_target_cap_and_user_param} shows that
\normRiskPerBid{\targetSite}{\bidIdx} generally decreased ($\pValue <
10^{-8}$) as \popularity decreased, meaning less popular sites are
fairly included by their peers and receive comparable risk reduction.
Of course, an unpopular \targetSite is penalized if its capacity is
set too low, as shown in the uptick of
\normRiskPerBid{\targetSite}{\bidIdx} when a \targetSite with
$\popularity = 0$ sets $\capacityCoeff{\targetSite} = 0.2$ in
\cref{fig:varying_target_cap_and_user_param}.

\begin{figure}[t]
  \begin{center}
    \resizebox{\columnwidth}{!}{
    \begin{tabular}{c|c|*{4}{A}|*{4}{B}}
      & & \multicolumn{4}{c|}{\improvementFreq{\configVar}} &
      \multicolumn{4}{c}{\fracImprovement{\configVar}} \\
      \hline \hline
      \multirow{2}{*}{\cutLine{\targetSite}} &
      \multirow{2}{*}{\foresight{\targetSite}} 
      & \multicolumn{4}{c|}{\slack} & \multicolumn{4}{c}{\slack} \\ 
      & & \multicolumn{1}{c}{1} & \multicolumn{1}{c}{2} & \multicolumn{1}{c}{3} & \multicolumn{1}{c|}{$\infty$} 
      & \multicolumn{1}{c}{1} & \multicolumn{1}{c}{2} & \multicolumn{1}{c}{3} & \multicolumn{1}{c}{$\infty$} \\ \hline
      \multirow{2}{*}{\boolFalse} & 0 
      & 0.379 & 0.356 & 0.343 & 0.339 
      & 0.075 & 0.073 & 0.069 & 0.068 \\
      & 1 
      & 0.448 & 0.431 & 0.431 & 0.452
      & 0.080 & 0.080 & 0.075 & 0.080 \\ \hline
      \multirow{2}{*}{\boolTrue} & 0 
      & 0.515 & 0.480 & 0.462 & 0.452
      & 0.098 & 0.097 & 0.094 & 0.093 \\
      & 1 
      & 0.602 & 0.579 & 0.549 & 0.536
      & 0.134 & 0.125 & 0.114 & 0.110 \\
    \end{tabular}
    }
  \end{center}
  \caption{Comparison of \exhaustive and \prop strategies across
    \configVar constraints defined by combinations of
    \cutLine{\targetSite[\exhaustiveLabel]},
    \foresight{\targetSite[\exhaustiveLabel]}, and \slack. The left
    table shows \improvementFreq{\configVar} (\cref{eqn:improvement-freq}) and the right
    table shows \fracImprovement{\configVar}
    (\cref{eqn:frac-improvement}), both diminished when 
    $\configVar \gets \cutLine{\targetSite[\exhaustiveLabel]} =
    \boolFalse$, $\foresight{\targetSite[\exhaustiveLabel]} = 0$,
    $\slack = \infty$.}
  \label{fig:exhaustive_vs_prop_cfs_psi}
\end{figure}

\subsubsection{Limiting when a site can bid, and how much knowledge a
    site has of the bidding sequence, minimizes the risk improvement
    \exhaustive provides over \prop}
\label{sec:auction:eval:bidding}

To limit the advantage of an \exhaustive bidder and so reduce a \prop
bidder's incentive to change strategies, we focus on parameters
enforceable by global ecosystem settings:
\cutLine{\targetSite[\exhaustiveLabel]},
\foresight{\targetSite[\exhaustiveLabel]}, and \slack. Here we
identify values that minimize this advantage, deferring enforcement
details to \cref{sec:discussion:enforcement}.

Let \pairedAuctionsSet denote a set of auction pairs such that
$(\auction[\propLabel], \auction[\exhaustiveLabel]) \in
\pairedAuctionsSet$ implies that \auction[\propLabel] and
\auction[\exhaustiveLabel] were conducted with identical user
placements, site capacities, \slack, \smoothingFactor{\site}, and
bidding sequence (or template—i.e., the sequence excluding bids issued
by \targetSite when $\cutLine{\targetSite[\exhaustiveLabel]} =
\boolTrue$) but that \targetSite bid according to the \exhaustive
strategy in \auction[\exhaustiveLabel] and according to the \prop
strategy in \auction[\propLabel]. Let $\pairedAuctionsSet[\configVar]
\subseteq \pairedAuctionsSet$ be the subset of such pairs that satisfy
some further constraints, specified as \configVar. For various
conditions \configVar, we seek to quantify the fraction
\improvementFreq{\configVar} of bid indices \bidIdx in auction pairs
in \pairedAuctionsSet[\configVar] for which
\normRiskPerBid{\targetSite[\exhaustiveLabel]}{\bidIdx} is less than
(i.e., improved on) \normRiskPerBid{\targetSite[\propLabel]}{\bidIdx}
and, in those cases, the median absolute improvement
\absImprovement{\configVar} and the median relative improvement
\fracImprovement{\configVar}. Letting $\auction[][\bidIdx] =
\normRiskPerBid{\targetSite}{\bidIdx}$ at bid index \bidIdx in an
\nmbrBids-bid auction \auction (i.e., $\firstbidIdx \le \bidIdx \le
\nmbrBids$), then:
\begin{align}
  \improvementFreq{\configVar} &= \left(\frac{1}{\nmbrBids \setSize{\pairedAuctionsSet[\configVar]}}\right)
  \setSize{\cset{\Bigg}{(\auction[\propLabel][\bidIdx], \auction[\exhaustiveLabel][\bidIdx])}{\begin{array}{@{}l@{}} (\auction[\propLabel],\auction[\exhaustiveLabel]) \in \pairedAuctionsSet[\configVar]~\wedge \\
      \auction[\exhaustiveLabel][\bidIdx] < \auction[\propLabel][\bidIdx]
  \end{array}}} \label{eqn:improvement-freq} \\
  \absImprovement{\configVar} &= \median \cset{\Bigg}{\auction[\propLabel][\bidIdx] - \auction[\exhaustiveLabel][\bidIdx]}{\begin{array}{@{}l@{}} (\auction[\propLabel],\auction[\exhaustiveLabel]) \in \pairedAuctionsSet[\configVar]~\wedge \\
      \auction[\exhaustiveLabel][\bidIdx] < \auction[\propLabel][\bidIdx]
  \end{array}} \label{eqn:abs-improvement} \\
  \fracImprovement{\configVar} &= \median \cset{\Bigg}{\frac{\auction[\propLabel][\bidIdx] - \auction[\exhaustiveLabel][\bidIdx]}{\auction[\propLabel][\bidIdx]}}{\begin{array}{@{}l@{}} (\auction[\propLabel],\auction[\exhaustiveLabel]) \in \pairedAuctionsSet[\configVar]~\wedge \\
      \auction[\exhaustiveLabel][\bidIdx] < \auction[\propLabel][\bidIdx]
  \end{array}} \label{eqn:frac-improvement}
\end{align}

Among the globally enforceable parameters,
\cutLine{\targetSite[\exhaustiveLabel]} determines whether
\targetSite[\exhaustiveLabel] can control the timing of its bids. We
found that by disabling this choice (i.e., $\configVar \gets
\cutLine{\targetSite[\exhaustiveLabel]} = \boolFalse$), we reduced
\improvementFreq{\configVar} and reduced ($\pValue < 10^{-8}$)
\fracImprovement{\configVar} per
\foresight{\targetSite[\exhaustiveLabel]} and \slack, as shown in
\cref{fig:exhaustive_vs_prop_cfs_psi}.

The \foresight{\targetSite[\exhaustiveLabel]} parameter provides
\targetSite[\exhaustiveLabel] with information about the future
bidding sequence, and disabling it (i.e., $\configVar \gets
\foresight{\targetSite[\exhaustiveLabel]} = 0$) reduced
\improvementFreq{\configVar} and reduced ($\pValue < 10^{-8}$)
\fracImprovement{\configVar}, per $\cutLine{\targetSite}$ and
$\slack$.

The \slack parameter provides \targetSite[\exhaustiveLabel] with a
distribution over future bidders, and when \slack is tight,
\targetSite[\exhaustiveLabel] often identifies the next bidder
exactly. This trend is reflected in \cref{fig:exhaustive_vs_prop_cfs_psi},
where when $\configVar \gets \cutLine{\targetSite[\exhaustiveLabel]} =
\boolFalse \wedge \foresight{\targetSite[\exhaustiveLabel]} = 0$,
decreasing \slack resulted in higher \improvementFreq{\configVar}, and
also raised ($\pValue < 10^{-8}$) \fracImprovement{\configVar}. To
mitigate this advantage, we recommend setting no restrictions on slack
(i.e., $\slack = \infty$).

In summary, our findings suggest a system deployment where a site
cannot choose when to bid, cannot predict who will bid next, and
otherwise imposes no restrictions on possible bidding sequences.  We
denote these specific constraints as $\constrainedConfig \gets
\cutLine{\targetSite} = \boolFalse \wedge \foresight{\targetSite} = 0
\wedge \slack = \infty$. When considering the corresponding bids in
\pairedAuctionsSet[\constrainedConfig], we found
$\improvementFreq{\constrainedConfig} < 0.34$,
$\fracImprovement{\constrainedConfig} = 0.068$, and
\begin{align}
  \median \cset{\Bigg}{\frac{\auction[\propLabel][\bidIdx] -
    \auction[\exhaustiveLabel][\bidIdx]}{\auction[\propLabel][\bidIdx]}}{(\auction[\propLabel],\auction[\exhaustiveLabel])
    \in \pairedAuctionsSet[\constrainedConfig]} & = 0
\end{align}

This suggests that when all $\site \in \peerSet$ use the \prop
strategy, \targetSite has little incentive to deviate. While we cannot
enforce values for the remaining parameters (\smoothingFactor{\site},
\smoothingFactor{\targetSite[\propLabel]},
\lookahead{\targetSite[\exhaustiveLabel]}, \capacityCoeff{\site},
\capacityCoeff{\targetSite}) our evaulation under \constrainedConfig
shows that varying them either has negligible impact on
\targetSite[\propLabel]'s risk (see \cref{sec:auction:eval:past-bids})
or yields impractical gains for \targetSite[\exhaustiveLabel] (see
\cref{sec:auction:eval:impractical}).

\subsubsection{Keeping track of past allocations does not decrease a
  site's risk if it cannot choose when to bid and has limited
  knowledge about the next bidder}
\label{sec:auction:eval:past-bids}
Each \prop bidder uses \smoothingFactor{\site} to track average
allocations received per peer. However, under \constrainedConfig
constraints, we found that \smoothingFactor{\site} value employed by
\textit{other} sites does not affect
\normRiskPerBid{\targetSite}{\bidIdx} (\pValue = 0.566). We also found
that allowing \targetSite[\propLabel] to set $\smoothingFactor{\site}$
independently of its peers did not affect
\normRiskPerBid{\targetSite[\propLabel]}{\bidIdx} ($\pValue > 1 -
10^{-8}$).

\begin{figure}[t]
  \begin{center}
\begin{subfigure}[b]{0.75\columnwidth}
   \FPeval{\MinNumber}{0.00132}
   \FPeval{\MaxNumber}{0.00732}
   \resizebox{\textwidth}{!}{
   \begin{tabular}{@{\hspace{2pt}}r@{\hspace{2pt}}|*{5}{X}|*{3}{X}}
     & \multicolumn{5}{c|}{\capacityCoeff{\site}} & \multicolumn{3}{c}{\lookahead{\targetSite}} \\ 
     \multicolumn{1}{@{\hspace{2pt}}c@{\hspace{2pt}}|}{\capacityCoeff{\targetSite}} 
     & \multicolumn{1}{c}{0.2}
     & \multicolumn{1}{c}{0.4}
     & \multicolumn{1}{c}{0.6}
     & \multicolumn{1}{c}{0.8}
     & \multicolumn{1}{c|}{1.0}
     & \multicolumn{1}{c}{1}
     & \multicolumn{1}{c}{2}
     & \multicolumn{1}{c}{3} \\ \hline
1 &   0.00212 & 0.00212 & 0.00173 & 0.00150 & 0.00132 & 0.00170 & 0.00200 & 0.00202 \\
0.8 & 0.00309 & 0.00211 & 0.00200 & 0.00139 & 0.00202 & 0.00200 & 0.00203 & 0.00203 \\
0.6 & 0.00211 & 0.00201 & 0.00134 & 0.00227 & 0.00182 & 0.00210 & 0.00210 & 0.00210 \\
0.4 & 0.00408 & 0.00293 & 0.00295 & 0.00372 & 0.00277 & 0.00296 & 0.00296 & 0.00296 \\
0.2 & 0.00563 & 0.00489 & 0.00653 & 0.00732 & 0.00591 & 0.00578 & 0.00582 & 0.00592 \\
   \end{tabular}
   }
   \caption{$\ultraConstrainedConfig \gets \constrainedConfig \wedge
     \popularity = 1$
     ($\improvementFreq{\ultraConstrainedConfig} = 0.513$)}
   \label{fig:exhaustive_vs_prop_cl:0_psica}
 \end{subfigure}
\\ \vspace{0.25cm}
\begin{subfigure}[b]{0.75\columnwidth}
   \FPeval{\MinNumber}{0.00101}
   \FPeval{\MaxNumber}{0.01345}
   \resizebox{\textwidth}{!}{
   \begin{tabular}{@{\hspace{2pt}}r@{\hspace{2pt}}|*{5}{X}|*{3}{X}}
     & \multicolumn{5}{c|}{\capacityCoeff{\site}} & \multicolumn{3}{c}{\lookahead{\targetSite}} \\ 
     \multicolumn{1}{@{\hspace{2pt}}c@{\hspace{2pt}}|}{\capacityCoeff{\targetSite}} 
     & \multicolumn{1}{c}{0.2}
     & \multicolumn{1}{c}{0.4}
     & \multicolumn{1}{c}{0.6}
     & \multicolumn{1}{c}{0.8}
     & \multicolumn{1}{c|}{1.0}
     & \multicolumn{1}{c}{1}
     & \multicolumn{1}{c}{2}
     & \multicolumn{1}{c}{3} \\ \hline
1   & 0.00241 & 0.00158 & 0.00113 & 0.00101 & 0.00110 & 0.00129 & 0.00129 & 0.00129 \\
0.8 & 0.00308 & 0.00211 & 0.00127 & 0.00119 & 0.00157 & 0.00159 & 0.00159 & 0.00159 \\
0.6 & 0.00309 & 0.00212 & 0.00170 & 0.00139 & 0.00162 & 0.00172 & 0.00172 & 0.00172 \\
0.4 & 0.00308 & 0.00239 & 0.00170 & 0.00202 & 0.00279 & 0.00215 & 0.00212 & 0.00212 \\
0.2 & 0.00564 & 0.00355 & 0.00423 & 0.00654 & 0.01345 & 0.00539 & 0.00540 & 0.00540 \\
   \end{tabular}
   }
   \caption{$\ultraConstrainedConfig \gets \constrainedConfig \wedge
     \popularity = 0.75$
     ($\improvementFreq{\ultraConstrainedConfig} =
     0.442$)} \label{fig:exhaustive_vs_prop_cl:0.25_psica}
 \end{subfigure}
 
  \end{center}
  \caption{Each cell shows \absImprovement{\ultraConstrainedConfig} with lighter cells indicating less \exhaustive
    advantage. We omit $\popularity < 0.75$ due to space, though
    trends are similar. \absImprovement{\ultraConstrainedConfig} peaked when
    $\capacityCoeff{\targetSite} \ll \capacityCoeff{\site}$, and
    minimally improved at higher
    \lookahead{\targetSite[\exhaustiveLabel]}.}
  \label{fig:exhaustive_vs_prop_cl}
\end{figure}

\subsubsection{The costs of computing \exhaustive bids render it
  inferior to \prop, even in the rare cases where it can reduce
  \targetSite's risk}
\label{sec:auction:eval:impractical}

\cref{fig:exhaustive_vs_prop_cl} shows that per \popularity,
\targetSite[\exhaustiveLabel] maximized
\absImprovement{\constrainedConfig} when its capacity was scaled much
lower than its peers' capacities (i.e., $\capacityCoeff{\targetSite}
\ll \capacityCoeff{\site}$). This suggests
\targetSite[\exhaustiveLabel] can reduce its risk moderately relative
to \targetSite[\propLabel], but only by severely restricting
monitoring costs. Yet, computing the \exhaustive strategy is a
considerable common-case cost---far more than simply increasing
\capacityCoeff{\targetSite} to match its peers, as shown in
\cref{sec:performance}. Our efforts to efficiently approximate
\exhaustive to actualize this advantage have been unsuccessful,
leaving this direction as possible future research.

While increasing \lookahead{\targetSite[\exhaustiveLabel]} reduced
($\pValue < 10^{-8}$) \absImprovement{\constrainedConfig}, the effect
size ($4.17 \times 10^{-6}$) was negligible. We attribute significance
to dataset size and conclude \lookahead{\targetSite[\exhaustiveLabel]}
has little practical impact in decreasing
\targetSite[\exhaustiveLabel]'s risk, as
\Cref{fig:exhaustive_vs_prop_cl} shows no notable increase in
\absImprovement{\constrainedConfig} per \popularity. Since computing
the optimal allocation grows exponentially in
\lookahead{\targetSite[\exhaustiveLabel]}, simply raising
\capacityCoeff{\targetSite} is a more practical way to reduce risk. We
conclude that \prop is the preferred strategy.

\section{Attacker Model}
\label{sec:attacker}

We now extend the analysis in \cref{sec:auction} to assess the
security of the allocations produced by the \prop strategy, against an
attacker attempting to harvest a site's accounts by first stuffing
sweetwords stolen from that site at peer sites.  We restrict our
attention to sites using the \prop strategy, since, as concluded in
\cref{sec:auction:eval}, preventing a site from predicting forthcoming
bidders ($\foresight{\targetSite[\exhaustiveLabel]} = 0$), from
choosing when to bid ($\cutLine{\targetSite[\exhaustiveLabel]} =
\boolFalse$), and imposing no restrictions on bidding sequences
($\slack = \infty$) leaves a site little reason to deviate from \prop,
if trying to minimize its risk.

\subsection{Attacker Threat Model}
\label{sec:attacker:model}

The attacker who breaches a site \targetSite's password database and
stuffs these credentials elsewhere is rationally motivated, seeking to
harvest as many accounts as possible at \targetSite while dodging
\targetSite's monitoring efforts.  Sites---in particular, \targetSite,
despite having its credential database stolen---continue to behave as
assumed in \cref{sec:auction:model} and, in particular, follow \prop
bidding. This threat model characterizes common practice while also
focusing on the challenge of \textit{detecting} the passive breach of
the database. If, instead, we allowed the attacker to actively corrupt
\targetSite to behave differently (in our algorithm or elsewhere),
then this would provide additional features by which \targetSite's
administrators might detect the breach had occurred. So, we do not
consider Byzantine behaviors here.

We make two (conservative) allowances for the attacker.
\begin{itemize}[nosep,leftmargin=1em,labelwidth=*,align=left]
\item If the attacker dodges \targetSite's monitoring at \site, then
  the attacker has successfully harvested all those user accounts it
  stuffed at \site, and therefore at \targetSite (and every $\site \in
  \peerSet$).  This allowance is optimistic for the attacker, but not
  unreasonably so, given users' tendencies to reuse the same or
  similar passwords across sites~\cite{Das:2014:Tangled,
  Pearman:2017:Habitat, Wang:2018:Domino}.  That is, once the attacker
  has harvested these accounts at \site, it can easily harvest the
  same user's accounts at \targetSite, too~\cite{Huang:2024:Exposed}.
\item We allow the attacker to know all parameters of \prop per $\site
  \in \siteSet$ (i.e., \slack, \smoothingFactor{\site}, and
  \capacity{\site}, as these could inferred from bidding times and
  allocations) and to know \userSet[\site] for every site $\site \in
  \siteSet$ and so \siteSet[\user] for every user $\user \in
  \userSet$, reflecting the routine leakage of email-based identifiers
  through breach aggregators.
\end{itemize}
Despite these allowances, the attacker cannot observe for which users a site
deploys monitoring requests, as requests exist as ephemeral
state protected by the underlying protocol~\cite{Wang:2021:Amnesia,
Wang:2024:Bernoulli}.

\subsection{Attacker's Stuffing Strategy}
\label{sec:attacker:strat}

Much like the \exhaustive \targetSite detailed in
\cref{sec:auction:exhaustive}, the attacker \attacker
is characterized by certain parameter choices.
\begin{itemize}[nosep,leftmargin=1em,labelwidth=*,align=left]
\item \foresight{\attacker}: A natural number denoting how many
forthcoming bidders the attacker can predict correctly.
\item \lookahead{\attacker}: A natural number indicating how many
steps beyond \foresight{\attacker} the attacker can explore. The attacker
evaluates all sequences of this length, treating eligible
bidders as equiprobable. We always set $\lookahead{\attacker} \ge 1$.
\item \aggressionLevel{\attacker}: A float $\in [0,1]$ that tunes the
  attacker's \textit{aggression level}. The attacker only mounts
  stuffing attempts when one minus the cumulative dodge
  probability (from the beginning of the attack) does not exceed
  \aggressionLevel{\attacker}.
\end{itemize}

Recall $\firstbidIdx$ denotes the index of the first
\ref{alg:prop-rsp:next-bid} bid issued by any $\site \in \siteSet$. We
assume the attacker begins stuffing only at bid indices $\bidIdx \ge
\firstbidIdx$; otherwise, it would trivially capture users at sites
that have not yet had the chance to allocate monitoring capacity to
$\targetSite$. At each such $\bidIdx$, the attacker observes the
allocations in effect---i.e.,
$\{\allocTo{\site}{\targetSite}\}_{\site \in \peerSet[\targetSite]}$
from bids prior to $\bidIdx$---and seeks to maximize
$\costPerBid{\targetSite}{\bidIdx}$, the expected number of users in
$\userSet[\targetSite]$ for which it can stuff some account at some
site in $\peerSet[\targetSite]$, subject to
$\aggressionLevel{\attacker}$.

To do so, the attacker examines the tree of possible monitoring
allocations based on the known \foresight{\attacker} bidders and
equiprobable bidders \lookahead{\attacker} after that, for fixed
values of \foresight{\attacker} and \lookahead{\attacker}.  The
attacker also generates all possible stuffing strategies across those
$\foresight{\attacker} + \lookahead{\attacker}$ bids consisting of
unharvested users in $\userSet[\targetSite]$ and that would satisfy
the $\aggressionLevel{\attacker}$ constraint.  From this tree, the
attacker identifies the leaf that maximizes the expected number of
users in \userSet[\targetSite] at which the attacker has captured an
account at some \site (and so presumably at all $\siteAlt \in
\peerSet[\site]$, as well).  It then performs stuffing attempts per
$\site \in \peerSet[\targetSite]$ that are prescribed in the first
step of the path to that leaf. The attacker continues building
$\foresight{\attacker} + \lookahead{\attacker}$-depth trees until the
\aggressionLevel{\attacker} parameter no longer permits stuffing
attempts or once the accounts of all users in \userSet[\targetSite]
have been harvested.

\subsection{Evaluation}
\label{sec:attacker:eval}

Our evaluation of the attacker's strategy aims to establish whether:
(1) \targetSite's risk is a good predictor of the number of users that
an attacker can harvest at the beginning of its attack, and so
minimizing risk is fruitful for a site to maximize its detection
ability, and (2) an attacker can arbitrarily increase its ability to
harvest users in \userSet[\targetSite]. We investigate these trends
via model-checking experiments with $\nmbrSites = 4$ sites and
$\nmbrUsers = 400$ users. Per \cref{sec:auction:eval:bidding}, we set
$\slack = \infty$; per \cref{sec:auction:eval:past-bids}, we set
$\smoothingFactor{\site} = 1$ and $\smoothingFactor{\targetSite} = 1$
for the \prop strategy.

Due to computational costs, we set $\lookahead{\attacker} +
\foresight{\attacker} \leq 2$ with $\lookahead{\attacker} \geq 1$, and
varied $\aggressionLevel{\attacker} \in \{0.25, 0.5, 0.75\}$.  We
varied each of \popularity, \capacityCoeff{\site}, and
\capacityCoeff{\targetSite} $\in [0,0.25, 0.5, 0.75, 1]$, generating
30 user configurations (and resulting capacities) per combination. We
also generated 30 bidding sequences that were distinct over
$[\firstbidIdx, \firstbidIdx + \nmbrBids)$; throughout
\cref{sec:attacker:eval} we restrict bid index \bidIdx to this range.
We define $\vulnSet[\targetSite]$ as users in \userSet[\targetSite]
susceptible to stuffing via shared credentials with some $\site \in
\peerSet[\targetSite]$. Under our current assumption of a password
reuse rate of 1 (optimistic for the attacker), $\vulnSet[\targetSite]$
includes every user in \userSet[\targetSite] who has at least one
account at $\site \in \peerSet[\targetSite]$.  To compare
\costPerBid{\targetSite}{\bidIdx} across \targetSite with varying
\popularity we normalize
\begin{align}
  \normCostPerBid{\targetSite}{\bidIdx} &= \frac{\costPerBid{\targetSite}{\bidIdx}}{\setSize{\vulnSet[\targetSite]}}
  \label{eqn:normCostPerBid}
\end{align}
since higher \popularity implies larger
\setSize{\vulnSet[\targetSite]}.

We used the Kruskal-Wallis \hStat test and post-hoc Dunn test with
Bonferroni correction to analyze the effect of various parameters on
\normCostPerBid{\targetSite}{\bidIdx} and
$\normRiskPerBid{\targetSite}{\bidIdx} -
\normCostPerBid{\targetSite}{\bidIdx}$. The latter value emphasizes
the predictive accuracy of \normRiskPerBid{\targetSite}{\bidIdx} for
\normCostPerBid{\targetSite}{\bidIdx}, with values closer to 0
indicating better predictions.

\begin{wrapfigure}{r}{0.5\columnwidth}
  \vspace{-3ex}
  \centering
  \resizebox{0.5\columnwidth}{!}{
    \input{norm_risk_minus_norm_cost_first_round_psi.tex}
  }
  \caption{\normRiskPerBid{\targetSite}{\firstbidIdx}
    $-$ \normCostPerBid{\targetSite}{\firstbidIdx} by \popularity and
    \aggressionLevel{\attacker}. Boxes span 25--75 \percentile;
    whiskers span 5--95 \percentile; diamonds are means; red lines
    are medians.}
  \label{fig:norm_risk_minus_norm_cost_round_1}
  \vspace{-2ex}
\end{wrapfigure}

\subsubsection{Risk predicts cost more accurately at early rounds of
  an attack} 
  \label{sec:attacker:eval:early}
  Since sites are unaware of when an attack begins,
  \normRiskPerBid{\targetSite}{\bidIdx} is computed assuming no prior
  compromise. As a result, at later bid indices
  \normRiskPerBid{\targetSite}{\bidIdx} overestimates
  \normCostPerBid{\targetSite}{\bidIdx} because the accounts of some
  subset of \userSet[\targetSite] were already compromised. Still,
  \normRiskPerBid{\targetSite}{\firstbidIdx} remains a good predictor
  of \normCostPerBid{\targetSite}{\firstbidIdx} with the median
  difference (over all auctions) being $0.0222$ of vulnerable users at
  \targetSite.

\subsubsection{Risk predicts cost more accurately for less popular sites}
\label{sec:attacker:eval:popular}

Site \targetSite calculates \normRiskPerBid{\targetSite}{\bidIdx} in a
pairwise fashion, i.e., summing
\normRisk{\targetSite}{\site}{\allocationVal} incurred from a site
\site due to the allocation $\allocationVal =
\allocTo{\site}{\targetSite}$, over all $\site \in
\peerSet[\targetSite]$ (\cref{eqn:perBidNormRisk}). Since the set of
users that \targetSite shares with one site \site can overlap with
those it shares with another site \siteAlt, calculating
\normRiskPerBid{\targetSite}{\bidIdx} in this way overestimates the
number of its accounts that an attacker can expect to harvest at other
sites. \cref{fig:norm_risk_minus_norm_cost_round_1} shows this trend:
the distribution of $\normRiskPerBid{\targetSite}{\firstbidIdx} -
\normCostPerBid{\targetSite}{\firstbidIdx}$ decreased ($\pValue <
10^{-8}$) with lower \popularity per \aggressionLevel{\attacker},
since such a \targetSite shares less users with its peers.
\Cref{sec:discussion:challenges} explores if risk can be measured
independently of a site's popularity.

\subsubsection{Risk predicts cost more accurately when the attacker is
  more aggressive}
\label{sec:attacker:eval:aggression}

Since \targetSite lacks knowledge of the attacker, it calculates
\normRiskPerBid{\targetSite}{\bidIdx} assuming maximal aggression.
Therefore, \normRiskPerBid{\targetSite}{\firstbidIdx} best estimates
\normCostPerBid{\targetSite}{\firstbidIdx} when
\aggressionLevel{\attacker} is high.
\cref{fig:norm_risk_minus_norm_cost_round_1} confirms that per
\popularity, the distribution of
$\normRiskPerBid{\targetSite}{\firstbidIdx} -
\normCostPerBid{\targetSite}{\firstbidIdx}$ decreased ($\pValue <
10^{-8}$) as \aggressionLevel{\attacker} increased. However, this
effect decreased as \popularity decreased, due to less vulnerable
users at an unpopular \targetSite.

\subsubsection{The attacker's ability to capture users increases
only modestly with additional knowledge of the bidding sequence}
\label{sec:attacker:eval:foresight}

We confirmed that an attacker with nonzero \foresight{\attacker} can
benefit ($\pValue <10^{-8}$) from its ability to predict the next
bidder. Regardless, \cref{sec:discussion:enforcement} outlines
mechanisms to enforce $\foresight{\attacker} = 0$, eliminating this
predictive advantage.  
When $\foresight{\attacker}= 0$, examining the distribution of
potential next bidders for $\lookahead{\attacker}$ steps ahead has no
effect ($\pValue = 0.265$) on $\normCostPerBid{\targetSite}{\bidIdx}$.
So we adopt $\foresight{\attacker} = 0$, $\lookahead{\attacker} = 1$
for larger experiments in \cref{sec:attacker:larger}, enabled by the
model-checker implementation described in
\cref{sec:model_checking:attacker}.

\subsection{Larger systems}
\label{sec:attacker:larger}

We now analyze the effectiveness of the attacker's strategy against
\prop bidders in a model-checking experiment with $\nmbrSites = 10$
sites and $\nmbrUsers = 1000$ users. While previously we enabled the
attacker to begin stuffing at bid index \firstbidIdx, we now require
the attacker to wait until \secondbidIdx, the bid index at which a
site first issues a second \ref{alg:prop-rsp:next-bid} bid. This delay
was a concession to the scalability of these experiments, allowing the
attacker to invest its stuffing attempts only after bids incorporate
sites' responses to earlier \ref{alg:prop-rsp:next-bid} allocations.
Starting at bid index \secondbidIdx, we evaluate the attacker over a
window of \nmbrBids bids. We define the total cost incurred over the
$\nmbrBids$-bid evaluation window as:
\begin{align} 
  \cost{\targetSite} = \sum_{\bidIdxAlt = 0}^{\nmbrBids - 1} \costPerBid{\targetSite}{\secondbidIdx + \bidIdxAlt}
  \label{eqn:totalCost}
\end{align}

We similarly define the normalized total cost as:
\begin{align}
  \normCost{\targetSite} &= \frac{\cost{\targetSite}}{\setSize{\vulnSet[\targetSite]}}
  \label{eqn:normRelCost}
\end{align}

We largely adopt the remaining setup from \cref{sec:attacker:eval},
setting $\slack = \infty$, $\smoothingFactor{\site} = 1$ and
$\smoothingFactor{\site} = 1$, and varying
$\aggressionLevel{\attacker} \in \{0.25, 0.5, 0.75\}$, The finding in
\cref{sec:attacker:eval:foresight} and anticipated deployment
described in \cref{sec:discussion:enforcement} justify setting
$\foresight{\attacker} = 0$ and $\lookahead{\attacker} = 1$. For
$\popularity \in \{0, 0.25, 0.5, 0.75, 1\}$, $\capacityCoeff{\site}
\in \{1, 2, 3, 4\}$, and $\capacityCoeff{\targetSite} \in \{1, 2, 3,
4, 100\}$ (using $\capacityCoeff{\targetSite} = 100$ to understand
\targetSite's security in the limit) we generated 30 user
configurations (and resulting capacities) per combination. We also
generated 30 bidding sequences that were distinct over the
$[\secondbidIdx, \secondbidIdx + \nmbrBids)$ window. Effects on
$\normCost{\targetSite}$ were analyzed using the Kruskal-Wallis \hStat
test.

\begin{figure}[t]
  \begin{subfigure}[t]{0.48\columnwidth}
    \centering
    \begin{tabular}{@{}c@{}}
      \FPeval{\MinNumber}{0.00689}
      \FPeval{\MaxNumber}{0.02888}
      \resizebox{\columnwidth}{!}{
        \begin{tabular}{@{\hspace{2pt}}r@{\hspace{2pt}}|*{5}{X}}
          & \multicolumn{5}{c}{\capacityCoeff{\targetSite}} \\
           \multicolumn{1}{@{\hspace{2pt}}c@{\hspace{2pt}}|}{\popularity}
           & \multicolumn{1}{c}{1}
           & \multicolumn{1}{c}{2}
           & \multicolumn{1}{c}{3}
           & \multicolumn{1}{c}{4}
           & \multicolumn{1}{c}{100}
           \\ \hline
1.0 & 0.02888 & 0.02754 & 0.02754 & 0.02754 & 0.02754 \\
0.8 & 0.02675 & 0.02651 & 0.02651 & 0.02651 & 0.02651 \\
0.6 & 0.02633 & 0.02611 & 0.02611 & 0.02611 & 0.02611 \\
0.4 & 0.01565 & 0.01565 & 0.01565 & 0.01565 & 0.01565 \\
0.2 & 0.00689 & 0.00689 & 0.00689 & 0.00689 & 0.00689 \\
        \end{tabular}
      }
    \end{tabular}
    \caption{$\capacityCoeff{\site} = 1$} \label{fig:resource_rich:cap1:psi}
  \end{subfigure}
  \hfill
  \begin{subfigure}[t]{0.48\columnwidth}
    \centering
    \begin{tabular}{@{}c@{}}  
      \FPeval{\MinNumber}{0.00109}
      \FPeval{\MaxNumber}{0.02704}
      \resizebox{\columnwidth}{!}{
        \begin{tabular}{@{\hspace{2pt}}r@{\hspace{2pt}}|*{5}{X}}
          & \multicolumn{5}{c}{\capacityCoeff{\targetSite}} \\
           \multicolumn{1}{@{\hspace{2pt}}c@{\hspace{2pt}}|}{\popularity}
           & \multicolumn{1}{c}{1}
           & \multicolumn{1}{c}{2}
           & \multicolumn{1}{c}{3}
           & \multicolumn{1}{c}{4}
           & \multicolumn{1}{c}{100}
           \\ \hline
1.0 & 0.02704 & 0.01269 & 0.01205 & 0.01205 & 0.01205 \\
0.8 & 0.02402 & 0.01138 & 0.01129 & 0.01120 & 0.01120 \\
0.6 & 0.01551 & 0.00898 & 0.00898 & 0.00883 & 0.00883 \\
0.4 & 0.00592 & 0.00476 & 0.00440 & 0.00440 & 0.00440 \\
0.2 & 0.00196 & 0.00109 & 0.00109 & 0.00109 & 0.00109 \\
        \end{tabular}
      }
    \end{tabular}
    \caption{$\capacityCoeff{\site} = 2$} \label{fig:resource_rich:cap2:psi}
  \end{subfigure}
\\
\begin{subfigure}[t]{0.48\columnwidth}
    \centering
    \begin{tabular}{@{}c@{}}
      \FPeval{\MinNumber}{0}
      \FPeval{\MaxNumber}{0.02958}
      \resizebox{\columnwidth}{!}{
        \begin{tabular}{@{\hspace{2pt}}r@{\hspace{2pt}}|*{5}{X}}
          & \multicolumn{5}{c}{\capacityCoeff{\targetSite}} \\
           \multicolumn{1}{@{\hspace{2pt}}c@{\hspace{2pt}}|}{\popularity}
           & \multicolumn{1}{c}{1}
           & \multicolumn{1}{c}{2}
           & \multicolumn{1}{c}{3}
           & \multicolumn{1}{c}{4}
           & \multicolumn{1}{c}{100}
           \\ \hline
1.0 & 0.01700 & 0.01148 & 0.00625 & 0.00550 & 0.00550 \\
0.8 & 0.01618 & 0.00720 & 0.00579 & 0.00572 & 0.00572 \\
0.6 & 0.02958 & 0.00585 & 0.00442 & 0.00442 & 0.00431 \\
0.4 & 0.02299 & 0.00218 & 0.00166 & 0.00166 & 0.00154 \\
0.2 & 0.00236 & 0.00012 & 0.00000 & 0.00000 & 0.00000 \\
        \end{tabular}
      }
    \end{tabular}
    \caption{$\capacityCoeff{\site} = 3$} \label{fig:resource_rich:cap3:psi}
  \end{subfigure}
  \hfill
  \begin{subfigure}[t]{0.48\columnwidth}
    \centering
    \begin{tabular}{@{}c@{}}  
      \FPeval{\MinNumber}{0}
      \FPeval{\MaxNumber}{0.14976}
      \resizebox{\columnwidth}{!}{
        \begin{tabular}{@{\hspace{2pt}}r@{\hspace{2pt}}|*{5}{X}}
          & \multicolumn{5}{c}{\capacityCoeff{\targetSite}} \\
           \multicolumn{1}{@{\hspace{2pt}}c@{\hspace{2pt}}|}{\popularity}
           & \multicolumn{1}{c}{1}
           & \multicolumn{1}{c}{2}
           & \multicolumn{1}{c}{3}
           & \multicolumn{1}{c}{4}
           & \multicolumn{1}{c}{100}
           \\ \hline
1.0 & 0.13266 & 0.01201 & 0.00773 & 0.00396 & 0.00319 \\
0.8 & 0.14976 & 0.01072 & 0.00702 & 0.00321 & 0.00321 \\
0.6 & 0.02338 & 0.00782 & 0.00240 & 0.00167 & 0.00159 \\
0.4 & 0.02235 & 0.00413 & 0.00048 & 0.00031 & 0.00009 \\
0.2 & 0.00795 & 0.00076 & 0.00021 & 0.00000 & 0.00000 \\
        \end{tabular}
      }
    \end{tabular}
    \caption{$\capacityCoeff{\site} = 4$} \label{fig:resource_rich:cap4:psi}
  \end{subfigure}
  \caption{Mean $\normCost{\targetSite}$ for
  $\aggressionLevel{\attacker} = 0.75$. 
  Lighter cells show
  lower cost.
  }
  \label{fig:resource_rich}
\end{figure}

\subsubsection{Sites can maximize their security by increasing their
capacities}
\label{sec:attacker:larger:scap}
\Cref{fig:resource_rich} shows that \targetSite can typically reduce
its \normCost{\targetSite} by raising its own
\capacityCoeff{\targetSite} above the \capacityCoeff{\site} of its
peers ($\pValue < 10^{-8}$). This not only improves its own security
but nudges the ecosystem toward a higher-capacity equilibrium (i.e.,
higher \capacityCoeff{\site}), making peers more likely to
reciprocate. We see this in \cref{fig:resource_rich}: when
$\capacityCoeff{\site} = 2$, every \targetSite is incentivized to
raise its $\capacityCoeff{\targetSite} \geq 3$
(\cref{fig:resource_rich:cap2:psi}), prompting peers to adopt
$\capacityCoeff{\site} = 3$ in \cref{fig:resource_rich:cap3:psi}. This
iteration continues, as each \targetSite again finds it beneficial to
raise $\capacityCoeff{\targetSite} \geq 4$, reinforcing a positive
feedback loop.

\subsubsection{Popular sites drive capacity escalation}
\label{sec:attacker:larger:popular_ratchet}
When peers are resource constrained
(\cref{fig:resource_rich:cap1:psi}), popular sites drive capacity
escalation due to their higher capacity (which scales with
\setSize{\userSet[\targetSite]}) and higher demand for their slots
(due to shared users with many peers). They prioritize popular peers,
limiting reciprocity for unpopular ones; \targetSite with $\popularity
= 0.2$ sees \normCost{\targetSite} plateau despite increasing
\capacityCoeff{\targetSite}. Still, moderately popular sites
($\popularity \geq 0.6$) remain incentivized to increase capacity,
sustaining cyclic reciprocation.

\subsubsection{Popular sites benefit from their peers} 
\label{sec:attacker:larger:popular}

While popular sites may offer more slots than they receive due to
their higher monitoring capacity, they still enjoy significant
($\pValue < 10^{-8}$) reductions in their \normCost{\targetSite} when
their peers increase their capacities.
\Cref{fig:resource_rich:cap2:psi} shows that a \targetSite with
$\popularity = 1$ has diminished ability to reduce its
\normCost{\targetSite} beyond 0.01205, suggesting its less popular
peers face resource constraints and cannot reciprocate at scale.
However, when these peers increase their capacity to
$\capacityCoeff{\site} = 3$—and \cref{fig:resource_rich:cap2:psi}
confirms that they are incentivized to do so—the most popular
\targetSite benefits, as evidenced from the reduction in
\normCost{\targetSite} plateauing to 0.00550 for $\popularity = 1$.
Thus, popular sites not only drive ecosystem improvements but also
benefit from peer investments that relieve capacity bottlenecks.

\subsubsection{Unpopular sites receive comparable security}
\label{sec:attacker:larger:unpopular}
\Cref{fig:resource_rich} shows that less popular sites typically incur
lower \normCost{\targetSite} than popular ones ($\pValue < 10^{-8}$),
indicating the \prop bidding ecosystem favors unpopular sites. This is
because popular sites provide surplus monitoring slots that often go
to less popular peers, ensuring strong baseline security for unpopular
sites.

\section{Data-Driven Simulations}
\label{sec:data-driven}

Next, we evaluate \prop bidding using the Cit0day
dataset~\cite{Hunt:2020:Inside}. The dataset has been widely
reported~\cite{Cimpanu:2020:23600,Endler:2021:How} and integrated into
breach alerting services~\cite{Hunt:2020:Inside}, limiting the risk of
harm in our analysis. Following prior
work~\cite{Kim:2024:PassREfinder}, we anonymized all email addresses
and stored the data on an isolated machine to preserve privacy. 

We cleaned the dataset by removing hashed and duplicate records and
retained only the largest connected component of the user-site graph,
since sites that share no users with peers face no risk from
credential stuffing. The resulting dataset had over 108 million
credential entries across nearly $8{,}000$ websites from 74 million
users, identified by their email addresses. For any pair of sites
sharing users, we defined the password reuse rate as the fraction of
shared users using the same password on both sites. Reuse was
pervasive, with a median rate above $0.94$. \Cref{sec:cit0day} has
more details.

\subsection{Attacker's \Greedy Stuffing Strategy}
The optimal attacker from \cref{sec:attacker} is infeasible to
evaluate at this dataset's scale, so we simulate a greedy attacker
that makes locally optimal choices. Like in
\cref{sec:attacker:larger}, we allow the attacker to start stuffing at
\secondbidIdx over a window of \nmbrBids bids. Setting
$\foresight{\attacker} = 0$ and $\lookahead{\attacker} = 1$, at each
bid index $\bidIdx \in [\secondbidIdx, \secondbidIdx + \nmbrBids)$ the
attacker emulates the next \prop bid to project how many slots each
$\site \in \peerSet$ will offer \targetSite. Based on this slot
projection, shared users, and empirical (rather than perfect)
password-reuse rate, the attacker then computes the marginal increase
in expected cost from one stuffing attempt for each site. It chooses
the site \site with the highest marginal cost and an unstuffed user
that has accounts at both \site and \targetSite but has the fewest
accounts elsewhere, to preserve future stuffing opportunities, and
breaks ties in site or user selection at random. The attacker repeats
this process, recomputing marginal improvements to \cost{\targetSite},
until no further expected gain remains.

\subsection{Evaluation}
\label{sec:data-driven:eval}

We adopted the optimal parameters for \prop identified in
\cref{sec:auction:eval}, setting $\smoothingFactor{\targetSite} =
\smoothingFactor{\site} = 1.0$ and $\slack = \infty$. We generated 10
bidding sequences that are distinct over $[\secondbidIdx,
\secondbidIdx+\nmbrBids]$, setting $\nmbrBids = 10$. To evaluate
security for sites of varying popularites, we grouped sites into
quartiles based on increasing $\setSize{\userSet[\site]}$ and sampled
10 sites per quartile; small (\siteSize{S}), medium (\siteSize{M}),
large (\siteSize{L}), and extra large (\siteSize{XL}). As before, we
scaled capacity by user set size, setting $\capacity{\site} \gets
\capacityCoeff{\site} \times \setSize{\userSet[\site]}$, and varied
$\capacityCoeff{\site} \in \{1, 2, 3, 4\}$ and
$\capacityCoeff{\targetSite} \in \{0.1, 1, 10, 100\}$; the last range
reflects order-of-magnitude variation in $\setSize{\userSet[\site]}$,
since a small \targetSite may need much higher capacity to bid
comparably to large peers. We simulated 10 \greedy attackers,
differing only in the random choices they make to break ties, and set
$\aggressionLevel{\attacker} = 1$ to simulate the worst-case scenario.

\begin{figure}[t]
  \begin{subfigure}[t]{0.48\columnwidth}
    \centering
    \begin{tabular}{@{}c@{}}
      \FPeval{\MinNumber}{0.0004}
      \FPeval{\MaxNumber}{0.0019}
      \resizebox{\columnwidth}{!}{
        \begin{tabular}{@{\hspace{2pt}}c@{\hspace{2pt}}|*{4}{X}}
          & \multicolumn{4}{c}{\capacityCoeff{\targetSite}} \\
          \setSize{\userSet[\targetSite]}
          & \multicolumn{1}{c}{0.1}
          & \multicolumn{1}{c}{1}
          & \multicolumn{1}{c}{10}
          & \multicolumn{1}{c}{100}
          \\ \hline
          \siteSize{XL} & 0.0007 & 0.0005 & 0.0005 & 0.0005 \\
          \siteSize{L} & 0.0015 & 0.0012 & 0.0013 & 0.0013 \\
          \siteSize{M} & 0.0011 & 0.0008 & 0.0008 & 0.0008 \\
          \siteSize{S} & 0.0019 & 0.0015 & 0.0014 & 0.0014 \\  
        \end{tabular}
      }
    \end{tabular}
    \caption{$\capacityCoeff{\site} = 2$}
    \label{fig:datadriven:cap2:psi}
  \end{subfigure}
  \hfill
  \begin{subfigure}[t]{0.48\columnwidth}
    \centering
    \begin{tabular}{@{}c@{}}
      \FPeval{\MinNumber}{0.0004}
      \FPeval{\MaxNumber}{0.0019}
      \resizebox{\columnwidth}{!}{
        \begin{tabular}{@{\hspace{2pt}}c@{\hspace{2pt}}|*{4}{X}}
          & \multicolumn{4}{c}{\capacityCoeff{\targetSite}} \\
          \setSize{\userSet[\targetSite]}
          & \multicolumn{1}{c}{0.1}
          & \multicolumn{1}{c}{1}
          & \multicolumn{1}{c}{10}
          & \multicolumn{1}{c}{100}
          \\ \hline
          \siteSize{XL} & 0.0006 & 0.0004 & 0.0004 & 0.0004 \\
          \siteSize{L} & 0.0013 & 0.0010 & 0.0010 & 0.0010 \\
          \siteSize{M} & 0.0010 & 0.0006 & 0.0005 & 0.0005 \\
          \siteSize{S} & 0.0018 & 0.0013 & 0.0012 & 0.0012 \\
        \end{tabular}
      }
    \end{tabular}
    \caption{$\capacityCoeff{\site} = 4$}
    \label{fig:datadriven:cap4:psi}
  \end{subfigure}
  \caption{Mean \normCost{\targetSite} for
    $\aggressionLevel{\attacker} = 1.0$. Lighter cells indicate lower
    \normCost{\targetSite}.}
  \label{fig:datadriven}
\end{figure}

\Cref{fig:datadriven} summarizes the results. We omit
$\capacityCoeff{\site} \in \{1, 3\}$ for space, but these follow the
same trends as $\capacityCoeff{\site} \in \{ 2, 4\}$. We organize our
findings by those that confirm the conclusions in
\cref{sec:attacker:larger}, and those that provide new nuance.

\subsubsection{Validation of model-checking results} 
\Cref{fig:datadriven} supports several claims in
\cref{sec:attacker:larger}. First, increasing $\capacity{\targetSite}$
reduces \normCost{\targetSite}, so a site improves its \textit{own}
security by increasing capacity for \textit{others} (cf.,
\cref{sec:attacker:larger:scap}). Second, \siteSize{XL}
\normCost{\targetSite} decreases as \capacityCoeff{\site} rises, so
popular sites benefit from less-popular peers (cf.,
\cref{sec:attacker:larger:popular}). Third, \normCost{\targetSite} is
similar for \siteSize{S} and \siteSize{XL} sites, so less-popular
sites achieve comparable protection (cf.,
\cref{sec:attacker:larger:unpopular}).

\subsubsection{Insights beyond model-checking results}
In Cit0day, popularity decouples from shared-user overlap and password
reuse, so the highest-\normCost{\targetSite} tier in
\cref{fig:datadriven} is \siteSize{S} rather than \siteSize{XL},
unlike \cref{fig:resource_rich} where popular sites consistently incur
the greatest cost. Regardless, the same capacity escalation mechanism
of \cref{sec:attacker:larger:popular_ratchet} applies: sites with the
highest \normCost{\targetSite} and largest cost reduction from
increasing \capacityCoeff{\targetSite} drive escalation---the most
popular sites in \cref{fig:resource_rich}, but \siteSize{S} sites in
\cref{fig:datadriven}.

This suggests that smaller but at-risk sites may be more active
drivers of \capacityCoeff{\site} than
\cref{sec:attacker:larger:popular_ratchet} alone would predict, and
that capacity planning at a site \targetSite would benefit from
information beyond \popularity, such as \siteSet[\user] for each
$\user \in \userSet[\targetSite]$. In the \PSILabel setting this can
be computed directly; otherwise, a privacy-preserving variant could
compute \siteSet[\user] while keeping \user anonymous. Comparing
\cref{fig:datadriven:cap2:psi} with \cref{fig:resource_rich:cap2:psi}
and \cref{fig:datadriven:cap4:psi} with
\cref{fig:resource_rich:cap4:psi}, the data-driven setting shows
better outcomes overall, reflecting the empirical (rather than
perfect) reuse rates.

\section{Discussion}
\label{sec:discussion}

\subsection{Community Formation}
\label{sec:discussion:community}

Our framework relies on a community of sites that monitor for each
other, due to users' tendency to reuse passwords across those sites.
Prior work has shown that password reuse is influenced by site
characteristics, such as function (e.g., shopping or
email~\cite{Wang:2018:Domino}), affiliation (e.g.,
universities~\cite{Wash:2016:Understanding}), geographic
location~\cite{AlSabah:2018:Culture, Mayer:2017:Second}, and security
posture~\cite{Stobert:2018:Password, Gaw:2006:Password}.  Existing
organizations could bootstrap such communities; e.g., ISACs already
coordinate threat-intelligence exchange per industry~\cite{ISAC:2025}
among vetted members~\cite{RENISAC:Membership}.

A site might still wish to verify that it shares users with community
members before joining. Computing pairwise \PSILabel/\PSICALabel would
confirm shared users (see~\cite{Ion:2020:Deploying, IAB:2025:ADMAP}
for industry-scale deployments), and tools like
PassREfinder~\cite{Kim:2024:PassREfinder} could estimate password
reuse rates. We stress, however, that the utility of our approach
depends on sites joining such communities liberally; sites that do not
participate in collective monitoring---the status quo today---create
blind spots that attackers can exploit by stuffing stolen credentials
at them.

Once a site joins a community, it needs to perform
\PSILabel/\PSICALabel computations with members to start bidding.  It
may recompute these overlaps periodically, but need not do so
frequently; e.g., 2023 industry data suggests annual user churn of
just 3.5--6.9\% across sectors~\cite{Recurly:2024:Business},
indicating largely stable user bases.

\subsection{Enforcing ecosystem-wide parameters}
\label{sec:discussion:enforcement}

\Cref{sec:auction} found setting $\cutLine{\targetSite} = \boolFalse$
and $\foresight{\targetSite} = 0$ limited the risk reduction an
\targetSite[\exhaustiveLabel] achieves over a \targetSite[\propLabel],
dissuading a \targetSite[\propLabel] from switching strategies. To
achieve this in practice, we observe a \targetSite cannot choose when
to bid ($\cutLine{\targetSite} = \boolFalse$) if the next bidder is
assigned outside its control, and when placing its bid, \targetSite
cannot predict the next bidder ($\foresight{\targetSite} = 0$) if it
is assigned randomly (and only after \targetSite has either placed its
bid or been eliminated from bidding due to its delay).  These
requirements can be met if the next bidder is assigned using a
\textit{randomness beacon}, which produces a random value (in our
case, naming the next bidder) at predictable times.

Randomness beacons are deployed with a range of trust assumptions. For
example, NIST's randomness beacon~\cite{Kelsey:2018:NIST} requires
consumers of its random values to trust NIST, whereas
DFINITY~\cite{Hanke:2018:DFINITY} and
drand~\cite{LeagueOfEntropy:2024:drand} offer distributed
implementations using threshold signatures that prevent a small
fraction of participants from biasing or learning the next random
value before honest participants~\cite{Malkhi:2000:Survivable,
Cachin:2000:Constantinople}.  Such schemes require a trusted party to
distribute secret key shares; distributed key-generation protocols
(e.g.,~\cite{Gennaro:2007:DKG, Kate:2009:DKG}) can alleviate this
need, and more recent designs avoid threshold signing altogether
(e.g.,~\cite{Kokoris-Kogias:2020:Asynch, Abraham:2021:DKG,
Das:2022:DKG, Gao:2022:BA, deSouza:2022:Approx, Abraham:2023:Bingo,
Das:2023:DKG, Bandarupalli:2024:Beacons}). While a trusted beacon
could support the collaboration we propose, we are agnostic to its
design.

Enforcing $\foresight{\attacker} = 0$, as required in
\cref{sec:attacker:larger} and \cref{sec:data-driven}, is more complex
since an attacker could learn the assigned bidder from the randomness
beacon even before the assigned bidder learns it has been chosen,
resulting in $\foresight{\attacker} = 1$.  This possibility can be
prevented if the participants implement the randomness beacon among
themselves and do not share their foreknowledge of the next bidder
with the attacker. Alternatively, the randomness beacon could emit a
cryptographic commitment to each random value, in place of the random
value itself, and then transmit the random value (and anything else
needed to open the commitment) to \textit{only} the next bidder that
it specifies. The selected bidder could then forward this random value
with its next bid, revealing to others that it is, in fact, the chosen
bidder, but eliminating the opportunity for the attacker to learn this
fact before the bid is already in transit.

\section{Conclusion}

In this paper, we proposed a novel algorithm to incentivize the
exchange of monitoring favors among sites. We systematically explored
a parameter space specifying defender-defender and attacker-defender
interactions and used model checking to conservatively estimate the
security offered by our algorithm. We found that sites of varying
popularity and resource constraints are incentivized to increase the
monitoring they provide to \textit{others} to improve their chance of
detecting their \textit{own} credential database breach. We validated
our design through simulations informed by a breached dataset
capturing user overlap and password reuse patterns. We expect that, if
deployed, our algorithm will enable a self sustaining credential
database breach detection ecosystem, and could serve as a foundation
for other cooperative security applications.

\bibliographystyle{IEEEtran}
\bibliography{full,bibliography}

@article{Abdallah:2020:Behavioral,
  title = {Behavioral and game-theoretic security investments in interdependent systems modeled by attack graphs},
  author = {M. Abdallah and P. Naghizadeh and A. R. Hota and T. Cason and S. Bagchi and S. Sundaram},
  journal = IEEETCNS,
  volume = 7,
  number = 4,
  pages = {1585-1596},
  year = 2020,
}

@inproceedings{Abdallah:2022:Tasharok,
  title = {Tasharok: Using mechanism design for enhancing security resource allocation in interdependent systems},
  author = {M. Abdallah and D. Woods and P. Naghizadeh and I. Khalil and T. Cason and S. Sundaram and S. Bagchi},
  booktitle = SSP,
  pages = {249-266},
  year = {2022}
}

@inproceedings{Abraham:2023:Bingo,
  author = {I. Abraham and P. Jovanovic and M. Maller and S. Meiklejohn and G. Stern},
  title = {Bingo: Adaptivity and asynchrony in verifiable secret sharing and distributed key generation},
  booktitle = CRYPTO # { 2023},
  series = LNCS,
  volume = 14081,
  month = aug,
  year = 2023
}

@inproceedings{Abraham:2021:DKG,
  author = {I. Abraham and P. Jovanovic and M. Maller and S. Meiklejogn and G. Stern and A. Tomescu},
  title = {Reaching consensus for asynchronous distributed key generation},
  booktitle = {40\textsuperscript{th} } # PODC,
  pages = {363-373},
  year = 2021
}

@article{Akshima:2019:Honeywords,
  author = {Akshima and D. Chang and A. Goel and S. Mishra and S. K. Sanadhya},
  title = {Generation of secure and reliable honeywords, preventing false detection},
  journal = IEEETDSC,
  pages = {757-769},
  volume = 16,
  number = 5,
  year = 2019
}

@article{Alsabah:2018:Culture,
    author = {M. AlSabah and G. Oligeri and R. Riley},
    title = {Your culture is in your password: An analysis of a demographically-diverse password dataset},
    journal = COSE,
    year = {2018},
    volume = {77},
    pages = {427--441}
}

@inproceedings{Bandarupalli:2024:Beacons,
  author = {A. Bandarupalli and A. Bhat and S. Bagchi and A. Kate and M. K. Reiter},
  title = {Random beacons in {Monte Carlo}: Efficient asynchronous random beacon without threshold cryptography},
  booktitle = {31\textsuperscript{st} } # CCS,
  year = 2024
}

@inproceedings{Branzei:2021:Proportional,
  title={Proportional dynamics in exchange economies},
  author={S. Br{\^a}nzei and N. Devanur and Y. Rabani},
  booktitle={22\textsuperscript{nd} } # ECONCOMP,
  pages={180--201},
  year={2021}
}

@inproceedings{Cachin:2000:Constantinople,
  title = {Random oracles in {Constantinople}: Practical asynchronous {Byzantine} agreement using cryptography},
  author = {C. Cachin and K. Kursawe and V. Shoup},
  booktitle = {19\textsuperscript{th} } # PODC,
  year = 2000
}

@article{Chakraborty:2022:Honeyword,
  title = {Honeyword-based authentication techniques for protecting passwords: A survey},
  author = {N. Chakraborty and J. Li and V. C. M. Leung and S. Mondal and Y. Pan and C. Luo and M. Mukherjee},
  journal = CSUR,
  volume = 55,
  pages = {1-37},
  year = 2022
}

@misc{Cimpanu:2020:23600,
  author = {C. Cimpanu},
  title = {23,600 hacked databases have leaked from a defunct 'data breach index' site},
  howpublished = {\url{https://www.zdnet.com/article/23600-hacked-databases-have-leaked-from-a-defunct-data-breach-index-site/}},
  month = nov,
  year = 2020
}

@misc{Cohen:2003:BitTorrent,
  title={Incentives build robustness in {BitTorrent}},
  author={B. Cohen},
  howpublished = {\url{http://bittorrent.org/bittorrentecon.pdf}},
  month = may,
  year={2003}
}

@misc{CTA:2025:Membership,
  author = {Cyber Threat Alliance},
  title = {Membership},
  howpublished = {\url{https://www.cyberthreatalliance.org/membership/}},
  year = 2025
}

@inproceedings{Das:2014:Tangled,
  title={The tangled web of password reuse},
  author={A. Das and J. Bonneau and M. Caesar and N. Borisov and X. Wang},
  booktitle={21\textsuperscript{st} } # NDSS,
  year=2014
}

@inproceedings{Das:2022:DKG,
  title = {Practical asynchronous distributed key generation},
  author = {S. Das and T. Yurek and Z. Xiang and A. Miller and L. Kokoris-Kogias and L. Ren},
  booktitle = {43\textsuperscript{rd} } # SSP,
  pages = {2518-2534},
  year = 2022
}

@inproceedings{Das:2023:DKG,
  title = {Practical asynchronous high-threshold distributed key generation and distributed polynomial sampling},
  author = {S. Das and Z. Xiang and L. Kokoris-Kogias and L. Ren},
  booktitle = {32\textsuperscript{nd} } # USESEC,
  month = aug,
  year = 2023
}

@inproceedings{Davidson:2017:PSICA,
  title = {An efficient toolkit for computing private set operations},
  author = {A. Davidson and C. Cid},
  booktitle = {22\textsuperscript{nd} } # ACISP,
  series = LNCS,
  volume = 10343,
  pages = {261-278},
  month = jul,
  year = 2017
}

@inproceedings{DeBlasio:2017:Tripwire,
  title = {Tripwire: Inferring internet site compromise},
  author = {J. {DeBlasio} and S. Savage and G. M. Voelker and A. C. Snoeren},
  booktitle = {17\textsuperscript{th} } # IMC,
  pages = {341-354},
  year = 2017
}

@inproceedings{Debnath:2015:PSICA,
  title = {Secure and efficient private set intersection cardinality using {Bloom} filter},
  author = {S. K. Debnath and R. Dutta},
  booktitle = {18\textsuperscript{th} } # ISC,
  pages = {209-226},
  year = 2015,
  series= LNCS,
  volume = 9290,
  month = sep
}

@inproceedings{DeCristofaro:2012:PSICA,
  author = {E. {De Cristofaro} and P. Gasti and G. Tsudik},
  title = {Fast and private computation of cardinality of set intersection and union},
  booktitle = {11\textsuperscript{th} } # CANS,
  series = LNCS,
  volume = 7712,
  pages = {218-231},
  year = 2012
}

@inproceedings{deSouza:2022:Approx,
  author = {L. F. {de Souza} and P. Kuznetsov and A. Tonkikh},
  title = {Distributed randomness from approximate agreement},
  booktitle = {36\textsuperscript{th} } # DISC,
  month = oct,
  year = 2022
}

@inproceedings{Dionysiou:2021:Honeygen,
  title = {Honeygen: generating honeywords using representation learning},
  author = {A. Dionysiou and V. Vassiliades and E. Athanasopoulos},
  booktitle = {16\textsuperscript{th} } # ASIACCS,
  year = 2021
}

@inproceedings{Duma:2006:Trust,
  title={A trust-aware, {P2P}-based overlay for intrusion detection},
  author={C. Duma and M. Karresand and N. Shahmehri and G. Caronni},
  booktitle={17\textsuperscript{th} } # DEXA,
  pages={692-697},
  year={2006}
}

@inproceedings{Egert:2015:PSICA,
  author = {R. Egert and M. Fischlin and D. Gens and S. Jacob and M. Senker and J. Tillmanns},
  title = {Privately computing set-union and set-intersection cardinality via {Bloom} filters},
  booktitle = {20\textsuperscript{th} } # ACISP,
  series = LNCS,
  volume = 9144,
  year = 2015
}

@misc{Endler:2021:How,
  author = {D. Endler},
  title = {How Much Data Was Leaked To Cybercriminals In 2020 — And What They're Doing With It},
  howpublished = {\url{https://www.forbes.com/councils/forbestechcouncil/2021/04/20/how-much-data-was-leaked-to-cybercriminals-in-2020---and-what-theyre-doing-with-it/}},
  month = apr,
  year = 2021
}

@article{Erguler:2016:Flatness,
  author = {I. Erguler},
  title = {Achieving flatness: Selecting the honeywords from existing user passwords},
  journal = IEEETPDS,
  volume = 13,
  number = 2,
  year = 2016
}

@inproceedings{Gao:2022:BA,
  author = {Y. Gao and Y. Lu and Z. Lu and Q. Tang and J. Xu and Z. Zhang},
  title = {Efficient asynchronous {Byzantine} agreement without private setups},
  booktitle = {42\textsuperscript{nd} } # ICDCS,
  pages = {246-257},
  month = jul,
  year = 2022
}

@inproceedings{Gaw:2006:Password,
  title={Password management strategies for online accounts},
  author={S. Gaw and E. W. Felten},
  booktitle= {2\textsuperscript{nd}} # SOUPS, 
  year={2006},
  pages={44--55}
}

@article{Gennaro:2007:DKG,
  author = {R. Gennaro and S. Jarecki and H. Krawczyk and T. Rabin},
  title = {Secure distributed key generation for discrete-log based cryptosystems},
  journal = JCRYPT,
  volume = 20,
  pages = {51-83},
  year = 2007
}

@inproceedings{Habib:2018:User,
  title={User behaviors and attitudes under password expiration policies},
  author={H. Habib and E.P. Naeini and S. Devlin and M. Oates and C. Swoopes and L. Bauer and N. Christin and L.F. Cranor},
  booktitle = {14\textsuperscript{th} } # SOUPS,
  year={2018}
}

@misc{Hanke:2018:DFINITY,
  author = {T. Hanke and M. Movahedi and D. Williams},
  title = {{DFINITY} technology overview series, consensus system},
  year = 2018,
  howpublished = {arXiv:1805.04548 [cs.DC]}
}

@incollection{Hota:2018:Game,
  title = {A game-theoretic framework for securing interdependent assets in networks},
  author = {A. R. Hota and A. A. Clements and S. Bagchi and S. Sundaram},
  booktitle = {Game Theory for Security and Risk Management},
  pages = {157-184},
  year = 2018,
  publisher = {Springer}
}

@inproceedings{Huang:2024:Exposed,
  author = {Z. Huang and L. Bauer and M. K. Reiter},
  title = {The impact of exposed passwords on honeyword efficacy},
  booktitle = {33\textsuperscript{rd} } # USESEC,
  month = aug,
  year = 2024
}

@misc{Hunt:HaveIBeenPwned,
  author = {T. Hunt},
  title = {Have {I} Been Pwned?},
  howpublished = {\url{https://haveibeenpwned.com}}
}

@misc{Hunt:2020:Inside,
  author = {T. Hunt},
  title = {Inside the Cit0Day Breach Collection},
  howpublished = {\url{https://www.troyhunt.com/inside-the-cit0day-breach-collection/}},
  month = nov,
  year = 2020
}

@misc{IAB:2025:ADMAP,
  title={ADMAP: Attribution Data Matching Protocol},
  author={IAB Tech Laboratory},
  howpublished = {\url{https://iabtechlab.com/standards/addressability-and-pets/attribution-data-matching-protocol-admap-v1-0/}},
  year={2025}
}

@misc{IBM:2024:Cost,
  author = {{IBM}},
  title = {Cost of a Data Breach Report 2024},
  howpublished = {\url{https://www.ibm.com/reports/data-breach}},
  year = 2024
}

@inproceedings{Ion:2020:Deploying,
  title={On deploying secure computing: Private intersection-sum-with-cardinality},
  author={M. Ion and B. Kreuter and A.E. Nergiz and S. Patel and S. Saxena and K. Seth and M. Raykova and D. Shanahan and M. Yung},
  booktitle = {5\textsuperscript{th} } # EUROSP,
  year={2020},
}

@misc{ISAC:2025,
  title={About ISACs},
  author={National Council of ISACs},
  howpublished = {\url{https://www.nationalisacs.org/about-isacs}},
  year={2025}
}

@inproceedings{Juels:2013:Honeywords,
  author = {A. Juels and R. L. Rivest},
  title = {Honeywords: Making password-cracking detectable},
  booktitle = {20\textsuperscript{th} } # CCS,
  pages = {145-160},
  year = 2013
}

@inproceedings{Kate:2009:DKG,
  author = {A. Kate and I. Goldberg},
  title = {Distributed key generation for the {Internet}},
  booktitle = {29\textsuperscript{th} } # ICDCS,
  month = jun,
  year = 2009
}

@misc{Kelsey:2018:NIST,
  author = {J. Kelsey and L. T. A. N. Brand{\~a}o and R. Peralta and H. Booth},
  title = {A reference for randomness beacons: Format and protocol version 2},
  howpublished = {\url{https://doi.org/10.6028/NIST.IR.8213-draft}},
  month = may,
  year = 2019
}

@inproceedings{Kim:2024:PassREfinder,
  author = {J. Kim and M. Song and M. Seo and Y. Jin and S. Shin},
  title = {\textsc{PassREfinder}: Credential stuffing risk prediction by representing password reuse between websites on a graph},
  booktitle = {45\textsuperscript{th} } # SSP,
  year = 2024,
  month = may
}

@inproceedings{Kissner:2005:PSICA,
  title = {Privacy-preserving set operations},
  author = {L. Kissner and D. Song},
  booktitle = CRYPTO # { 2005},
  pages = {241-257},
  year = 2005,
  month = aug,
  series = LNCS,
  volume = 3621
}

@inproceedings{Kokoris-Kogias:2020:Asynch,
  title = {Asynchronous distributed key generation for computationally secure randomness, consensus, and threshold signatures},
  author = {E. Kokoris-Kogias and D. Malkhi and A. Spiegelman},
  booktitle = {27\textsuperscript{th} } # CCS,
  pages = {1751-1767},
  month = nov,
  year = 2020
}

@inproceedings{Kolumbus:2023:Asynchronous,
  title={Asynchronous proportional response dynamics: Convergence in markets with adversarial scheduling},
  author={Y. Kolumbus and M. Levy and N. Nisan},
  booktitle = {37\textsuperscript{th} } # NEURIPS,
  pages={25409-25434},
  year={2023}
}

@article{Kunreuther:2003:Interdependent,
  title = {Interdependent security},
  author = {H. Kunreuther and G. Heal},
  journal = JRU,
  volume = {26},
  pages = {231-249},
  year = {2003}
}

@misc{Lastpass:2022:Psychology,
  author = {{Lastpass}},
  title = {Psychology of Passwords},
  howpublished = {\url{https://www.lastpass.com/-/media/3c627ed089e84bc39ca2bf6bf1d7cdec.pdf}},
  year = 2022
}

@misc{Lauter:2021:Monitor,
  author = {K. Lauter and S. Kannepalli and K. Laine and R. C. Moreno},
  title = {{Password Monitor}: Safeguarding passwords in {Microsoft Edge}},
  howpublished = {\url{https://www.microsoft.com/en-us/research/blog/password-monitor-safeguarding-passwords-in-microsoft-edge/}},
  month = {21 } # jan,
  year = 2021
}

@misc{LeagueOfEntropy:2024:drand,
  author = {{The League of Entropy}},
  title = {{drand}: A distributed randomness beacon},
  year = 2024,
  howpublished = {\url{https://drand.cloudflare.com/}},
  note = {Accessed: 7 } # dec # { 2024}
}

@inproceedings{Lelarge:2008:Local,
  title = {A local mean field analysis of security investments in networks},
  author = {M. Lelarge and J. Bolot},
  booktitle = {3\textsuperscript{rd} } # NETECON,
  pages = {25--30},
  year = {2008}
}

@misc{Lemos:2021:Stuffing,
  title = {Credential stuffing reaches 193 billion login attempts annually},
  author = {R. Lemos},
  howpublished = {\url{https://www.darkreading.com/cloud-security/credential-stuffing-reaches-193-billion-login-attempts-annually}},
  month = {19 } # may,
  year = 2021
}

@inproceedings{Levin:2008:BitTorrent,
  title = {{BitTorrent} is an auction: Analyzing and improving {BitTorrent}'s incentives},
  author = {D. Levin and K. LaCurts and N. Spring and B. Bhattacharjee},
  booktitle = SIGCOMM,
  pages = {243-254},
  year = 2008
}

@article{Lou:2017:Multidefender,
  title = {Multidefender security games},
  author = {J. Lou and A. M. Smith and Y. Vorobeychik},
  journal = INTSYS,
  volume = {32},
  number = {1},
  pages = {50--60},
  year = {2017}
}

@article{Malkhi:2000:Survivable,
  title = {An architecture for survivable coordination in large distributed systems},
  author = {D. Malkhi and M. K. Reiter},
  journal = IEEETKDE,
  volume = 12,
  number = 2,
  month = mar # {/} # apr,
  year = 2000
}

@inproceedings{Mayer:2022:Managers,
  title = {Why users (don't) use password managers at a large educational institution},
  author = {P. Mayer and C. W. Munyendo and M. L. Mazurek and A. J. Aviv},
  booktitle = {31\textsuperscript{st} } # USESEC,
  month = aug,
  year = 2022
}

@inproceedings{Mayer:2017:Second,
  title={A second look at password composition policies in the wild: Comparing samples from 2010 and 2016},
  author={P. Mayer and J. Kirchner and M. Volkamer},
  booktitle={13\textsuperscript{th}} # SOUPS,
  pages={13--28},
  year={2017}
}

@misc{Meta:2025:ThreatExchange,
  author = {Meta},
  title = {ThreatExchange},
  howpublished = {\url{https://developers.facebook.com/docs/threat-exchange/}},
  year = 2025
}

@inproceedings{Miura:2008:Security,
  title={Security decision-making among interdependent organizations},
  author={R. A. Miura-Ko and B. Yolken and J. Mitchell and N. Bambos},
  booktitle={21\textsuperscript{st} } # CSF,
  pages={66--80},
  year={2008}
}

@inproceedings{Nguyen:2009:Stochastic,
  title = {Stochastic games for security in networks with interdependent nodes},
  author = {K. C. Nguyen and T. Alpcan and T. Basar},
  booktitle = {1\textsuperscript{st} } # GAMENETS,
  pages={697--703},
  year={2009}
}

@inproceedings{Nisenoff:2023:Two,
  title={A $\{$Two-Decade$\}$ Retrospective Analysis of a University's Vulnerability to Attacks Exploiting Reused Passwords},
  author={A. Nisenoff and M. Golla and M. Wei and J. Hainline and H. Szymanek and A. Braun and A. Hildebrandt and B. Christensen and D. Langenberg and B. Ur},
  booktitle = {32\textsuperscript{nd} } # USESEC,
  year={2023}
}

@misc{OneCloud:2024:Detection,
  author = {{OneCloud}},
  title = {What is the average response time to detect a cyber breach in 2024?},
  howpublished = {\url{https://www.onecloud.com.au/resources/what-is-the-average-response-time-to-detect-a-cyber-breach-in-2024/}},
  month = {4 } # sep,
  year = 2024
}

@article{Oruganti:2023:Impact,
  title = {The impact of network design interventions on the security of interdependent systems},
  author = {P. S. Oruganti and P. Naghizadeh and Q. Ahmed},
  journal = IEEETCNS,
  volume = {11},
  number = {1},
  pages = {173--184},
  year = {2023}
}

@book{Osborne:1994:Game,
  XXauthor = {M.J. Osborne and A. Rubinstein},
  author = {M.J. Osborne and A. Rubinstein},
  title = {A Course in Game Theory},
  publisher = {MIT Press},
  year = {1994}
}

@inproceedings{Pal:2019:Beyond,
  title={Beyond credential stuffing: Password similarity models using neural networks},
  author={B. Pal and T. Daniel and R. Chatterjee and T. Ristenpart},
  booktitle={IEEE Security and Privacy},
  year={2019},
}

@inproceedings{Pal:2022:MightIGetPwned,
  title={Might {I} Get Pwned: A Second Generation Compromised Credential Checking Service},
  author={B. Pal and M. Islam and M. Sanusi and N. Sullivan and L. Valenta and T. Whalen and C. Wood and T. Ristenpart and R. Chattejee},
  booktitle= {31\textsuperscript{st} } # USESEC,
  month = aug,
  year = 2022
}

@inproceedings{Pearman:2017:Habitat,
  author = {S. Pearman and J. Thomas and P. E. Naeini and H. Habib and L. Bauer and N. Christin and L. F. Cranor and S. Egelman and A. Forget},
  title = {Let's go in for a closer look: Observing passwords in their natural habitat},
  booktitle = {24\textsuperscript{th} } # CCS,
  month = oct,
  year = 2017
}

@article{Pinkas:2018:PSI,
  author = {B. Pinkas and T. Schneider and M. Zohner},
  title = {Scalable private set intersection based on {OT} extension},
  journal = ACMTOPS,
  volume = 21,
  number = 2,
  year = 2018
}

@misc{Pullman:2019:Checkup,
  author = {J. Pullman and K. Thomas and E. Bursztein},
  title = {Protect your accounts from data breaches with {Password Checkup}},
  howpublished = {\url{https://security.googleblog.com/2019/02/protect-your-accounts-from-data.html}},
  month = {5 } # feb,
  year = 2019
}

@misc{Radwan:2025:Password,
  author = {R. Radwan and S. Zejnilovic},
  title = {Password reuse is rampant: nearly half of observed user logins are compromised},
  howpublished = {\url{https://blog.cloudflare.com/password-reuse-rampant-half-user-logins-compromised/}},
  month = mar,
  year = 2025
}

@misc{Recurly:2024:Business,
  title={Business churn rate by industry},
  author={Recurly Research},
  howpublished = {\url{https://recurly.com/research/churn-rate-benchmarks/}},
  year={2024}
}

@misc{RENISAC:Membership,
  title={Membership},
  author={REN-ISAC},
  howpublished = {\url{https://www.ren-isac.net/membership/membertypes.html}}
}

@article{Robbins:1955:Remark,
  title={A remark on {Stirling's} formula},
  author={H. Robbins},
  journal={The American Mathematical Monthly},
  volume={62},
  number={1},
  pages={26--29},
  year={1955}
}

@article{Stobert:2018:Password,
  title={The password life cycle},
  author={E. Stobert and R. Biddle},
  journal= ACMTOPS,
  volume={21},
  number={3},
  pages={1--32},
  year={2018},
  publisher={ACM New York, NY, USA}
}

@misc{Terry:2025:Honey,
  title = {Honey accounts explained},
  author = {R. Terry},
  howpublished = {\url{https://www.crowdstrike.com/en-us/cybersecurity-101/identity-protection/honey-account/}},
  month = {7 } # jan,
  year = 2025
}

@misc{Verizon:2024:DBIR,
  author = {{Verizon Business}},
  title = {Verizon 2024 Data Breach Investigations Report},
  howpublished = {\url{https://verizon.com/dbir}},
  year = 2024
}

@inproceedings{Wang:2018:Domino,
  title = {The next domino to fall: Empirical analysis of user passwords across online services},
  author = {C. Wang and S. T. K. Jan and H. Hu and D. Bossart and G. Wang},
  booktitle = {8\textsuperscript{th} } # CODASPY,
  month = mar,
  year = 2018,
  pages = {196-203}
}

@inproceedings{Wang:2022:Honeywords,
  title = {How to attack and generate honeywords},
  author = {D. Wang and Y. Zou and Q. Dong and Y. Song and X. Huang},
  booktitle = {43\textsuperscript{rd} } # SSP,
  month = may,
  year = 2022
}

@inproceedings{Wang:2021:Amnesia,
  title = {Using {Amnesia} to detect credential database breaches},
  author={K. C. Wang and M. K. Reiter},
  booktitle= {30\textsuperscript{th} } # USESEC,
  month = aug,
  year = 2021
}

@inproceedings{Wang:2024:Bernoulli,
  title = {Bernoulli honeywords},
  author={K. C. Wang and M. K. Reiter},
  booktitle= {31\textsuperscript{st} } # NDSS,
  month = feb,
  year = 2024
}

@inproceedings{Wash:2016:Understanding,
  title={Understanding password choices: How frequently entered passwords are re-used across websites},
  author={R. Wash and E. Rader and R. Berman and Z. Wellmer},
  booktitle= {12\textsuperscript{th} } # SOUPS,
  year={2016},
  month= jun
}

@article{Zhang:2011:Proportional,
  title = {Proportional response dynamics in the {Fisher} market},
  author = {L. Zhang},
  journal = TCS,
  volume = {412},
  number = {24},
  pages={2691--2698},
  year = {2011}
}

@article{Zhu:2012:GUIDEX,
  author={Q. Zhu and C. Fung and R. Boutaba and T. Basar},
  journal= IEEEJSAC, 
  title={{GUIDEX}: A game-theoretic incentive-based mechanism for intrusion detection networks}, 
  year={2012},
  volume={30},
  number={11},
  pages={2220-2230}
}

@misc{Datatab:2025:Kwh,
  author = {{DATAtab Team}},
  title = {{Kruskal}-{Wallis}-Test},
  howpublished = {\url{https://datatab.net/tutorial/kruskal-wallis-test}}
}

@phdthesis{Fung:2013:IDN,
  author = {C. Fung},
  title = {Design and Management of Collaborative Intrusion Detection Networks},
  school = {University of Waterloo},
  year = 2013
}

@article{Fung:2016:Facid,
  title = {{FACID}: A trust-based collaborative decision framework for intrusion detection networks},
  author = {C. J. Fung and Q. Zhu},
  journal = ADHOCNETS,
  volume = {53},
  pages = {17-31},
  year = {2016}
}

@misc{Hunt:2018:Killer,
  author = {T. Hunt},
  title = {Here's why [insert thing here] is not a password killer},
  howpublished = {\url{https://www.troyhunt.com/heres-why-insert-thing-here-is-not-a-password-killer/}},
  month = {05 } # nov,
  year = 2018
}

@inproceedings{Janakiraman:2003:Indra,
  author = {R. W. Janakiraman and M. Waldvogel and Q. Zhang},
  title = {Indra: A peer-to-peer approach to network intrusion detection and prevention},
  booktitle = {12\textsuperscript{th} IEEE International Workshop on Enabling Technologies: Infrastructure for Collaborative Enterprises},
  month = jun,
  year = 2003
}

@article{Jiang:2010:Bad,
  title = {How bad are selfish investments in network security?},
  author = {L. Jiang and V. Anantharam and J. Walrand},
  journal = IEEETON,
  volume = {19},
  number = {2},
  pages = {549--560},
  year = {2010}
}

@inproceedings{Kwiatkowska:2011:Prism,
  author = {M. Kwiatkowska and G. Norman and D. Parker},
  title = {{PRISM 4.0}: Verification of probabilistic real-time systems},
  booktitle = {International Conference on Computer Aided Verification},
  year = 2011
}

@misc{Lomuscio:2021:Kwh,
  author = {{S. Lomuscio}},
  title = {Getting Started with the {Kruskal}-{Wallis}-Test},
  howpublished = {\url{https://library.virginia.edu/data/articles/getting-started-with-the-kruskal-wallis-test}}
}

@inproceedings{Seitz:2017:Differences,
  title={Do differences in password policies prevent password reuse?},
  author={T. Seitz and M. Hartmann and J. Pfab and S. Souque},
  booktitle={ACM Conference on Human Factors in Computing Systems},
  year={2017}
}

@inproceedings{Wu:2003:CIDS,
  author = {Y.-S. Wu and B. Foo and Y. Mei and S. Bagchi},
  title = {Collaborative intrusion detection system ({CIDS}): A framework for accurate and efficient {IDS}},
  booktitle = {19\textsuperscript{th} Annual Computer Security Applications Conference},
  month = dec,
  year = 2003
}

@inproceedings{Yegneswaran:2004:DOMINO,
  author = {V. Yegneswaran and P. Barford and S. Jha},
  title = {Global intrusion detection in the {DOMINO} overlay system},
  booktitle = {11\textsuperscript{th} } # NDSS,
  month = feb,
  year = 2004
}

@inproceedings{Wu:2007:PropResponse,
  title={Proportional response dynamics leads to market equilibrium},
  author={F. Wu and L. Zhang},
  booktitle={39\textsuperscript{th} } # STOC,
  pages={354--363},
  year={2007}
}

@string{ACISP = {Australasian Conference on Information Security and Privacy}}

@string{ADHOCNETS = {Ad Hoc Networks}}

@string{EUROSP = {IEEE European Symposium on Security and Privacy}}

@string{JCRYPT = {Journal of Cryptology}}

@string{FAST = {USENIX Conference on File and Storage Technologies}}

@string{PODC = {ACM Symposium on Principles of Distributed Computing}}

@string{ISC = {International Conference on Information Security}}

@string{ASIACCS = {ACM Symposium on Information, Computer and Communications Security}}

@string{CANS = {International Conference on Cryptology and Network Security}}

@string{CCS = {ACM Conference on Computer and Communications Security}}

@string{CODASPY = {ACM Conference on Data and Application Security and Privacy}}

@string{CRYPTO = {Advances in Cryptology -- CRYPTO}}

@string{CSF = {IEEE Computer Security Foundations Symposium}}

@string{DEXA = {International Workshop on Database and Expert Systems Applications}}

@string{DISC = {International Conference on Distributed Computing}}

@string{ECONCOMP = {ACM Conference on Economics and Computation}}

@string{ICDCS = {IEEE International Conference on Distributed Computing Systems}}

@string{IMC = {Internet Measurement Conference}}

@string{INTSYS = {IEEE Intelligent Systems}}

@string{JRU = {Journal of Risk and Uncertainty}}

@string{NDSS = {ISOC Network and Distributed System Security Symposium}}

@string{NEURIPS = {Conference on Neural Information Processing Systems}}

@string{SSP = {IEEE Symposium on Security and Privacy}}

@string{SIGCOMM = {ACM SIGCOMM Conference on Applications, Technologies, Architectures, and Protocols for Computer Communications}}

@string{USESEC = {USENIX Security Symposium}}

@string{STOC = {ACM Symposium on Theory of Computing}}

@string{TCS = {Theoretical Computer Science}}

@string{ACMTOPS = {ACM Transactions on Privacy and Security}}

@string{GAMENETS = {International Conference on Game Theory for Networks}}

@string{IEEEJSAC = {IEEE Journal on Selected Areas in Communications}}

@string{IEEETPDS = {IEEE Transactions on Parallel and Distributed Systems}}

@string{IEEETDSC = {IEEE Transactions on Dependable and Secure Computing}}

@string{IEEETKDE = {IEEE Transactions on Knowledge and Data Engineering}}

@string{IEEETCNS = {IEEE Transactions on Control of Network Systems}}

@string{IEEETON = {IEEE/ACM Transactions on Networking}}

@string{NETECON = {Workshop on Economics of Networked Systems}}

@string{SOUPS = {Symposium on Usable Privacy and Security}}

@string{CSUR = {ACM Computing Surveys}}

@string{LNCS = {Lecture Notes in Computer Science}}

@string{COSE = {Computers \& Security}}

\appendices
\crefalias{section}{appsec}

\FloatBarrier
\section{Proportional Response: \PSICALabel Variant}
\label{sec:prop-response-receiving-psica}

The main text presents the exploration in the \PSILabel setting, in
which each site \site knows the membership of $\userSet[\site] \cap
\userSet[\siteAlt]$ for all $\siteAlt \in \peerSet[\site]$. Now, we
present the corresponding \PSICALabel variant, which assumes only that
each \site knows the size of this intersection,
$\setSize{\userSet[\site] \cap \userSet[\siteAlt]}$, so that \site
estimates
\prob{\dodgeEvent{\site}{\siteAlt}{\allocationVal}{\stuffingAttempts}}
by
\begin{align}
  \prob{\dodgeEvent{\site}{\siteAlt}{\allocationVal}{\stuffingAttempts}}
  & \approx \frac{{\setSize{\userSet[\site]} - \stuffingAttempts \choose \min\{\setSize{\userSet[\site]}, \allocationVal\} }}{{\setSize{\userSet[\site]} \choose \min\{\setSize{\userSet[\site]}, \allocationVal\} }}
\label{eqn:psiCaCatchProb}
\end{align}
Since \site does not know $\userSet[\site] \cap \userSet[\siteAlt]$,
it must deploy monitor requests for accounts chosen from
\userSet[\site].  The probability is given by the number of ways \site
can choose accounts that the attacker does not stuff divided by the
total number of ways it can choose accounts. This modified risk
definition is used consistently by both the bidding
(\cref{alg:prop-rsp-bid}) and receiving procedures. In particular,
\cref{alg:prop-rsp-receive-psica} presents the \PSICALabel variant of
the \prop receiving algorithm, which mirrors
\cref{alg:prop-rsp-receive-psi} but differs only in how the effective
allocation \allocationValAdj is computed; since a site \site in the
\PSICALabel setting does not know the membership of $\userSet[\site]
\cap \userSet[\siteAlt]$, it caps \allocationVal by
\setSize{\userSet[\site]} rather than \setSize{\userSet[\site] \cap
\userSet[\siteAlt]}.

\begin{algorithm}[htbp]
\caption{Proportional Response: Receiving (\PSICALabel)}
\label{alg:prop-rsp-receive-psica}
$\avgAllocFrom{\site}{\siteAlt} \gets \bot$ \\

\vspace{1ex}

\Upon{site \site receiving allocation \allocationVal from \siteAlt}{
  
  $\allocationValAdj \gets \min\{\allocationVal,
  \setSize{\userSet[\site]}\}$
  
  \textbf{\rcvAlgoInitLabel:} \If{some $\siteAltAlt \in \siteSet$ has
    not yet placed a bid}{
    \algsteplabel{alg:prop-rsp:init-rcv-alloc}{\rcvAlgoInitLabel}
    $\avgRisk{\site}{\siteAlt} \gets
    \risk{\site}[\siteAlt][\allocationValAdj]$
    \label{line:init-rcv-alloc}
  } \textbf{\rcvAlgoLabel:} \Else{
    \algsteplabel{alg:prop-rsp:rcv-alloc}{\rcvAlgoLabel}
    $\allocationValPrev \gets \ternary{\avgAllocFrom{\site}{\siteAlt} =
      \bot}{\allocationValAdj}{\avgAllocFrom{\site}{\siteAlt}}$
    
    $\avgAllocFrom{\site}{\siteAlt} \gets \smoothingFactor{\site} \times
    \allocationValAdj + (1-\smoothingFactor{\site}) \times
    \allocationValPrev$
    
    $\avgRisk{\site}{\siteAlt} \gets
    \risk{\site}[\siteAlt][\avgAllocFrom{\site}{\siteAlt}]$
    \label{line:exp-smoothing}
  }
  \label{line:avg-risk}
}
\end{algorithm}

We emphasize that the \PSICALabel variant exhibits the same
qualitative behavior as reported throughout the main evaluation.
Accordingly, this appendix focuses on analyses that directly contrast
\PSILabel and \PSICALabel, isolating the effect of whether sites know
the membership of $\userSet[\site] \cap \userSet[\siteAlt]$ or only
its size $\setSize{\userSet[\site] \cap \userSet[\siteAlt]}$. 

\begin{figure}[t]
  \vspace{-3ex}
\centering
\resizebox{0.5\columnwidth}{!}{
\input{user_param_psi_ca.tex}
  } \caption{\normRiskPerBid{\targetSite}{\bidIdx} by \popularity and
\capacityCoeff{\targetSite} for \PSICALabel setting.  Boxes span
25--75 \percentile; whiskers span 5--95 \percentile; diamonds are
means; red lines are medians.}
\label{fig:varying_target_cap_and_user_param_appendix}
\vspace{-2ex}
\end{figure}

\begin{figure}[t]
  \begin{center}
    \resizebox{\columnwidth}{!}{
    \begin{tabular}{c|c|*{4}{C}|*{4}{D}}
      & & \multicolumn{4}{c|}{\improvementFreq{\configVar}} &
      \multicolumn{4}{c}{\fracImprovement{\configVar}} \\
      \hline \hline
      \multirow{2}{*}{\cutLine{\targetSite}} &
      \multirow{2}{*}{\foresight{\targetSite}} 
      & \multicolumn{4}{c|}{\slack} & \multicolumn{4}{c}{\slack} \\ 
      & & \multicolumn{1}{c}{1} & \multicolumn{1}{c}{2} & \multicolumn{1}{c}{3} & \multicolumn{1}{c|}{$\infty$} 
      & \multicolumn{1}{c}{1} & \multicolumn{1}{c}{2} & \multicolumn{1}{c}{3} & \multicolumn{1}{c}{$\infty$} \\ \hline
      \multirow{2}{*}{\boolFalse} & 0 
      & 0.638 & 0.625 & 0.620 & 0.621 
      & 0.258 & 0.231 & 0.220 & 0.213 \\
      & 1 
      & 0.650 & 0.636 & 0.631 & 0.637
      & 0.258 & 0.231 & 0.222 & 0.214 \\ \hline
      \multirow{2}{*}{\boolTrue} & 0 
      & 0.672 & 0.665 & 0.661 & 0.664
      & 0.259 & 0.242 & 0.236 & 0.237 \\
      & 1 
      & 0.681 & 0.676 & 0.670 & 0.673
      & 0.263 & 0.245 & 0.236 & 0.237 \\
    \end{tabular}
    }
  \end{center}
  \caption{Comparison of \exhaustive and \prop strategies across
    \configVar constraints defined by combinations of
    \cutLine{\targetSite[\exhaustiveLabel]},
    \foresight{\targetSite[\exhaustiveLabel]}, and \slack in the
    \PSICALabel setting. The left
    table shows \improvementFreq{\configVar} (\cref{eqn:improvement-freq}) and the right
    table shows \fracImprovement{\configVar}
    (\cref{eqn:frac-improvement}), both diminished when 
    $\configVar \gets \cutLine{\targetSite[\exhaustiveLabel]} =
    \boolFalse$, $\foresight{\targetSite[\exhaustiveLabel]} = 0$,
    $\slack = \infty$.}
  \label{fig:exhaustive_vs_prop_cfs_psi_ca}
\end{figure}

\subsection{Risk outcomes in the \PSICALabel setting}
\label{sec:psica:normrisk}

Using the parameter space defined in \cref{sec:auction:eval}, we
confirmed under \PSICALabel sites retain incentives to increase
capacity to decrease their $\normRiskPerBid{\targetSite}{\bidIdx}$,
and unpopular sites can still mitigate their
$\normRiskPerBid{\targetSite}{\bidIdx}$, as shown in
\cref{fig:varying_target_cap_and_user_param_appendix}. In the
comparison between \PSILabel and \PSICALabel we found the \targetSite
incurred less ($\pValue < 10^{-8}$)
\normRiskPerBid{\targetSite}{\bidIdx} in the \PSILabel setting versus
the \PSICALabel setting. In the \PSILabel setting, \targetSite
monitors only its vulnerable users, $\userSet[\targetSite] \cap
\userSet[\site]$, at peer \site. In contrast, under the \PSICALabel
setting, \targetSite may deploy monitoring requests for users outside
this intersection. While \PSILabel reduces
\normRiskPerBid{\targetSite}{\bidIdx}, it discloses cross-site account
memberships, posing a privacy risk to users.

Also similar to \targetSite[\propLabel], \targetSite using the \prop
strategy, in the \PSILabel setting, a \targetSite[\propLabel] in the
\PSICALabel setting maintains competitiveness with a
\targetSite[\exhaustiveLabel] strategy when $\slack = \infty$,
$\cutLine{\targetSite} = \boolFalse$, and $\foresight{\targetSite} =
0$, as confirmed in
\cref{fig:varying_target_cap_and_user_param_appendix}. However, the
absolute values for \improvementFreq{\configVar} and
\fracImprovement{\configVar} are elevated in the \PSICALabel setting
because it intrinsically generates higher demand for monitoring
slots--—sites would ideally deploy requests for their entire user base
since they cannot know shared users. Although \exhaustive and \prop
use the same imprecise risk estimates in \PSICALabel, this higher
demand creates more scenarios where \exhaustive's secondary
advantages—--access to peer capacities, peer allocations,
\cutLine{\targetSite[\exhaustiveLabel]},
\foresight{\targetSite[\exhaustiveLabel]},
\lookahead{\targetSite[\exhaustiveLabel]}—--can be leveraged to
outperform \prop. Still, we found in that in the \PSICALabel setting
$\improvementFreq{\constrainedConfig} < 0.63$ and
\begin{align*}
  \median \cset{\Bigg}{\frac{\auction[\propLabel][\bidIdx] -
    \auction[\exhaustiveLabel][\bidIdx]}{\auction[\propLabel][\bidIdx]}}{(\auction[\propLabel],\auction[\exhaustiveLabel])
    \in \pairedAuctionsSet[\constrainedConfig]} & = 0.085
\end{align*}

\subsection{Risk predictiveness of cost under the \PSICALabel setting}
\label{sec:psica:risk_prediction}

\begin{figure}[t]
  \begin{subfigure}[b]{0.49\columnwidth}
    \resizebox{\textwidth}{!}{\input{norm_risk_minus_norm_cost_first_round_psi.tex}}
    \caption{\PSILabel}
    \label{fig:norm_risk_minus_norm_cost_round_1:psi_appendix}
  \end{subfigure}
  \hfill
  \begin{subfigure}[b]{0.49\columnwidth}
    \resizebox{\textwidth}{!}{\input{norm_risk_minus_norm_cost_first_round_psi_ca.tex}}
    \caption{\PSICALabel}
    \label{fig:norm_risk_minus_norm_cost_round_1:psica}
  \end{subfigure}
  \caption{Distribution of $\normRiskPerBid{\targetSite}{\firstbidIdx} -
    \normCostPerBid{\targetSite}{\firstbidIdx}$ per
    $\popularity$ and per $\aggressionLevel{\attacker}$. Each boxplot's
    whiskers span the 5th-95th percentile; the diamond is the mean;
    red line is the median. $\normRiskPerBid{\targetSite}{\firstbidIdx}
    - \normCostPerBid{\targetSite}{\firstbidIdx}$ decreased as
    \aggressionLevel{\attacker} and \popularity decreased, and was
    lower for \PSILabel than \PSICALabel.}
  \label{fig:norm_risk_minus_norm_cost_appendix}
\end{figure}

Using the parameter space defined in \cref{sec:attacker:eval}, we
confirmed under \PSICALabel \normRiskPerBid{\targetSite}{\firstbidIdx}
is still a good predictor of
\normCostPerBid{\targetSite}{\firstbidIdx}
(\cref{sec:attacker:eval:early}). As shown in
\cref{fig:norm_risk_minus_norm_cost_round_1:psica}, we also confirmed
under the \PSICALabel setting
$\normRiskPerBid{\targetSite}{\firstbidIdx} -
\normCostPerBid{\targetSite}{\firstbidIdx}$ was decreased for less
popular \targetSite (\cref{sec:attacker:eval:popular}) and for lower
\aggressionLevel{\attacker} (\cref{sec:attacker:eval:aggression}).

In the comparison between \PSILabel and \PSICALabel we found that for
a fixed \popularity and \aggressionLevel{\attacker},
$\normRiskPerBid{\targetSite}{\firstbidIdx} -
\normCostPerBid{\targetSite}{\firstbidIdx}$ was lower under \PSILabel
than under \PSICALabel ($\pValue < 10^{-8}$), as confirmed in
\cref{fig:norm_risk_minus_norm_cost_appendix}. The reduced predictive
ability of \normRiskPerBid{\targetSite}{\bidIdx} in the \PSICALabel
setting is a shortcoming further discussed in
\cref{sec:discussion:challenges}.

\subsection{Cost outcomes in the \PSICALabel setting} 
\label{sec:psica:normcost}

Using the parameter space defined in \cref{sec:attacker:larger}, we
confirmed under the \PSICALabel setting sites minimize their cost by
increasing their capacites (\cref{sec:attacker:larger:scap}), popular
sites still benefit from their peers
(\cref{sec:attacker:larger:popular}) and unpopular sites retain
comparable security (\cref{sec:attacker:larger:unpopular}). We also
confirmed these trends using the parameter space defined in
\cref{sec:data-driven:eval} as evidenced by
\cref{fig:datadriven_appendix}.

In both the model-checking and data-driven validation, we confirmed
that \normCost{\targetSite} was significantly lower ($\pValue <
10^{-8}$) in the \PSILabel setting than in \PSICALabel. Lacking
knowledge of the membership of $\userSet[\targetSite] \cap
\userSet[\site]$ causes \targetSite to deploy requests for users
outside this intersection—users the attacker knows not to stuff at
\site. The effect of using \PSICALabel versus \PSILabel on
\normCost{\targetSite} was more pronounced at lower
\capacityCoeff{\site}, since when \targetSite receives limited slots,
strategic deployment becomes even more important.

\begin{figure}[htbp]
  \begin{subfigure}[t]{0.48\columnwidth}
    \centering
    \begin{tabular}{@{}c@{}}
      \FPeval{\MinNumber}{0.0004}
      \FPeval{\MaxNumber}{0.0019}
      \resizebox{\columnwidth}{!}{
        \begin{tabular}{@{\hspace{2pt}}c@{\hspace{2pt}}|*{4}{X}}
          & \multicolumn{4}{c}{\capacityCoeff{\targetSite}} \\
          \setSize{\userSet[\targetSite]}
          & \multicolumn{1}{c}{0.1}
          & \multicolumn{1}{c}{1}
          & \multicolumn{1}{c}{10}
          & \multicolumn{1}{c}{100}
          \\ \hline
          XL & 0.0007 & 0.0005 & 0.0005 & 0.0005 \\
          L & 0.0015 & 0.0012 & 0.0013 & 0.0013 \\
          M & 0.0011 & 0.0008 & 0.0008 & 0.0008 \\
          S & 0.0019 & 0.0015 & 0.0014 & 0.0014 \\  
        \end{tabular}
      }
    \end{tabular}
    \caption{$\capacityCoeff{\site} = 2$, \PSILabel}
    \label{fig:datadriven:cap2:psi_appendix}
  \end{subfigure}
  \hfill
  \begin{subfigure}[t]{0.48\columnwidth}
    \centering
    \begin{tabular}{@{}c@{}}
      \FPeval{\MinNumber}{0.0004}
      \FPeval{\MaxNumber}{0.0019}
      \resizebox{\columnwidth}{!}{
        \begin{tabular}{@{\hspace{2pt}}c@{\hspace{2pt}}|*{4}{X}}
          & \multicolumn{4}{c}{\capacityCoeff{\targetSite}} \\
          \setSize{\userSet[\targetSite]}
          & \multicolumn{1}{c}{0.1}
          & \multicolumn{1}{c}{1}
          & \multicolumn{1}{c}{10}
          & \multicolumn{1}{c}{100}
          \\ \hline
          XL & 0.0006 & 0.0004 & 0.0004 & 0.0004 \\
          L & 0.0013 & 0.0010 & 0.0010 & 0.0010 \\
          M & 0.0010 & 0.0006 & 0.0005 & 0.0005 \\
          S & 0.0018 & 0.0013 & 0.0012 & 0.0012 \\
        \end{tabular}
      }
    \end{tabular}
    \caption{$\capacityCoeff{\site} = 4, \PSILabel$}
    \label{fig:datadriven:cap4:psi_appendix}
  \end{subfigure}
  \\
  \begin{subfigure}[t]{0.48\columnwidth}
    \centering
    \begin{tabular}{@{}c@{}}
      \FPeval{\MinNumber}{0.0022}
      \FPeval{\MaxNumber}{0.0034}
      \resizebox{\columnwidth}{!}{
        \begin{tabular}{@{\hspace{2pt}}c@{\hspace{2pt}}|*{4}{X}}
          & \multicolumn{4}{c}{\capacityCoeff{\targetSite}} \\
          \setSize{\userSet[\targetSite]}
          & \multicolumn{1}{c}{0.1}
          & \multicolumn{1}{c}{1}
          & \multicolumn{1}{c}{10}
          & \multicolumn{1}{c}{100}
          \\ \hline
          XL & 0.0027 & 0.0024 & 0.0024 & 0.0024 \\
          L & 0.0025 & 0.0022 & 0.0022 & 0.0022 \\
          M & 0.0033 & 0.0030 & 0.0030 & 0.0029 \\
          S & 0.0034 & 0.0033 & 0.0032 & 0.0032 \\
        \end{tabular}
      }
    \end{tabular}
    \caption{$\capacityCoeff{\site} = 2, 
      \PSICALabel$} \label{fig:datadriven:cap2:psica}
  \end{subfigure}
  \hfill
  \begin{subfigure}[t]{0.48\columnwidth}
    \centering
    \begin{tabular}{@{}c@{}}
      \FPeval{\MinNumber}{0.0020}
      \FPeval{\MaxNumber}{0.0030}
      \resizebox{\columnwidth}{!}{
        \begin{tabular}{@{\hspace{2pt}}c@{\hspace{2pt}}|*{4}{X}}
          & \multicolumn{4}{c}{\capacityCoeff{\targetSite}} \\
          \setSize{\userSet[\targetSite]}
          & \multicolumn{1}{c}{0.1}
          & \multicolumn{1}{c}{1}
          & \multicolumn{1}{c}{10}
          & \multicolumn{1}{c}{100}
          \\ \hline
          XL & 0.0023 & 0.0021 & 0.0020 & 0.0020 \\
          L & 0.0022 & 0.0021 & 0.0021 & 0.0021 \\
          M & 0.0027 & 0.0025 & 0.0024 & 0.0024 \\
          S & 0.0030 & 0.0030 & 0.0030 & 0.0030 \\
        \end{tabular}
      }
    \end{tabular}
    \caption{$\capacityCoeff{\site} = 4,  
      \PSICALabel$} \label{fig:datadriven:cap4:psica}
  \end{subfigure}
  \caption{Mean \normCost{\targetSite} for
    $\aggressionLevel{\attacker} = 1.0$. Lighter cells indicate lower
    \normCost{\targetSite}.}
  \label{fig:datadriven_appendix}
\end{figure}

\section{Proportional Response: Statements and Proofs}
\label{sec:prop-response-proofs}

Here we provide the formal statements and proofs of properties
satisfied by \prop as defined in
\cref{alg:prop-rsp-bid,alg:prop-rsp-receive-psi}.

\noindent\textbf{Allocation feasibility.} We claim:
\[\allocTo{\site}{\siteAlt} \ge 0 \text{ for all } \siteAlt \in
\peerSet[\site]\]
and
\[\sum_{\siteAlt \in
\peerSet[\site]} \allocTo{\site}{\siteAlt} \le \capacity{\site}\]

\noindent\emph{Bootstrap bidding} (\ref{alg:prop-rsp:init-bid}). 

\noindent\emph{Step 1: Baseline weights are non-negative.} 
Since $\risk{\site}[\siteAlt][1] \ge 0$ and $\risk{\site}[\siteAlt][1] \le \sum_{\siteAltAlt \in \peerSet[\site]} \risk{\site}[\siteAltAlt][1]$:
\[\baselineWeight{\site}{\siteAlt} = \frac{1}{\nmbrSites-2} \left(1 -
\frac{\risk{\site}[\siteAlt][1]}{\sum_{\siteAltAlt \in
\peerSet[\site]} \risk{\site}[\siteAltAlt][1]} \right) \ge 0\]

\noindent\emph{Step 2: Baseline weights sum to 1.}
\begin{align*}
\sum_{\siteAlt \in \peerSet[\site]} \baselineWeight{\site}{\siteAlt} 
&= \frac{1}{\nmbrSites-2} \left(\sum_{\siteAlt \in \peerSet[\site]} 1 - \frac{\sum_{\siteAlt \in \peerSet[\site]} \risk{\site}[\siteAlt][1]}{\sum_{\siteAltAlt \in \peerSet[\site]} \risk{\site}[\siteAltAlt][1]} \right) \\
&= \frac{1}{\nmbrSites-2}\bigl((\nmbrSites-1) - 1\bigr) = 1
\end{align*}

\noindent\emph{Step 3: Leftover is non-negative.}
\[\sum_{\siteAlt \in \peerSet[\site]} \left\lfloor \capacity{\site} \times \baselineWeight{\site}{\siteAlt} \right\rfloor \le \capacity{\site} \sum_{\siteAlt \in \peerSet[\site]} \baselineWeight{\site}{\siteAlt} = \capacity{\site}\]
so $\capacity{\site} - \sum_{\siteAlt \in \peerSet[\site]} \lfloor \capacity{\site} \times \baselineWeight{\site}{\siteAlt} \rfloor \ge 0$.

\noindent\emph{Step 4: All allocations are non-negative.}
Let $\siteAltAltAlt = \argmin_{\siteAlt \in \peerSet[\site]} \baselineWeight{\site}{\siteAlt}$. 
For $\siteAlt \neq \siteAltAltAlt$:
\[\allocTo{\site}{\siteAlt} = \left\lfloor \capacity{\site} \times \baselineWeight{\site}{\siteAlt} \right\rfloor \ge 0\]
For $\siteAltAltAlt$, using Step 3:
\begin{align*}
\allocTo{\site}{\siteAltAltAlt} &= \left\lfloor \capacity{\site} \times \baselineWeight{\site}{\siteAltAltAlt} \right\rfloor \\
&\quad + \left(\capacity{\site} - \sum_{\siteAlt \in \peerSet[\site]} \left\lfloor \capacity{\site} \times \baselineWeight{\site}{\siteAlt} \right\rfloor\right) \ge 0
\end{align*}

\noindent\emph{Step 5: Sum equals capacity.}
\begin{align*}
\sum_{\siteAlt \in \peerSet[\site]} \allocTo{\site}{\siteAlt} 
&= \sum_{\siteAlt \in \peerSet[\site]} \left\lfloor \capacity{\site} \times \baselineWeight{\site}{\siteAlt} \right\rfloor \\
&\quad + \capacity{\site} - \sum_{\siteAlt \in \peerSet[\site]} \left\lfloor \capacity{\site} \times \baselineWeight{\site}{\siteAlt} \right\rfloor \\
&= \capacity{\site}
\end{align*}

\noindent\emph{Steady-state bidding} (\ref{alg:prop-rsp:next-bid}). 

\noindent\emph{Steps 1--2: Weights are non-negative and sum to 1.}
By the same argument with $\avgRisk{\site}{\siteAlt}$ in place of $\risk{\site}[\siteAlt][1]$, $\weight{\site}{\siteAlt} \ge 0$ and $\sum_{\siteAlt \in \peerSet[\site]} \weight{\site}{\siteAlt} = 1$. Since $\blendedWeight{\site}{\siteAlt}$ is a convex combination:
\begin{align*}
\sum_{\siteAlt \in \peerSet[\site]} \blendedWeight{\site}{\siteAlt} 
&= \frac{1}{1+\maxRisk} \sum_{\siteAlt \in \peerSet[\site]} \baselineWeight{\site}{\siteAlt} \\
&\quad + \frac{\maxRisk}{1+\maxRisk} \sum_{\siteAlt \in \peerSet[\site]} \weight{\site}{\siteAlt} \\
&= \frac{1}{1+\maxRisk} \times 1 + \frac{\maxRisk}{1+\maxRisk} \times 1 = 1
\end{align*}

\noindent\emph{Steps 3--5: Leftover is non-negative, allocations are non-negative, sum equals capacity.}
By the same argument as bootstrap bidding with $\blendedWeight{\site}{\siteAlt}$ in place of $\baselineWeight{\site}{\siteAlt}$, $\sum_{\siteAlt \in \peerSet[\site]} \allocTo{\site}{\siteAlt} = \capacity{\site}$.
\hfill$\square$

\noindent\textbf{Capacity exhaustion.} We claim:
\[\sum_{\siteAlt \in \peerSet[\site]} \allocTo{\site}{\siteAlt} = \capacity{\site}\]

This follows directly from Step 5 of the allocation feasibility proof.
\hfill$\square$

\noindent\textbf{Monotonicity in risk.} We claim allocations are weakly decreasing in risk, up to at most $\nmbrSites-1$ slots:

\noindent\emph{Bootstrap bidding} (\ref{alg:prop-rsp:init-bid}). For $\siteAlt, \siteAltAlt \in \peerSet[\site]$:
\[\risk{\site}[\siteAlt][1] < \risk{\site}[\siteAltAlt][1] \implies \allocTo{\site}{\siteAlt} \ge \allocTo{\site}{\siteAltAlt} - (\nmbrSites-1)\]

\noindent\emph{Step 1: Lower risk implies higher baseline weight.}
For $\siteAlt, \siteAltAlt \in \peerSet[\site]$ with $\risk{\site}[\siteAlt][1] < \risk{\site}[\siteAltAlt][1]$:
\begin{align*}
\frac{\risk{\site}[\siteAlt][1]} {\sum_{\siteAltAltAlt \in \peerSet[\site]} \risk{\site}[\siteAltAltAlt][1]} &< \frac{\risk{\site}[\siteAltAlt][1]} {\sum_{\siteAltAltAlt \in \peerSet[\site]} \risk{\site}[\siteAltAltAlt][1]} \\
\implies 1 - \frac{\risk{\site}[\siteAlt][1]}{\sum_{\siteAltAltAlt \in \peerSet[\site]} \risk{\site}[\siteAltAltAlt][1]} &> 1 - \frac{\risk{\site}[\siteAltAlt][1]}{\sum_{\siteAltAltAlt \in \peerSet[\site]} \risk{\site}[\siteAltAltAlt][1]} \\
\implies \baselineWeight{\site}{\siteAlt} &> \baselineWeight{\site}{\siteAltAlt}
\end{align*}

\noindent\emph{Step 2: Higher baseline weight implies higher allocation up to $\nmbrSites-1$ slots.}
Let $\siteAltAltAlt = \argmin_{\siteAlt \in \peerSet[\site]} \baselineWeight{\site}{\siteAlt}$. 

\noindent\emph{Case 1:} $\siteAltAlt \neq \siteAltAltAlt$. Then:
\begin{align*}
\allocTo{\site}{\siteAlt} &= \lfloor \capacity{\site} \times \baselineWeight{\site}{\siteAlt} \rfloor \\
&\ge \lfloor \capacity{\site} \times \baselineWeight{\site}{\siteAltAlt} \rfloor = \allocTo{\site}{\siteAltAlt}
\end{align*}

\noindent\emph{Case 2:} $\siteAltAlt = \siteAltAltAlt$. Then
$\allocTo{\site}{\siteAltAlt}$ includes leftover at most
$\nmbrSites-1$ slots:
\begin{align*}
\allocTo{\site}{\siteAlt} &= \lfloor \capacity{\site} \times \baselineWeight{\site}{\siteAlt} \rfloor \\
&\ge \lfloor \capacity{\site} \times \baselineWeight{\site}{\siteAltAlt} \rfloor \\
&\ge \allocTo{\site}{\siteAltAlt} - (\nmbrSites-1)
\end{align*}

\noindent\emph{Steady-state bidding} (\ref{alg:prop-rsp:next-bid}). For $\siteAlt, \siteAltAlt \in \peerSet[\site]$, if both $\avgRisk{\site}{\siteAlt} < \avgRisk{\site}{\siteAltAlt}$ and $\risk{\site}[\siteAlt][1] \le \risk{\site}[\siteAltAlt][1]$:
\[\allocTo{\site}{\siteAlt} \ge \allocTo{\site}{\siteAltAlt} - (\nmbrSites-1)\]

\noindent\emph{Step 1: Lower average risk implies higher weight.} 
By the same argument with $\avgRisk{\site}{\siteAlt}$ in place of $\risk{\site}[\siteAlt][1]$, $\weight{\site}{\siteAlt} > \weight{\site}{\siteAltAlt}$. 

\noindent\emph{Step 2: Higher weight with ordered baseline weights implies higher allocation up to $\nmbrSites-1$ slots.}
From the condition $\risk{\site}[\siteAlt][1] \le \risk{\site}[\siteAltAlt][1]$ and the bootstrap case analysis, $\baselineWeight{\site}{\siteAlt} \ge \baselineWeight{\site}{\siteAltAlt}$. Therefore:
\begin{align*}
\blendedWeight{\site}{\siteAlt} &= \frac{1}{1+\maxRisk}\baselineWeight{\site}{\siteAlt} + \frac{\maxRisk}{1+\maxRisk}\weight{\site}{\siteAlt} \\
&> \frac{1}{1+\maxRisk}\baselineWeight{\site}{\siteAltAlt} + \frac{\maxRisk}{1+\maxRisk}\weight{\site}{\siteAltAlt} \\
&= \blendedWeight{\site}{\siteAltAlt}
\end{align*}
By the same case analysis as bootstrap bidding, $\allocTo{\site}{\siteAlt} \ge \allocTo{\site}{\siteAltAlt} - (\nmbrSites-1)$.
\hfill$\square$

\noindent\textbf{Allocation smoothness.} We claim that similar risks yield similar allocations:

\noindent\emph{Bootstrap bidding} (\ref{alg:prop-rsp:init-bid}). For $\siteAlt, \siteAltAlt \in \peerSet[\site]$, if
\[\left|\frac{\risk{\site}[\siteAlt][1]}{\sum_{\siteAltAltAlt} \risk{\site}[\siteAltAltAlt][1]} - \frac{\risk{\site}[\siteAltAlt][1]}{\sum_{\siteAltAltAlt} \risk{\site}[\siteAltAltAlt][1]}\right| \le \mathEpsilon\]
then $|\allocTo{\site}{\siteAlt} - \allocTo{\site}{\siteAltAlt}| = O(\mathEpsilon \times \capacity{\site})$.

\noindent\emph{Step 1: Similar normalized risks imply similar baseline weights.}
\begin{align*}
&|\baselineWeight{\site}{\siteAlt} - \baselineWeight{\site}{\siteAltAlt}| \\
&= \frac{1}{\nmbrSites-2} \left|\frac{\risk{\site}[\siteAltAlt][1]}{\sum_{\siteAltAltAlt} \risk{\site}[\siteAltAltAlt][1]} - \frac{\risk{\site}[\siteAlt][1]}{\sum_{\siteAltAltAlt} \risk{\site}[\siteAltAltAlt][1]}\right| \\
&\le \frac{\mathEpsilon}{\nmbrSites-2}
\end{align*}

\noindent\emph{Step 2: Similar baseline weights imply similar allocations.}
\begin{align*}
&|\allocTo{\site}{\siteAlt} - \allocTo{\site}{\siteAltAlt}| \\
&= \left|\left\lfloor \capacity{\site} \times \baselineWeight{\site}{\siteAlt} \right\rfloor - \left\lfloor \capacity{\site} \times \baselineWeight{\site}{\siteAltAlt} \right\rfloor\right| \\
&\le \capacity{\site} \times |\baselineWeight{\site}{\siteAlt} - \baselineWeight{\site}{\siteAltAlt}| + 1 \\
&\le \capacity{\site} \times \frac{\mathEpsilon}{\nmbrSites-2} + 1 \\
&= O(\mathEpsilon \times \capacity{\site})
\end{align*}

\noindent\emph{Steady-state bidding} (\ref{alg:prop-rsp:next-bid}). For $\siteAlt, \siteAltAlt \in \peerSet[\site]$, if both
\[\left|\frac{\risk{\site}[\siteAlt][1]}{\sum_{\siteAltAltAlt} \risk{\site}[\siteAltAltAlt][1]} - \frac{\risk{\site}[\siteAltAlt][1]}{\sum_{\siteAltAltAlt} \risk{\site}[\siteAltAltAlt][1]}\right| \le \mathEpsilon\]
and
\[\left|\frac{\avgRisk{\site}{\siteAlt}}{\sum_{\siteAltAltAlt} \avgRisk{\site}{\siteAltAltAlt}} - \frac{\avgRisk{\site}{\siteAltAlt}}{\sum_{\siteAltAltAlt} \avgRisk{\site}{\siteAltAltAlt}}\right| \le \mathEpsilon\]
then $|\allocTo{\site}{\siteAlt} - \allocTo{\site}{\siteAltAlt}| = O(\mathEpsilon \times \capacity{\site})$.

\noindent\emph{Step 1: Similar normalized average risks imply similar weights.}
By the same argument with $\avgRisk{\site}{\siteAlt}$ in place of $\risk{\site}[\siteAlt][1]$:
\[|\weight{\site}{\siteAlt} - \weight{\site}{\siteAltAlt}| \le \frac{\mathEpsilon}{\nmbrSites-2}\]

\noindent\emph{Step 2: Similar baseline weights and similar weights imply similar allocations.}
From the bootstrap case with the first condition, $|\baselineWeight{\site}{\siteAlt} - \baselineWeight{\site}{\siteAltAlt}| \le \frac{\mathEpsilon}{\nmbrSites-2}$. Therefore:
\begin{align*}
&|\allocTo{\site}{\siteAlt} - \allocTo{\site}{\siteAltAlt}| \\
&\le \capacity{\site} \times |\blendedWeight{\site}{\siteAlt} - \blendedWeight{\site}{\siteAltAlt}| + 1 \\
&\le \capacity{\site} \times \biggl(\frac{1}{1+\maxRisk} |\baselineWeight{\site}{\siteAlt} - \baselineWeight{\site}{\siteAltAlt}| \\
&\quad + \frac{\maxRisk}{1+\maxRisk} |\weight{\site}{\siteAlt} - \weight{\site}{\siteAltAlt}|\biggr) + 1 \\
&\le \capacity{\site} \times \frac{\mathEpsilon}{\nmbrSites-2} \biggl(\frac{1}{1+\maxRisk} + \frac{\maxRisk}{1+\maxRisk}\biggr) + 1 \\
&= \capacity{\site} \times \frac{\mathEpsilon}{\nmbrSites-2} + 1 = O(\mathEpsilon \times \capacity{\site})
\end{align*}
\hfill$\square$

\noindent\textbf{Allocation stability.}
\noindent\emph{Steady-state bidding} (\ref{alg:prop-rsp:next-bid}). For $\siteAlt, \siteAltAlt \in \peerSet[\site]$ with similar baseline weights
\[\left|\baselineWeight{\site}{\siteAlt} - \baselineWeight{\site}{\siteAltAlt}\right| \le \frac{\mathEpsilon}{\nmbrSites-2}\]
and small maximum observed risk $\maxRisk \le \mathDelta$ for $\mathDelta = O(\mathEpsilon)$, the allocations satisfy
\[\left|\allocTo{\site}{\siteAlt} - \allocTo{\site}{\siteAltAlt}\right| = O(\mathEpsilon \times \capacity{\site})\]

\noindent\emph{Step 1: When max risk is small, blended weights are anchored by baseline weights.}
\begin{align*}
&|\blendedWeight{\site}{\siteAlt} - \blendedWeight{\site}{\siteAltAlt}| \\
&\le \frac{1}{1+\maxRisk} |\baselineWeight{\site}{\siteAlt} - \baselineWeight{\site}{\siteAltAlt}| \\
&\quad + \frac{\maxRisk}{1+\maxRisk} |\weight{\site}{\siteAlt} - \weight{\site}{\siteAltAlt}| \\
&\le \frac{1}{1+\mathDelta} \times \frac{\mathEpsilon}{\nmbrSites-2} + \frac{\mathDelta}{1+\mathDelta} \times \frac{1}{\nmbrSites-2} \\
&\le \frac{\mathEpsilon}{\nmbrSites-2} + \frac{\mathDelta}{\nmbrSites-2} = O\left(\frac{\mathEpsilon}{\nmbrSites-2}\right)
\end{align*}

\noindent\emph{Step 2: Similar blended weights imply similar allocations.}
\begin{align*}
&|\allocTo{\site}{\siteAlt} - \allocTo{\site}{\siteAltAlt}| \\
&\le \capacity{\site} \times |\blendedWeight{\site}{\siteAlt} - \blendedWeight{\site}{\siteAltAlt}| + 1 \\
&\le \capacity{\site} \times O\left(\frac{\mathEpsilon}{\nmbrSites-2}\right) + 1 = O(\mathEpsilon \times \capacity{\site})
\end{align*}
\hfill$\square$

\noindent\textbf{Risk persistence.} We claim past allocations cannot
permanently inflate a peer's perceived contribution. Assuming we are
in \ref{alg:prop-rsp:rcv-alloc}, if $\smoothingFactor{\site} \in
(0,1]$ and peer $\siteAlt$ stops allocating to site $\site$ after bid
index $\bidIdx$, then for any $\bidIdxAlt > \bidIdx$:
\[\avgAllocFrom{\site}{\siteAlt} = O\!\left((1 -
\smoothingFactor{\site})^{\bidIdxAlt - \bidIdx}\right)\]

\noindent\emph{Step 1: Average contribution decays when allocations stop.}
Let $\avgAllocFrom{\site}{\siteAlt}_{\bidIdxAlt}$ be the
allocation $\site$ recieved from $\siteAlt$ at bid index $\bidIdx$.
Suppose $\siteAlt$ stops allocating to $\site$ after bid index
$\bidIdx$, so $\avgAllocFrom{\site}{\siteAlt} = 0$ for all $\bidIdxAlt
> \bidIdx$. 

\[\avgAllocFrom{\site}{\siteAlt}_{\bidIdxAlt} = (1 - \smoothingFactor{\site}) \times \avgAllocFrom{\site}{\siteAlt}_{\bidIdxAlt-1}\]

\noindent\emph{Step 2: Unrolling the recurrence gives exponential decay.}
Unrolling for $\bidIdxAlt - \bidIdx$ steps:
\begin{align*}
\avgAllocFrom{\site}{\siteAlt}_{\bidIdxAlt} 
&= (1 - \smoothingFactor{\site})^{\bidIdxAlt - \bidIdx} \times \avgAllocFrom{\site}{\siteAlt}_{\bidIdx}
\end{align*}
Since $(1 - \smoothingFactor{\site}) \in [0,1)$ and $\avgAllocFrom{\site}{\siteAlt}_{\bidIdx}$ is constant:
\[\avgAllocFrom{\site}{\siteAlt}_{\bidIdxAlt} = O\left((1 - \smoothingFactor{\site})^{\bidIdxAlt - \bidIdx}\right)\]
\hfill$\square$

\section{Performance}
\label{sec:performance}
Here we evaluate the performance of our design. The time a site spends
participating in a breach-detection ecosystem can be attributed to
either the overhead imposed by our bidding infrastructure, or to the
time of making and responding to monitoring requests according to
previously developed interactive breach detection
protocols~\cite{Wang:2021:Amnesia, Wang:2024:Bernoulli}.

For the sake of this analysis, we consider the Amnesia
protocol~\cite{Wang:2021:Amnesia}, and we evaluate the time to perform
a single bidding step in our design.  Specifically, the time incurred
by a bidding step that is directly attributable to our design includes
the time a site \site takes to allocate its capacity in step
\ref{alg:prop-rsp:next-bid}, denoted
\runtime{\ref{alg:prop-rsp:next-bid}},\footnote{A naive implementation
would cause \ref{alg:prop-rsp:next-bid} to scale with sites and users;
\cref{sec:performance:optimization} shows how we reduce this to sites
only.} and the time each peer $\siteAlt \in \peerSet[\site]$ takes to
update its slot allocations from \site in step
\ref{alg:prop-rsp:rcv-alloc}, denoted
\runtime{\ref{alg:prop-rsp:rcv-alloc}}, accumulated over all
peers---so, $(\nmbrSites-1) \times
\runtime{\ref{alg:prop-rsp:rcv-alloc}}$. This bid induces additional
computation on the peer sites \peerSet[\site], however, to create
monitoring requests per the Amnesia protocol, to fulfill the
allocation each receives in this bidding step.  We denote the time to
generate one monitoring request in this protocol as
\runtime{\monReqGenOp}, and accumulate this time per monitoring slot
that \site allocates, i.e., $\capacity{\site} \times
\runtime{\monReqGenOp}$ in total.  Note that this time is invariant to
the actual bids or how they are computed, and so is best attributed to
Amnesia itself.  We report the ratio of ``bidding time'' to the
``Amnesia time'', or in other words
\begin{align}
\frac{\runtime{\ref{alg:prop-rsp:next-bid}} + (\nmbrSites-1) \times
  \runtime{\ref{alg:prop-rsp:rcv-alloc}}}{\capacity{\site} \times \runtime{\monReqGenOp}}
\label{eqn:timeRatio}
\end{align}

We used the implementation of the Amnesia protocol due to Wang et
al.~\cite{Wang:2021:Amnesia}. To minimize \runtime{\monReqGenOp}, we
conservatively set the number of honeywords monitored per account to
$16$, the lowest number they reported, and adopted the remaining
recommended parameters from their work.  We implemented our \prop
bidding strategy in Python, and compiled performance-critical
functions (such as \cref{eqn:risk}) to machine code using NUMBA. We
conducted experiments on a single machine running Ubuntu 22.04.5 LTS,
with an Intel Xeon Gold 6226 processor (2.7~GHz), 768~GB of RAM, and a
fixed configuration of two threads.

\begin{figure}[t]
\begin{subfigure}[b]{0.48\columnwidth}
  \vspace*{6ex}
  \resizebox{\textwidth}{!}{
    \begin{tabular}{@{}c@{\hspace{2pt}}|*{5}{c}}
      & \multicolumn{5}{c}{\capacity{\site}} \\
      \nmbrSites & $10^{1}$ & $10^{2}$ & $10^{3}$ & $10^{4}$ & $10^{5}$
      \\ \hline
      $10^{5}$ & 0.0484 & 0.0166 & 0.0135 & 0.0132 & 0.0131 \\
      $10^{4}$ & 0.0042 & 0.0014 & 0.0011 & 0.0011 & 0.0011 \\
      $10^{3}$ & 0.0013 & 0.0004 & 0.0003 & 0.0003 & 0.0003 \\
      $10^{2}$ & 0.0006 & 0.0002 & 0.0002 & 0.0002 & 0.0002 \\
      $10^{1}$ & 0.0043 & 0.0006 & 0.0002 & 0.0002 & 0.0002 \\
    \end{tabular}
  }
  \vspace*{4ex}
  \caption{Ratio of bidding time to monitoring request
    generation time (\cref{eqn:timeRatio}).  Values $<1.0$ mean
    bidding is cheaper.}
  \label{fig:performance:relative}
\end{subfigure}
\hfill
\begin{subfigure}[b]{0.48\columnwidth}
  \centering
  \resizebox{\textwidth}{!}{
    \begin{tikzpicture}
\begin{semilogyaxis}[
    width=12cm,
    height=8cm,
    grid=both,
    minor grid style={line width=0.25pt,draw=gray!10},
    major grid style={line width=0.5pt,draw=gray!30},
    xlabel={\nmbrSites},
    ylabel={Time (ms)},
    legend style={at={(0.98,0.02)}, anchor=south east, font=\Large},
    label style={font=\Large},
    tick label style={font=\Large},
    log basis y=10,
    xmin=10000,
    scaled x ticks=base 10:6,
    xtick scale label code/.code={},
    xtick={10000,100000,500000,1000000,5000000,10000000},
    xticklabels={$10^4$,$10^5$,$5 \times 10^5$,$10^6$,$5 \times 10^6$,
    $10^7$},
    ymax=500,
    ymin = 0.001,
    cycle list={
        {black, solid, line width=2pt},
        {black, dotted, line width=2pt},
        {blue, solid, line width=1.5pt},
        {blue, dotted, line width=1.5pt}
    }
]

\addplot coordinates {
  (10,0.03355979919433594)
  (50,0.03205299377441406)
  (100,0.030870437622070312)
  (500,0.046987533569334224)
  (1000,0.04925251007080078)
  (5000,0.04292011260986328)
  (10000,0.054526329040527344)
  (50000,0.514960289001414)
  (100000,1.046390533447218)
  (500000,3.6697888374328183)
  (1000000,8.238291740417429)
};

\addplot coordinates {
  (10,1.1321091651916286)
  (50,0.11007690429682232)
  (100,0.11513185501094408)
  (500,0.18712043762202302)
  (1000,0.267312526702835)
  (5000,0.47260379791254975)
  (10000,0.8454513549804241)
  (50000,4.154674530029248)
  (100000,9.703459739685009)
  (500000,49.01626443862911)
  (1000000,104.95905590057369)
};

\addplot coordinates {
  (10,7.551102638244628)
  (50,0.00843048095703125)
  (100,0.008702278137207031)
  (500,0.011129379272460938)
  (1000,0.011725425720214844)
  (5000,0.017547607421875)
  (10000,0.02601146697998047)
  (50000,0.1496362686156777)
  (100000,0.347385406494086)
  (500000,1.418466567993111)
  (1000000,4.192185401916452)
};

\addplot coordinates {
  (10,0.004466533660888672)
  (50,0.004374027252197266)
  (100,0.00440216064453125)
  (500,0.006808280944824219)
  (1000,0.007806301116943359)
  (5000,0.018019676208496094)
  (10000,0.030243396759033203)
  (50000,0.142571926116895)
  (100000,0.3611102104186524)
  (500000,1.9939696788787369)
  (1000000,4.56371307373042)
};

\legend{\bidAlgoBoostrapLabel, \bidAlgoLabel,\rcvAlgoInitLabel,\rcvAlgoLabel}
\end{semilogyaxis}
\end{tikzpicture}
  } 
  \caption{Time to compute bids
  (\ref{alg:prop-rsp:init-bid}, \ref{alg:prop-rsp:next-bid}) and
  cumulative time for peers to process bids
  (\ref{alg:prop-rsp:init-rcv-alloc}, \ref{alg:prop-rsp:rcv-alloc}).}
  \label{fig:performance:absolute}
\end{subfigure}
\caption{Performance of \prop bidding}
\label{fig:performance}
\end{figure}

\cref{fig:performance:relative} shows the values of
\cref{eqn:timeRatio} as a function of \capacity{\site} and \nmbrSites.
As illustrated by the very small values, the timing cost of our
bidding algorithm is overwhelmed by the time for Amnesia to generate
monitoring requests in response to our bids.  That is, the timing
costs of our \prop algorithm are a tiny fraction of the time needed to
deploy monitoring requests in total.  By contrast, \exhaustive bidding
dominates the time to generate monitoring requests; e.g., the analog
of \cref{eqn:timeRatio} for \exhaustive bidding is $>7.18$ for
$\nmbrSites = 5$ and $\capacity{\site} = 10$.  For completeness, in
\cref{fig:performance:absolute} we show the times for all four steps
of \prop bidding, ignoring time attributable to Amnesia.

While \prop bidding is very efficient, we highlight that bidding and
monitoring-request deployment are \textit{not} the most important
performance costs in a breach-detection ecosystem, as they are not
costs that the adversary can induce (in our threat model).  In
contrast, the adversary \textit{can} induce the generation of
monitoring responses by making login attempts, though our bidding
algorithm plays no role in these costs; we refer the reader to Wang et
al.~\cite[Sec.~6.5]{Wang:2021:Amnesia} for a discussion of these costs
for Amnesia.

\section{Cit0day Data Exploration}
\label{sec:cit0day}

The processed Cit0day dataset includes 74,268,368 users across 7,914
sites and 53,241,884 unique passwords. Site sizes vary widely (see
\cref{fig:cit0day:cdf:totalusers}). However, a site's popularity does
not directly translate to greater risk. From the site's own
perspective, the number of users it shares with other sites—its only
observable signal of stuffing risk—correlates only weakly with site
size (r = 0.197; \cref{fig:cit0day:totalVsShared}). This suggests that
larger sites do not necessarily have a proportionally higher number of
users with accounts elsewhere. In contrast, the actual number of
vulnerable users (those who reuse passwords across sites) shows only a
moderate correlation with site size (r = 0.610;
\cref{fig:cit0day:totalVsCaptureable}), indicating that overall risk
does not scale directly with popularity.

\begin{wrapfigure}{r}{0.3\columnwidth}
  \vspace{-2ex}
  \centering
  \resizebox{0.29\columnwidth}{!}{
    \input{reuse_rates.tex}
  }
  \caption{Password reuse rate}
  \label{fig:cit0day:cdf:reuse}
  \vspace{-2ex}
\end{wrapfigure}

\Cref{fig:cit0day:cdf:totalusers,fig:cit0day:cdf:sharedusers} show
heavy-tailed distributions: most sites are small, and most site pairs
share few users. Still, password reuse is rampant among users with
multiple accounts. Among the 9.3\% of site pairs that share users,
reuse is nearly universal; the median reuse rate in
\cref{fig:cit0day:cdf:reuse} is 94.5\%.

\begin{figure}[htbp]
\begin{subfigure}[b]{0.48\columnwidth}
  \centering
  \resizebox{\textwidth}{2.5cm}{
    \input{total_vs_captureable.tex}
  }
  \caption{Total vs.\ capturable users. We randomly sample 1,000 sites to illustrate trends clearly.}
  \label{fig:cit0day:totalVsCaptureable}
\end{subfigure}
\begin{subfigure}[b]{0.48\columnwidth}
  \centering
  \resizebox{\textwidth}{2.5cm}{
    \input{total_vs_shared.tex}
  }
  \caption{Total vs.\ shared users. We randomly sample 1,000 sites to illustrate trends clearly.}
  \label{fig:cit0day:totalVsShared}
\end{subfigure}
\\
\begin{subfigure}[b]{0.48\columnwidth}
  \centering
  \resizebox{\textwidth}{2.25cm}{
    \input{users_per_site.tex}
  }
  \caption{Users per site}
  \label{fig:cit0day:cdf:totalusers}
\end{subfigure}
\begin{subfigure}[b]{0.48\columnwidth}
  \centering
  \resizebox{\textwidth}{2.25cm}{
    \input{shared_users.tex}
  }
  \caption{Shared users per site}
  \label{fig:cit0day:cdf:sharedusers}
\end{subfigure}
\caption{Exploration of Cit0day Dataset}
\label{fig:cit0day}
\end{figure}

\section{Model Checking Implementation}
\label{sec:model_checking}

In the inter-organization model (\cref{sec:auction}), \targetSite's
goal is to minimize its risk (\cref{eqn:risk}). Assuming the
\targetSite's credential database is breached, the attacker's
objective (\cref{sec:attacker}) is to harvest as many users as
possible by stuffing the \targetSite's compromised passwords at its
peers. We model each of these interactions as Markov Decision
Processes (MDPs) and use probabilistic model checking to analyze them,
providing a worst-case assessment of \targetSite's security.

Probabilistic model checking requires exhaustively searching the
entire state space, which imposes significant computational and memory
constraints. As a result, off-the-shelf tools like PRISM
\cite{Kwiatkowska:2011:Prism} struggle to scale to the number of sites
and users in our analysis
(\cref{sec:auction:eval,sec:attacker:eval,sec:attacker:larger}). To
address this, we developed custom model checkers in Python with
application-specific optimizations, detailed below. We will release
our model checker artifacts upon paper acceptance. We validated our
implementations by comparing results from small-scale
experiments---limited in site and user count and excluding
combinations of parameters that PRISM cannot handle---against PRISM's
output.

\subsection{Inter-organization Model Implementation}
\label{sec:model_checking:sites}

A state in the inter-organization model consists of (1) the bid
number, (2) the average number of slots each site has received from
its peers, (3) the latest bid each site has placed for \targetSite. If
$\cutLine{\targetSite} = \boolTrue$, we additionally store the number
of bids placed by $\site \in \peerSet$, since this determines which of
\targetSite's peers is slated to bid next. Additionally, if $\slack <
\infty$,  we maintain an \nmbrSites-sized array that tracks how many
more bids ahead each site has placed relative to its peers; each
element in this array is bounded by \slack.

To implement the \exhaustive strategy when it bids, the
\targetSite[\exhaustiveLabel] computes the allocation that minimizes
\risk{\targetSite} over \foresight{\targetSite[\exhaustiveLabel]} +
\lookahead{\targetSite[\exhaustiveLabel]} future steps. Our
model-checker uses breadth-first search (BFS) to build a tree of this
depth, with each node stores the state, as described above. Within the
\foresight{\targetSite[\exhaustiveLabel]} depth, child nodes
correspond to the next bidder specified in the bidding sequence and,
if $\cutLine{\targetSite[\exhaustiveLabel]} = \boolTrue$, the
\targetSite assuming it had not bid previously. Beyond
\foresight{\targetSite[\exhaustiveLabel]}, nodes branch on all valid
next bidders: if $\cutLine{\targetSite[\exhaustiveLabel]} =
\boolFalse$, this includes every $\site \in \siteSet$; if
$\cutLine{\targetSite[\exhaustiveLabel]} =  \boolTrue$ it includes
every $\site \in \peerSet$, and $\targetSite[\exhaustiveLabel]$,
provided it was not the previous bidder. When
\targetSite[\exhaustiveLabel] bids, nodes further branch on its
possible allocations; to control state explosion, our implementation
only allows \targetSite[\exhaustiveLabel] to allocate
\capacity{\targetSite} in fixed, tunable increments.

We cap the tree depth at $\foresight{\targetSite[\exhaustiveLabel]} +
\lookahead{\targetSite[\exhaustiveLabel]}$ to reflect the realistic
assumption that \targetSite cannot predict bids indefinitely, and to
bound memory usage during BFS. Since PRISM lacks support to easily
restrict tree depth below auction length, \nmbrBids, we validate our
model checker against PRISM only in cases where
$\foresight{\targetSite[\exhaustiveLabel]} +
\lookahead{\targetSite[\exhaustiveLabel]} = \nmbrBids$. 

After building the tree, we back-propagate optimal bids and cumulative
risk from leaves to root, yielding \targetSite[\exhaustiveLabel]'s
best move. During this process, we cache the optimal rewards and
policies of nodes at depths
(\foresight{\targetSite[\exhaustiveLabel]},
\foresight{\targetSite[\exhaustiveLabel]} +
\lookahead{\targetSite[\exhaustiveLabel]}) for reuse in later bids or
bidding sequences, assuming all other parameters remain fixed.

\subsection{Attacker Model Implementation}
\label{sec:model_checking:attacker}

A state in the attacker model consists of (1) the bid number, (2) each
$\site \in \peerSet$ allocation to \targetSite for that round, (3) the
attacker's current stuffing strategy—i.e., which users in
\userSet[\targetSite] are stuffed and at which $\site \in \peerSet$,
and (4) the attacker's cumulative dodge probability.

To implement worst-case attacker, our model checker  uses depth-first
search (DFS) to build a tree of depth \foresight{\attacker} +
\lookahead{\attacker}, enumerating all attacker strategies feasible
within \aggressionLevel{\attacker}. We use DFS, in lieu of BFS, due to
memory-constraints imposed by the explosive number of attacker
strategies. Within the \foresight{\attacker} depth, the model checker
generates attacker stuffing strategies assuming allocations
corresponding to the next bidder specified in the bidding sequence.
Beyond \foresight{\attacker}, the model checker also branches on the
possible next bidders, as described in
\cref{sec:model_checking:sites}, generating attacker strategies that
satisfies \aggressionLevel{\attacker} assuming the largest number of
slots due to possible bids. To reduce the number of nodes generated in
the tree, we cache nodes with few stuffing attempts, as these are
likely to satisfy the \aggressionLevel{\attacker} and thus be
regenerated at multiple tree depths. For fixed $\allocationVal =
\allocTo{\targetSite}{\site}$, we also bound stuffing attempts per
site using the unimodal optimization in
\cref{sec:performance:optimization:unimodal}.

Still, enumerating the attacker's stuffing strategies to depth
$\foresight{\attacker} + \lookahead{\attacker}$ is computationally
prohibitive. To this end, we underspecify the attacker's stuffing
strategy by only the number of stuffing attempts per $\site \in
\peerSet$. Not all such underspecified strategies correspond to a
valid set of uncaptured users in \userSet[\targetSite], so we
incrementally filter out invalid strategies as determined by a
constraint solver. 

We backpropagate from tree leaves to root, to compute the optimal
underspecified strategy—per-site stuffing counts—over the
\foresight{\attacker}+\lookahead{\attacker} depth. To concretize this
strategy with valid users, we retroactively search for a subset of
uncaptured users in \userSet[\targetSite] that satisfies the strategy.
To narrow this recursive search, we observe an attacker prioritizes
stuffing users with fewer accounts to preserve future stuffing
options.

Even enumerating underspecified strategies becomes computationally
prohibitive as the number of sites and users grows, since suboptimal
per-site stuffing counts (at a given depth) may yield high payoff
later if \allocTo{\targetSite}{\site} decreases at deeper levels.
However, when $\foresight{\attacker} + \lookahead{\attacker} = 1$, as
in \cref{sec:attacker:larger}, the attacker's goal is to immediately
maximize expected user captures. In this case, our model checker
iterates over site indices to find the maximal underspecified per-site
stuffing strategy according to
\cref{sec:performance:optimization:unimodal}, subject to
\aggressionLevel{\attacker}. Since this maximal strategy may not
correspond to a valid subset of uncaptured users in
\userSet[\targetSite], we track both (1) the best valid strategy found
so far and (2) invalid strategies that yield higher payoff but
currently violate constraints. These invalid strategies act as upper
bounds that may later become valid. If no such invalid strategies
remain, we conclude the optimal per-site stuffing strategy has been
found.

\section{Optimizations}
\label{sec:performance:optimization}
To place a bid according to \ref{alg:prop-rsp:next-bid}, \site must
determine \avgRisk{\site}{\siteAlt} from all $\siteAlt \in
\peerSet[\site]$. However, calculating \avgRisk{\site}{\siteAlt}
involves computing the optimal number \stuffingAttempts of stuffing
attempts by an attacker which maximizes \cref{eqn:risk} assuming $\allocationVal = \allocTo{\siteAlt}{\site}$. A naive
implementation could iterate over all the possible values of
$\stuffingAttempts$, which is upper-bounded by $\intersectionSize =
\setSize{\userSet[\site] \cap \userSet[\siteAlt]}$.  However, this
will quickly be computationally prohibitive as the number of users
increases. Instead, we observe \cref{eqn:risk} is unimodal, and the
optimal \stuffingAttempts is either $\lfloor(\intersectionSize -
\min\{\intersectionSize, \allocationVal\}) / (\min\{\intersectionSize,
\allocationVal\} + 1)\rfloor$ or $\lceil(\intersectionSize -
\min\{\intersectionSize, \allocationVal\}) / (\min\{\intersectionSize,
\allocationVal\} + 1)\rceil$ in the \PSILabel setting and either
$\lfloor(\setSize{\userSet[\siteAlt]} -
\min\{\setSize{\userSet[\siteAlt]}, \allocationVal\} ) /
(\min\{\setSize{\userSet[\siteAlt]}, \allocationVal\} + 1)\rfloor$ or
$\lceil(\setSize{\userSet[\siteAlt]} -
\min\{\setSize{\userSet[\siteAlt]}, \allocationVal\} ) /
(\min\{\setSize{\userSet[\siteAlt]}, \allocationVal\} + 1)\rceil$ in the
\PSICALabel setting. \Cref{sec:performance:optimization:unimodal}
contains the proof.

Despite reducing the number of calculations to compute
\cref{eqn:risk}, evaluating \avgRisk{\site}{\siteAlt} still scales
with $O(\setSize{\userSet[\site]})$. To compute it in constant time,
we use Stirling's approximation of log factorials.
\Cref{sec:performance:optimization:error} provides the error terms.
Combined, these two optimizations let us estimate \cref{eqn:risk} in
$O(1)$ time, implying \avgRisk{\site}{\siteAlt} in
\ref{alg:prop-rsp:next-bid} is also calculated in $O(1)$ time.
Therefore, computing a bid according to \ref{alg:prop-rsp:next-bid}
only depends on the number of bids that \site is receiving, which
scales with the number of sites, $O(\nmbrSites)$.

\subsection{\cref{eqn:risk} is unimodal}
\label{sec:performance:optimization:unimodal}
We prove the \PSILabel case; \PSICALabel is analogous.
\Cref{eqn:risk} in the \PSILabel setting is:
\begin{equation}
\risk{\site}[\siteAlt][\allocationVal] = \max_{0 \leq \stuffingAttempts \leq \intersectionSize}  \stuffingAttempts \times  \frac{{\intersectionSize - \stuffingAttempts \choose \minSlotIntersectionVal}}{{\intersectionSize \choose \minSlotIntersectionVal}}
\end{equation}
where $\minSlotIntersectionVal = \min\{\intersectionSize, \allocationVal\}$.
To find the maximum, we check when $\expv{\attackerGainRV{\stuffingAttempts + 1}{\allocationVal}} > \expv{\attackerGainRV{\stuffingAttempts}{\allocationVal}}$:
\begin{align}
\frac{\expv{\attackerGainRV{\stuffingAttempts + 1}{\allocationVal}}}{\expv{\attackerGainRV{\stuffingAttempts}{\allocationVal}}} > 1 &\iff \frac{\stuffingAttempts+1}{\stuffingAttempts} \cdot \frac{{\intersectionSize - \stuffingAttempts - 1 \choose \minSlotIntersectionVal}}{{\intersectionSize - \stuffingAttempts \choose \minSlotIntersectionVal}} > 1 \\
&\iff \frac{\stuffingAttempts+1}{\stuffingAttempts} \cdot \frac{\intersectionSize-\stuffingAttempts-\minSlotIntersectionVal}{\intersectionSize-\stuffingAttempts} > 1 \\
&\iff \stuffingAttempts < \frac{\intersectionSize -\minSlotIntersectionVal}{\minSlotIntersectionVal+1}
\end{align}

Therefore, the maximum occurs at $\lfloor(\intersectionSize - \minSlotIntersectionVal) / (\minSlotIntersectionVal + 1)\rfloor$ or $\lceil(\intersectionSize - \minSlotIntersectionVal) / (\minSlotIntersectionVal + 1)\rceil$.

\subsection{Stirling approximation of log factorial error term}
\label{sec:performance:optimization:error}

We derive the error bounds for approximating \cref{eqn:psiCatchProb}
(\PSILabel); the derivation for \cref{eqn:psiCaCatchProb}
(\PSICALabel) is analogous. When $\allocationVal > \intersectionSize$
and $\stuffingAttempts = 0$, \cref{eqn:psiCatchProb} evaluates to 1.
When $\allocationVal > \intersectionSize$ and $\stuffingAttempts > 0$,
\cref{eqn:psiCatchProb} evaluates to 0. The interesting case is when
$\allocationVal \leq \intersectionSize$. We have:
\begin{align}
  \frac{{\intersectionSize - \stuffingAttempts \choose \allocationVal}}{{\intersectionSize \choose \allocationVal}}
  &= \exp\left(\begin{array}{@{}r@{}}
         \ln((\intersectionSize-\stuffingAttempts)!) + \ln((\intersectionSize-\allocationVal)!) \\
         - \ln((\intersectionSize-\stuffingAttempts-\allocationVal)!) - \ln(\intersectionSize!)
     \end{array}\right)
\label{eqn:chooseRatio}
\end{align}

Let $\StirlingApprox$ denote the value of \cref{eqn:chooseRatio}.
Using Stirling's approximation $\ln(\stirlingTerm!) \approx
\stirlingTerm\ln(\stirlingTerm) - \stirlingTerm +
\frac{1}{2}\ln(2\pi\stirlingTerm) + \varepsilon_\stirlingTerm$ with
Robbins bounds $\frac{1}{12\stirlingTerm+1} <
\varepsilon_\stirlingTerm < \frac{1}{12\stirlingTerm}$
\cite{Robbins:1955:Remark}, we can approximate \cref{eqn:chooseRatio}
by $\StirlingApprox \cdot \exp(\ErrorTerm)$ where:

\begin{equation}
\ErrorTerm = \varepsilon_{\intersectionSize-\stuffingAttempts} + \varepsilon_{\intersectionSize-\allocationVal} - \varepsilon_{\intersectionSize-\stuffingAttempts-\allocationVal} - \varepsilon_{\intersectionSize}
\end{equation}

The error bounds are:
\begin{align}
\ErrorTerm[\mathrm{lower}] &= \frac{1}{12(\intersectionSize-\stuffingAttempts)+1} + \frac{1}{12(\intersectionSize-\allocationVal)+1} \notag \\
&\quad - \frac{1}{12(\intersectionSize-\stuffingAttempts-\allocationVal)} - \frac{1}{12\intersectionSize} \\
\ErrorTerm[\mathrm{upper}] &= \frac{1}{12(\intersectionSize-\stuffingAttempts)} + \frac{1}{12(\intersectionSize-\allocationVal)} \notag \\
&\quad - \frac{1}{12(\intersectionSize-\stuffingAttempts-\allocationVal)+1} - \frac{1}{12\intersectionSize+1}
\end{align}

Therefore:
\begin{equation}
\frac{{\intersectionSize - \stuffingAttempts \choose \allocationVal}}{{\intersectionSize \choose \allocationVal}} \in \left[\StirlingApprox \cdot \exp(\ErrorTerm[\mathrm{lower}]), \StirlingApprox \cdot \exp(\ErrorTerm[\mathrm{upper}])\right]
\end{equation}

\section{Challenges in Predicting Cost}
\label{sec:discussion:challenges}

In adapting proportional response from other domains
(e.g.,~\cite{Levin:2008:BitTorrent, Wu:2007:PropResponse,
Branzei:2021:Proportional}) to our setting, we considered many
variants of \cref{eqn:risk} and the \prop strategy
(\cref{alg:prop-rsp-bid,alg:prop-rsp-receive-psi,alg:prop-rsp-receive-psica}).
Below, we summarize two of these versions and outline why did not
prefer them.

\begin{itemize}[nosep,leftmargin=1em,labelwidth=*,align=left]
\item \normRiskPerBid{\targetSite}{\bidIdx} is a less accurate
  predictor of \normCostPerBid{\targetSite}{\bidIdx} for sites \targetSite that
  are popular (see \cref{sec:attacker:eval}).  In this case, the
  overlap between $\userSet[\targetSite] \cap \userSet[\site]$ and
  $\userSet[\targetSite] \cap \userSet[\siteAlt]$ for other sites
  \site, \siteAlt tends to be large, meaning that computing
  \normRiskPerBid{\targetSite}{\bidIdx} as a sum of
  \normRisk{\targetSite}{\site}{\allocationVal} for $\site \in
  \peerSet[\targetSite]$ (see \cref{eqn:normRisk}) ``double counts''
  the large number of users in that overlap.  This double counting
  might be avoided by calculating
  \normRiskPerBid{\targetSite}{\bidIdx} holistically, using the
  portfolio of allocations $\{\allocTo{\site}{\targetSite}\}_{\site
  \in \peerSet[\targetSite]}$, versus as a simple sum of per-site
  contributions.  While such a ``portfolio'' approach is possible (at
  least in the \PSILabel case), we nevertheless found that attributing
  risk to each peer individually (i.e.,
  \normRisk{\targetSite}{\site}{\allocTo{\site}{\targetSite}} is
  particularly useful because the lever available to incentivize a
  peer is adjusting its individual allocation; i.e., it is useful to
  be able to assign blame individually, so that we can incentivize
  each peer individually.

  \item We explored various other measures to predict
  \normCostPerBid{\targetSite}{\bidIdx}, besides
  \normRiskPerBid{\targetSite}{\bidIdx}.  Most were measures of
  \textit{utility}, expressed as a function of desired allocations of
  slots from each $\site \in \peerSet[\targetSite]$ (itself computed
  using \setSize{\userSet[\targetSite] \cap \userSet[\site]}) and the
  actual allocation of slots from \site.  Measures of utility that
  grow linearly the allocation from \site suffered from the fact that
  incrementing (or decrementing) allocations tended to affect
  \normCostPerBid{\targetSite}{\bidIdx} much more (respectively, less) if the
  allocation was already small (respectively, large).  Nonlinear
  functions that we considered introduced additional tuning parameters
  that we found difficult to fit to the myriad other parameter
  settings we explored in our model-checking experiments.
\end{itemize}

An important direction for future work is therefore to refine
\normRiskPerBid{\targetSite}{\bidIdx} to better predict
\normCostPerBid{\targetSite}{\bidIdx} when \targetSite is popular.
Perhaps an even greater challenge is to improve
\normRiskPerBid{\targetSite}{\bidIdx} to better predict
\normCostPerBid{\targetSite}{\bidIdx} in the \PSICALabel setting or
when \aggressionLevel{\attacker} is low. Both scenarios involve an
inherent information imbalance: in the former, \targetSite lacks
knowledge of shared users across peers, while in the latter, it lacks
insight into the attacker's aggression.

\begin{table*}[p]
\centering
\caption{Notation}
\label{tab:notation}
\footnotesize
\begin{tabular}{@{}p{2em}@{\hspace{1em}}p{8.5em}@{\hspace{1em}}p{0.8\textwidth}@{}}
\toprule

& \textbf{Notation} & \textbf{Meaning} \\

\midrule

\multirow{5}{*}{\rotatebox[origin=c]{90}{User}} 
& $\user$ & A user\\

& $\nmbrUsers$ & Number of users across all sites \\

& $\intersectionSize$ & Number of users shared between two sites \\

& $\vulnSet[\targetSite]$ & Users that have accounts at $\targetSite$
that are vulnerable to stuffing attempts at any peer site $\site$ \\

& $\userSet[\site]$ & Set of users with accounts at site $\site$ \\

\midrule

\multirow{7}{*}{\rotatebox[origin=c]{90}{Site}} 
& $\site$ & A site \\

& $\targetSite$ & A designated site for which we compare \exhaustive
bidding (\targetSite[\exhaustiveLabel]) and \prop bidding
(\targetSite[\propLabel]) in \cref{sec:auction}, and then whose
accounts the attacker attempts to harvest (after breaching
\targetSite, and while while \targetSite uses \prop bidding) in
\cref{sec:attacker} \\

& $\capacity{\site}$ & Capacity (monitoring slots) of site \site \\

& $\siteSet[\user]$ & Set of sites where $\user$ has an account \\

& $\allocTo{\site}{\siteAlt}$ & Slots allocated by site $\site$ to
$\siteAlt$ \\

& $\allocationVal$ & Slots received from peer in given bid \\

\midrule

\multirow{11}{*}{\rotatebox[origin=c]{90}{Auction}} 
& $\nmbrSites$ & Number of bidding sites \\

& $\nmbrBids$ & The number of bids in an auction \\

& $\bidIdx$ & The bid index into a bidding sequence \\

&$\firstbidIdx$ (resp., $\secondbidIdx$) & The bid index at which a site
first issues a \ref{alg:prop-rsp:next-bid} bid (resp., the bid index at
which a site first issues a second \ref{alg:prop-rsp:next-bid} bid.)
\\

& $\slack$ & Maximum number by which one bidder's number of bids can
exceed another's \\

& $\popularity$ & Controls set account distribution; in particular,
$\popularity = 1$ indicates \targetSite is most popular \\

& $\capacityCoeff{\site}$ & Capacity scaling coefficient;
$\capacity{\site} = \capacityCoeff{\site} \times
\setSize{\userSet[\site]}$ \\

& $\pairedAuctionsSet$ & The set of auction pairs where all else held
equal, $\targetSite$ uses the \exhaustive strategy in one and
\prop strategy in the other \\

& $\pairedAuctionsSet[\configVar]$ & The subset of auction pairs in
\pairedAuctionsSet that satisfy constraint $\configVar$ \\

& $\auction[\propLabel]$ & An auction where $\targetSite$ uses the
\prop bidding strategy \\

& $\auction[\exhaustiveLabel]$ & An auction where $\targetSite$ uses
the \exhaustive bidding strategy \\

& $\constrainedConfig$ & A configuration where $\cutLine{\targetSite}$
is $\boolFalse$, $\foresight{\targetSite}$ is 0 and $\slack$ is
unbounded \\

\midrule

\multirow{6}{*}{\rotatebox[origin=c]{90}{\parbox{10ex}{\centering
      Exhaustive Bidding}}} \\
& $\cutLine{\targetSite}$ & If \boolTrue, then \targetSite can choose
when to bid \\

& $\foresight{\targetSite}$ & The number of forthcoming bidders site
$\targetSite$ can predict \\

& $\lookahead{\targetSite}$ & The depth beyond
$\foresight{\targetSite}$ for which \targetSite examines possible
bidding sequences \\

& ~ \\

\midrule

\multirow{6}{*}{\rotatebox[origin=c]{90}{\parbox{10ex}{\centering
      Proportional Bidding}}}
& $\smoothingFactor{\site}$ & Factor for calculating
$\avgAllocFrom{\site}{\siteAlt}$ \\

& $\avgAllocFrom{\site}{\siteAlt}$ & Average slots $\siteAlt$ provided
to $\site$ \\

& $\baselineWeight{\site}{\siteAlt}$ & Assuming each peer allocates
one slot to \site, one minus the fraction of the total risk to \site
attributable to \siteAlt \\

& $\avgRisk{\site}{\siteAlt}$ & Average risk from $\siteAlt$ to
$\site$ \\

& $\maxRisk$ & Maximum average risk over all peers \\

& $\weight{\site}{\siteAlt}$ & One minus the fraction of the total
average risk to \site attributable to \siteAlt \\

\midrule

\multirow{5}{*}{\rotatebox[origin=c]{90}{Attacker}}
& $\dodgeEvent{\site}{\siteAlt}{\allocationVal}{\stuffingAttempts}$ &
Event that none of the $\stuffingAttempts$ stuffing attempts at site
$\siteAlt$ for site $\site$'s users were detected, assuming $\siteAlt$
provided $\allocationVal$ slots to \site \\

& $\stuffingAttempts$ & Number of accounts stuffed at a given site and
bid \\

& $\foresight{\attacker}$ & The number of forthcoming bidders the
attacker $\attacker$ can predict \\

& $\lookahead{\attacker}$ & The depth beyond $\foresight{\attacker}$
for which \attacker examines possible bidding sequences \\

& $\aggressionLevel{\attacker}$ & Thresholds the attacks an attacker
will launch according to its cumulative probability of dodging
detection \\

\midrule

\multirow{20}{*}{\rotatebox[origin=c]{90}{Measures}}
& $\risk{\site}[\siteAlt][\allocationVal]$ & The risk site $\site$
incurs due to the allocation $\allocationVal$ provided by site
$\siteAlt$. See \cref{eqn:risk} \\

& $\riskPerBid{\site}{\bidIdx}$ & The risk site $\site$ incurs from
all $\siteAlt$ at bid index
$\bidIdx$. See \cref{eqn:perBidRisk} \\

& $\normRisk{\site}{\siteAlt}{\allocationVal}$ & The risk site $\site$
incurs due to the allocation $\allocationVal$ provided by site
$\siteAlt$, normalized by the number of shared users between $\site$
and $\siteAlt$. See \cref{eqn:normRisk}\\

& $\normRiskPerBid{\site}{\bidIdx}$ & The risk site $\site$ incurs at
bid index $\bidIdx$ from all $\siteAlt$ normalized by the number of shared
users between $\site$ and $\siteAlt$. See \cref{eqn:perBidNormRisk} \\

& $\improvementFreq{\configVar}$ & The fraction of bid-pairs where,
all else held equal, the auction with an \exhaustive $\targetSite$
incurred less risk than the auction with the \prop
$\targetSite$. See \cref{eqn:improvement-freq} \\

& $\absImprovement{\configVar}$ & The median difference in risk
incurred per bid between an auction with an \prop $\targetSite$
and an auction with an \exhaustive $\targetSite$, all else held equal.
See \cref{eqn:abs-improvement}
\\

& $\fracImprovement{\configVar}$ & The median relative increase in
risk per bid between an auction with an \prop $\targetSite$ and
an auction with an \exhaustive $\targetSite$, all else held equal. See \cref{eqn:frac-improvement} \\

& $\costPerBid{\targetSite}{\bidIdx}$ & At bid index $\bidIdx \ge
\firstbidIdx$, the expected number of users at $\targetSite$ for which
an attacker can undetectably stuff some account at a peer site, given
the most recent allocations to \targetSite received in bids prior to
$\bidIdx$, subject to the attacker's aggression value \\

& $\normCostPerBid{\targetSite}{\bidIdx}$ & At bid index $\bidIdx \ge
\firstbidIdx$, the expected number of users at $\targetSite$ for which
an attacker can undetectably stuff some account at a peer site, given
the most recent allocations to \targetSite received in bids prior to
$\bidIdx$, subject to the attacker's aggression value and normalized
by the number of vulnerable users at \targetSite. See
\cref{eqn:normCostPerBid} \\

& $\cost{\targetSite}$ & The expected number of users at $\targetSite$
for which an attacker can undetectably stuff at a peer site, assuming
the attacker starts stuffing at bid index $\bidIdx \ge \secondbidIdx$
over a window of \nmbrBids bids. See \cref{eqn:totalCost} \\

& $\normCost{\targetSite}$ &  The expected number of users at
$\targetSite$ that an attacker can undetectably stuff at a peer site,
assuming the attacker starts stuffing at bid index $\bidIdx \ge
\secondbidIdx$ over a window of \nmbrBids bids, normalized by the
number of vulnerable users at $\targetSite$. See
\cref{eqn:normRelCost} \\

\midrule

& \runtime{op} & Run time of operation ``op'' \\

& \pValue & Statistical significance level \\

\bottomrule
\end{tabular}
\end{table*}

\end{document}